\documentclass[aps,prx,twocolumn,reprint, superscriptaddress,nofootinbib,notitlepage]{revtex4-1}
\usepackage{graphicx}
\usepackage{amsmath}
\usepackage{amstext}
\usepackage{amssymb}
\usepackage[colorlinks,citecolor=blue]{hyperref}
\usepackage{graphicx}
\usepackage{amsmath}
\usepackage{amstext}
\usepackage{longtable}
\usepackage{amssymb}
\usepackage{amsfonts}
\usepackage{longtable,booktabs}
\usepackage{hyperref}\usepackage{url}%Turn on when output to pdf file
\usepackage{subfigure}%
\usepackage{dsfont}

\usepackage{amsbsy}
\usepackage{dcolumn}
\usepackage{amsthm}
\usepackage{bm}
\usepackage{esint}
\usepackage{multirow}
\usepackage{hyperref}
\hypersetup{
    colorlinks=true,
    linkcolor=blue,
    filecolor=magenta,
    urlcolor=cyan,
}
\usepackage{cleveref}

\usepackage{mathrsfs}
\usepackage{amsfonts}
\usepackage{amsbsy}
\usepackage{dcolumn}
\usepackage{bm}
\usepackage{multirow}
\usepackage{color}
\def\Z{\mathbb{Z}}

  \usepackage{extarrows}
 \usepackage{graphicx}
    \usepackage{amssymb}
    \usepackage{amsthm}
    \usepackage{mathtools}
    \usepackage{adjustbox}

\newcommand{\beq}{\begin{eqnarray} }
\newcommand{\eeq}{\end{eqnarray} }
\newcommand{\Beq}{\begin{eqnarray*} }
\newcommand{\Eeq}{\end{eqnarray*} }
\newcommand{\Bmat}{\left(\begin{matrix}}
\newcommand{\Emat}{\end{matrix}\right)}

\usepackage{datetime}

\usepackage{float}

      \theoremstyle{remark}
      
\begin{document}
 
  \title{{\fontfamily{ptm}\fontseries{sb}\selectfont    Symmetry Enrichment  in Three-Dimensional Topological Phases}}
 
\author{Shang-Qiang Ning}
   \affiliation{Institute for Advanced Study, Tsinghua University, Beijing, China, 100084}
   \author{Zheng-Xin Liu}
   \affiliation{Department of Physics, Renmin University of China, Beijing, China, 100872}

 \author{Peng Ye}
\email[ Corresponding author: ]{yphysics@illinois.edu}
    \affiliation{Department of Physics, University of Illinois at Urbana-Champaign, IL 61801, USA}
  \affiliation{Institute for Condensed Matter Theory, University of Illinois at Urbana-Champaign, IL 61801, USA}

   \begin{abstract}  
While two-dimensional symmetry-enriched topological phases ($\mathsf{SET}$s) have been studied  intensively and systematically, three-dimensional ones are still open issues. We propose an algorithmic approach of  imposing  global symmetry $G_s$ on gauge theories (denoted by $\mathsf{GT}$) with gauge group $G_g$.  The resulting symmetric gauge theories   are dubbed ``symmetry-enriched gauge theories'' ($\mathsf{SEG}$), which may be served as low-energy effective   theories of three-dimensional symmetric topological quantum spin liquids. We focus on $\mathsf{SEG}$s with gauge group $G_g=\Z_{N_1}\times\Z_{N_2}\times\cdots$ and on-site unitary symmetry group $G_s=\Z_{K_1}\times\Z_{K_2}\times\cdots$ or $G_s=\mathrm{U(1)}\times \Z_{K_1}\times\cdots$.   Each $\mathsf{SEG}(G_g,G_s)$ is described in the path integral formalism associated with certain symmetry assignment. From the path-integral expression, we propose how to physically diagnose the ground state properties (i.e., $\mathsf{SET}$ orders) of $\mathsf{SEG}$s in  experiments of charge-loop braidings (patterns of symmetry fractionalization) and the \emph{mixed} multi-loop braidings among deconfined loop excitations and confined symmetry fluxes. From these symmetry-enriched properties, one can obtain  the map from $\mathsf{SEG}$s to $\mathsf{SET}$s.  By giving full dynamics to background gauge fields, $\mathsf{SEG}$s may be eventually promoted to a set of new gauge theories (denoted by $\mathsf{GT}^*$). Based on their gauge groups,   $\mathsf{GT}^*$s may be further regrouped into different classes each of which is labeled by a gauge group ${G}^*_g$. Finally, a web of gauge theories involving $\mathsf{GT}$, $\mathsf{SEG}$, $\mathsf{SET}$ and $\mathsf{GT}^*$ is achieved. We demonstrate the above symmetry-enrichment physics and the web of gauge theories through many concrete examples.
    \end{abstract}
%\date{{\currenttime, \small\today} }
%\date{{ \small\today}}
  \maketitle
%\tableofcontents 
\section{Introduction}\label{sec:introduction_main}

Recently, the field of gapped phases with symmetry has been drawing a lot of attentions in condensed matter physics. There are two kinds of symmetric gapped phases: symmetry-protected topological phases ($\mathsf{SPT}$) and symmetry-enriched topological phases ($\mathsf{SET}$). $\mathsf{SPT}$ phases are short-range entangled \cite{Chenlong} with a global symmetry and have been studied intensively in strongly-correlated bosonic systems \cite{1DSPT,Chenlong,pollmann2010,Chen_science,Chen10,LV1219,spt1,spt2,spt3,spt4,spt5,spt5.5,spt6,spt7,spt7.5,spt8,spt9,spt10,spt11,spt12,spt13,spt14,spt15,spt16,spt17,spt18,spt19,spt20,jiang_ran04,wang_wen_3loop,spt21,spt22,spt23,hehuan2016,wwh2015}.   Much progress has also been made in two-dimensional (2D) $\mathsf{SET}$s  \cite{Wen2002,Levin2012,Essin2013,Mesaros2013,Gu2014,Hung2013,Lu2013,Barkeshli14arxiv,heinrich,set1,set2,set3}, which  are  partially driven by tremendous efforts in  quantum spin liquids (QSL) \cite{balents_review,Wen2002} that respect a certain global symmetry (e.g., spatial reflection, time-reversal, Ising $\Z_2$, $\mathrm{U(1)}$ and $\mathrm{SU(2)}$ spin rotations, etc.).      In contrast to $\mathsf{SPT}$s, $\mathsf{SET}$s are long-range entangled \cite{Chenlong} and  support emergent excitations, such as  anyons in 2D systems.  Furthermore, quantum numbers  carried by emergent excitations may be fractionalized.  Experimentally, it is   of interest to   detect  patterns of  such symmetry fractionalization, which may help us   characterize   QSLs \cite{balents_review}. 
In addition to the usual global symmetry, there are also $\mathsf{SET}$s enriched by a new kind of symmetry dubbed  ``topological (anyonic)'' symmetry \cite{Kitaev2006,Teo2015,Teo2014,ran,barkeshli_wen,Teo2013,Barkeshli14arxiv,You2012,genon_1,genon_2,genon_3,genon_4,Bombin2010}. This symmetry denotes an automorphism of the topological data (braiding statistics, quantum dimensions, etc.). A typical example is that $\Z_2$ topological order in two dimensions is invariant under $e$-$m$ exchange operation, namely, an electric-magnetic duality in discrete gauge theories \cite{Kitaev2006,Bombin2010}. 

Despite much success in 2D $\mathsf{SET}$s,  three-dimensional (3D) $\mathsf{SET}$ physics, especially the underlying general framework,  is still poorly understood so far, partially due to the presence of spatially extended loop excitations \cite{Kong_wen}.   In physical literatures, some attempts  have been made, including 3D $\mathrm{U(1)}$ QSLs and $\Z_2$ QSLs with symmetry, e.g., in Ref.~\cite{3dset_wang,3dset_xu,gangchen_2014,kimchi_14}.  Field theories of 3D $\mathsf{SET}$s with either time-reversal   or  $180^{\circ}$ spin rotation about $y$-axis were studied  where the dynamical axion electromagnetic action term  is considered \cite{spt10,witten1}.  The boundary anomaly of some 3D $\mathsf{SET}$s was viewed as   2D anomalous $\mathsf{SET}$s with   anyonic symmetry \cite{3dset_fidkowski}. In Ref.~\cite{3dset_cheng,3dset_chen}, a dimension reduction point of view was proposed to   demonstrate how symmetry is fractionalized on  loop excitations. In Ref.~\cite{3dset_ye}, the notion of ``2D anyonic symmetry'' was generalized to 3D ``charge-loop excitation symmetry''  ($\mathsf{Charles}$) which is a permutation operation among  particle excitations  and among loop excitations. As   typical examples of 3D  $\mathsf{SET}$s with $\mathrm{U(1)}$ and time-reversal, fractional topological insulators   were   constructed via a parton construction with   gauge confinement \cite{3dset_ye}.

\begin{figure}[t]
\centering
\includegraphics[width=8.2cm]{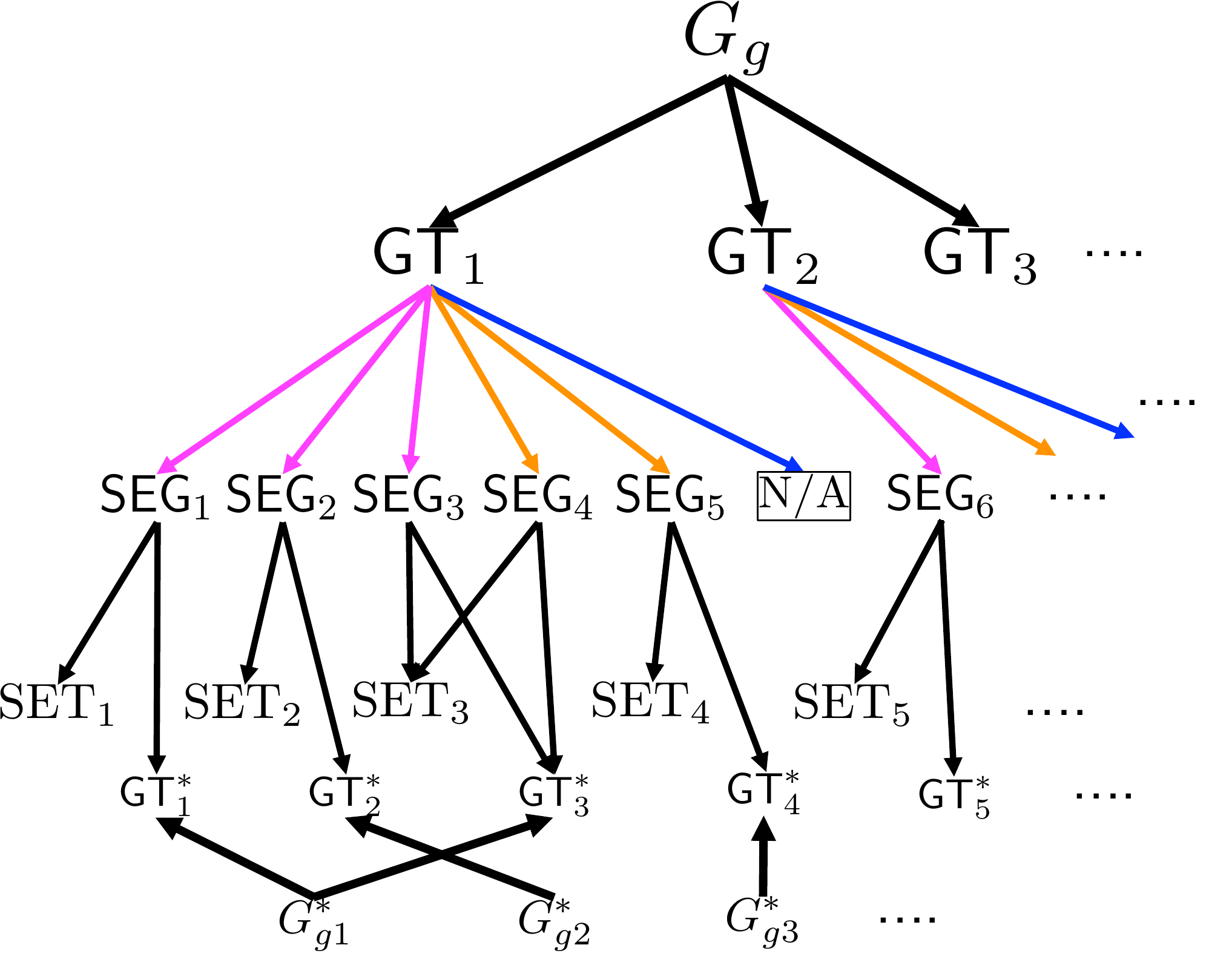}
\caption{(Color online) Schematic representation of a web of gauge theories with global symmetry. We start with a discrete gauge group $G_g$ that   generates several topologically distinct gauge theories ($\mathsf{GT}$s) one of which is untwisted.  We then   assign symmetry charge of  symmetry group $G_s$ to topological currents of gauge theories  through coupling to background gauge fields. There are usually many different ways of symmetry assignment, each of which is represented by a colored arrow. Within each specific symmetry assignment, we obtain many $\mathsf{SEG}$s. For example, $\mathsf{SEG}_1$, $\mathsf{SEG}_2$, and $\mathsf{SEG}_3$ belong to the same symmetry assignment (marked by magenta arrows) in $\mathsf{GT}_1$.  It is generically possible that some of symmetry assignment do not provide $\mathsf{SEG}$ descendants for $\mathsf{GT}$, which we mark as ``N/A''.   To identify $\mathsf{SET}$s, we need to further study ground state properties of $\mathsf{SEG}$s. We may externally insert symmetry fluxes into the system and perform Aharonov-Bohm experiments and the \emph{mixed} version of three-loop braiding   experiments. Two $\mathsf{SEG}$s (e.g.,  $\mathsf{SEG}_3$ and $\mathsf{SEG}_4$) may possibly describe the same $\mathsf{SET}$ phase.  Further, by giving full dynamics to the background gauge fields, the resulting new gauge theories  (denoted by $\mathsf{GT}^*$) are generated. Since basis transformations are allowed,    there should be in general many-to-one correspondence between $\mathsf{SEG}$s  and $\mathsf{GT}^*$s. Finally, all $\mathsf{GT}^*$s may be regrouped via identifying their gauge groups (denoted by $ {G}^*_g$).}
\label{figure_tree}
\end{figure}

In this paper, we   study   3D $\mathsf{SET}$s with Abelian topological orders \cite{wenbook} that are encoded by deconfined discrete Abelian gauge theories \cite{footnote_deconfined}. We     focus on discrete Abelian  gauge group $G_g=\Z_{N_1}\times\Z_{N_2}\times\cdots$ and on-site unitary Abelian symmetry group $G_s=\Z_{K_1}\times\Z_{K_2}\times\cdots$ or $G_s=\mathrm{U(1)}\times \Z_{K_1}\times\cdots$. Physically, these 3D $\mathsf{SET}$s can be viewed as 3D   gapped QSLs  that are  enriched by unbroken on-site symmetry $G_s$. Given a gauge group $G_g$, there are usually   many topologically distinct gauge theories (denoted by $\mathsf{GT}$) including one untwisted and several twisted ones \cite{spt15,dw1990}, as shown in Fig.~\ref{figure_tree}. After imposing global symmetry group, the resulting gauge field theory is called ``symmetry-enriched gauge theory'' ($\mathsf{SEG}$). Quantitatively, an $\mathsf{SEG}$ is defined through  two key ingredients:  
\begin{enumerate}
\item an action that   consists of topological terms (of one-form or two-form Abelian gauge fields) only;  
\item symmetry assignment via  a specific minimal coupling to background gauge fields  (denoted by $\{A^i\}$ with $i=1,2,\cdots$, where $A^i$ externally imposes symmetry fluxes in $\Z_{K_i}$ symmetry subgroup). 
\end{enumerate}
 We also stress that an anomaly-free $\mathsf{SEG}$ must  \emph{simultaneously} satisfy the following two stringent conditions \cite{footnote_anomaly}:
\begin{enumerate}
\item global symmetry is preserved; 
 \item gauge invariance  is guaranteed on a closed spacetime manifold. 
  \end{enumerate}
 We use the notation $\mathsf{SEG}(G_g,G_s)$ to denote such an $\mathsf{SEG}$. Then we try to provide answers to the following questions:
\begin{enumerate}
\item What is the path-integral formalism of an $\mathsf{SEG}$? And what is the ``parent''   $\mathsf{GT}$ of each $\mathsf{SEG}$? 
\item What is the relation between $\mathsf{SEG}$ and   $\mathsf{SET}$?  How can we probe symmetry-enriched properties  in experiments?
\item What is the resulting new gauge theory (denoted by $\mathsf{GT}^*$) after giving full dynamics \cite{footnote_gauging} to  $\{A^i\}$?
 \end{enumerate}
 To answer the first question is nothing but to look for anomaly-free  $\mathsf{SEG}$s  that meet the above definition and  conditions.
Following the 5-step general procedure (Sec.~\ref{sec:general_procedures}), the path-integral formalism of each $\mathsf{SEG}$  can be constructed, which is efficient for the practical purpose.  Each $\mathsf{SEG}$ can be  identified as a descendant of some $\mathsf{GT}$ (i.e., ``parent'').   Many concrete examples, including the  simplest case  $\mathsf{SEG}(\Z_2,\Z_2)$, are calculated explicitly in this paper. The method we will provide is doable for more general cases, some of which are collected in Appendix.

In the second question,  a complete description of an $\mathsf{SET}$ order requires the information of both topological orders and symmetry enrichment. In this sense, the total number of $\mathsf{SEG}$s is generically larger than that of distinct $\mathsf{SET}$ orders. For example, two anomaly-free $\mathsf{SEG}$s, may possibly give rise to the same  $\mathsf{SET}$ order.  If  two $\mathsf{SEG}$s have the same topological order, a practical way to probe symmetry enrichment is to insert symmetry fluxes into the 3D bulk and perform Aharonov-Bohm experiments between symmetry fluxes (flux loop formed by $A^i$) and bosons that are charged in the symmetry group. In addition, one should also perform  the \emph{mixed} version of three-loop braiding experiment \cite{spt18,footnote_4loop} among symmetry fluxes and gauge fluxes (i.e., loop excitations).  Through these thought experiments, one may find the relations between different $\mathsf{SEG}$s. If two $\mathsf{SEG}$s share the same bulk topological order data as well the same symmetry-enriched properties, they belong to the same $\mathsf{SET}$ ordered phase. Otherwise, they belong to two different $\mathsf{SET}$ phases (see Fig.~\ref{figure_tree}).

For the third question,  we note that in the action of an $\mathsf{SEG}$, $\{A^I\}$ is   a set of non-dynamical background gauge fields.  Symmetry fluxes formed by them are confined loop objects that are externally imposed into the bulk. These loop objects are fundamentally different from the gauge fluxes that are deconfined bulk loop excitations. Therefore, the usual basis transformations (mathematically represented by unimodular matrices of a general linear group) on gauge field variables are strictly prohibited \cite{spt2} \emph{if} the transformations  mix gauge fluxes and symmetry fluxes. However, if we give full dynamics to $\{A^I\}$ \cite{footnote_gauging}, then, the action actually represents a new gauge theory (denoted by $\mathsf{GT}^*$) and does not describe a $\mathsf{SEG}$ any more. In $\mathsf{GT}^*$s, symmetry fluxes are  legitimate deconfined bulk  loop excitations and arbitrary basis transformations are allowed. As a result, it is possible that the actions of two $\mathsf{SEG}$s may be  rigorously mapped to each other  via basis transformations,  both of which lead to the same $\mathsf{GT}^*$.  This set of gauge theories ``$\mathsf{GT}^*_1,\,\mathsf{GT}^*_2, \,\cdots$'' may be further regrouped by identifying their gauge groups (denoted by $ {G}^*_{g1}\,, {G}^*_{g2}\,,\cdots$). Finally, a web of gauge theories is obtained, as schematically shown in Fig.~\ref{figure_tree}.

The remainder of the paper is organized as follows.
Sec.~\ref{sec:general} is devoted to general discussions on $\mathsf{GT}$s, topological interactions and global symmetry. Especially, in Sec.~\ref{sec:general_procedures}, the 5-step general procedure is introduced in detail. Some calculation details in Sec.~\ref{cal_1},\ref{appendix_sub_n1n2_k1k2_1},\ref{appendix_sub_n1n2_k1k2_2} will be useful for quantitatively understanding the remaining sections, especially, Sec.~\ref{sec:examples}. For readers who are only interested in the final results, these details may be either skipped or gone through quickly.  In Sec.~\ref{sec:examples}, many simple examples are studied in details, including $\mathsf{SEG}(\Z_2,\Z_K)$,  $\mathsf{SEG}(\Z_2\times\Z_2,\Z_2)$,  and $\mathsf{SEG}(\Z_2\times\Z_2, \mathrm{U(1)} )$.    In Sec.~\ref{sec:fractionalization_3loop}, physical characterization of $\mathsf{SEG}$s is studied, including symmetry fractionalization and mixed three-loop braiding statistics among gauge fluxes and symmetry fluxes. In this way, we may achieve the map from $\mathsf{SEG}$ to $\mathsf{SET}$ as schematically shown in Fig.~\ref{figure_tree}. Simple examples are given, including $\mathsf{SEG}(\Z_2,\Z_K)$ with $K\in\Z_{\rm even}$ and $K\in\Z_{\rm odd}$.   In Sec.~\ref{sec:promotion}, full dynamics is given to the background gauge field, which promotes $\mathsf{SEG}$s to $\mathsf{GT}^*$s. Again, the discussions are followed by some simple examples including $\mathsf{SEG}(\Z_2,\Z_2)$, $\mathsf{SEG}(\Z_2,\Z_3)$ and $\mathsf{SEG}(\Z_2\times\Z_2,\Z_2)$.  Summary and outlook are made in Sec.~\ref{sec:conclusions}. More technical details and concrete examples are collected in Appendix.

 \section{Gauge theories, topological interactions, and global symmetry}\label{sec:general}

 \subsection{Inter-``layer'' topological interactions and addition of ``trivial'' layers}\label{sec:pure_gauge_theory}
In the continuum limit, gauge theories with discrete gauge groups can be written in terms of the following multi-component topological BF term \cite{horowitz}:
\begin{align}
S=i \sum_{I,J}\frac{\Lambda^{IJ}}{2\pi} \int_{\mathcal{M}^4}b^I\wedge d a^J\,,
 \end{align}
 where $\{b^I\}$ and $\{a^I\}$ are two sets of 2-form and 1-form U(1) gauge fields respectively. $I=1,2,\cdots, n$. $\Lambda^{IJ}$ is some $n\times n$ integer matrix, which may not be symmetric but the determinant of $\Lambda$ must be nonzero: $\mathrm{Det}\Lambda\neq 0$ \cite{footnote_zeromode}. In comparison to Horowitz's action term \cite{horowitz}, here we do not consider  $b^I\wedge b^J$. $\mathcal{M}^4$ is the 4D closed  spacetime (with imaginary time) manifold where our topological phases are defined.   In the following, the notation $\mathcal{M}^4$ will be neglected from the action for the sake of simplicity.

\begin{figure}[t]
\centering
\includegraphics[width=7.3cm]{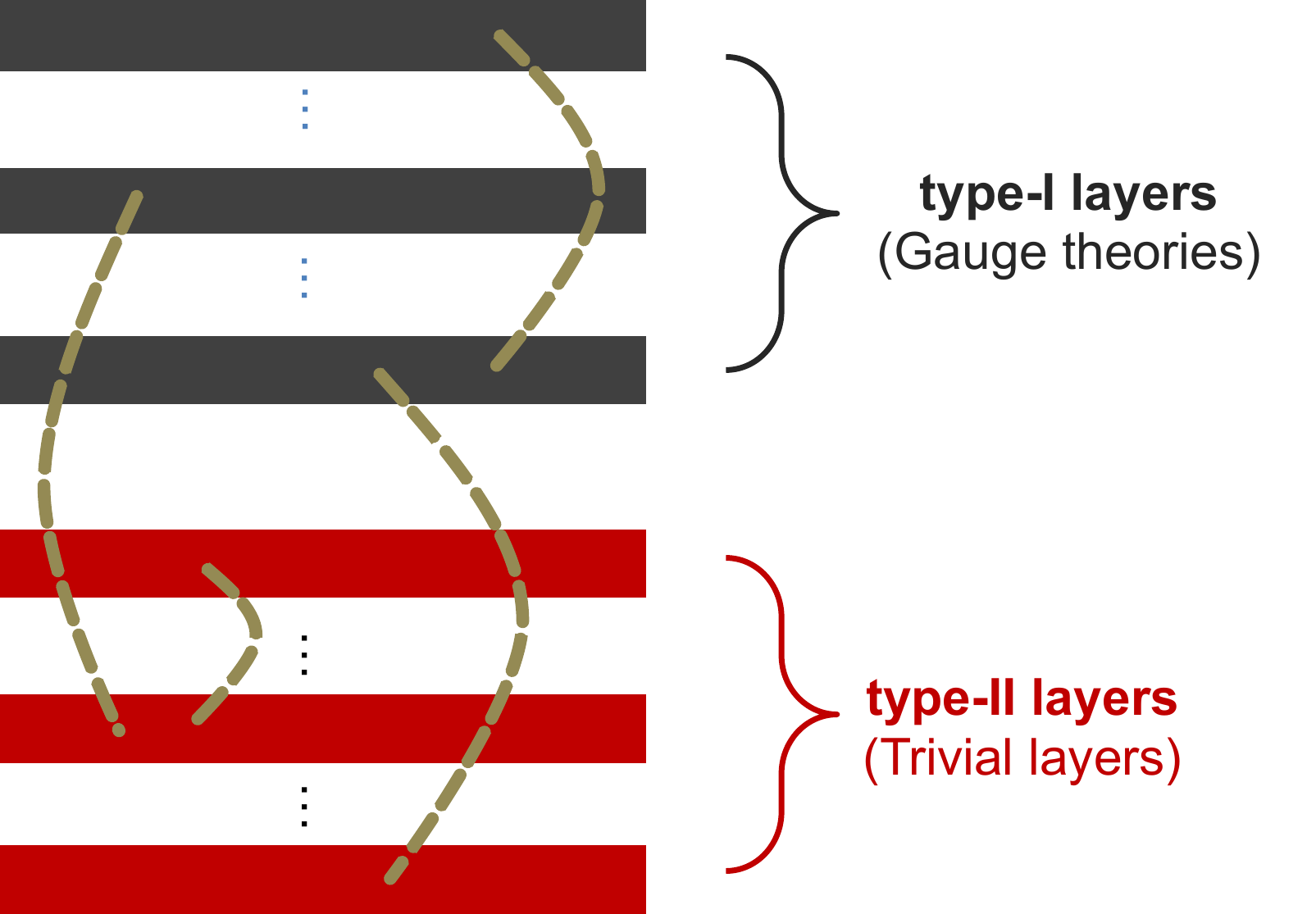}
\caption{(Color online) A schematic representation of  ``layers'' (Sec.~\ref{sec:pure_gauge_theory})  and the general procedure (Sec.~\ref{sec:general_procedures}). Each ``layer'' denotes a 3D system. It should   be noted that all layers are stacked together in the \emph{same} 3D spatial region although they are not so in this figure. $\mathsf{GT}$ before imposing symmetry resides in type-I layers. Type-II layers are   described by level-1 BF terms before imposing symmetry. By ``trivial'', we mean that these layers do not carry   gauge groups. The dashed curves represent topological interactions between layers. Actually, three-layer and four-layer topological interactions should also be considered.}
\label{figure_steps}
\end{figure}

 There are two \emph{independent} general linear transformations represented by two unimodular matrices $W,\Omega\in \mathbb{GL}(n,\Z)$ that ``rotate'' loop lattice and charge lattice respectively. Therefore,  $\Lambda$ can always be sent into its \emph{canonical form} via:
 \begin{align}
 W \Lambda \Omega^T=\text{diag} (N_1,N_2,\cdots,N_I,\cdots, N_n)\,,\label{eq:GL_transformation}
 \end{align}
 where $\{N_I\}$ are a set of \emph{positive} integers.   The superscript ``$T$'' denotes ``transpose''. 
 It is in sharp contrast to the multi-component  Chern-Simons theory \cite{wenbook} where $W=\Omega$ and the above diagonalized basis usually doesn't exist.   In the remaining parts of this paper, we work in this new basis unless otherwise specified.    In this new basis, each $I$ labels a ``layer system'' as schematically shown in the ``type-I layers'' in Fig.~\ref{figure_steps}  (N.B., the word  ``layer''    actually denotes a 3D spatial region). $N_I$ is the \emph{level} of the BF term in the $I$-th layer. 
 
$\{b^I\}$ and $\{a^I\}$, as two sets of gauge fields, are subject to the following Dirac quantization conditions:
\begin{align}
 \frac{1}{2\pi}\int_{\mathcal{M}^3}db^I&\in\Z\,, \label{eq:dirac_b}\\
 \frac{1}{2\pi}\int_{\mathcal{M}^2}da^I&\in\Z\,,\label{eq:dirac_a}
 \end{align}
 where $\mathcal{M}^3$ and $\mathcal{M}^2$ denote  3D and 2D  closed manifolds embedded in $\mathcal{M}^4$  respectively. These two equations will play important roles in the following discussions.

 The BF term in the canonical form is a  field theory of \emph{untwisted} $G_g=\Z_{N_1}\times\Z_{N_2}\times\cdots\times\Z_{N_n}$ gauge theory where layers are decoupled to each other.  However,  there are  topological interactions that can couple them together:
 \begin{align}
S=&i\sum_I \frac{N_I}{2\pi} \int b^I\wedge d a^I+i\sum_{IJK}\frac{q^{IJK}}{4\pi^2} \int  a^I\wedge a^J\wedge d a^K\nonumber\\
&+i\sum_{IJKL}\frac{ {t}^{IJKL}}{8\pi^3} \int a^I\wedge a^J\wedge a^K\wedge a^L\,,\label{eq:action_of_pure_gauge}
 \end{align}
where $\{q^{IJK}\}$ and $\{ t^{IJKL}\}$ are two sets of coefficients. These newly introduced  action terms are topological since their expressions are wedge products of differential forms. Recently a lot of progress has been made based on these topological terms in gauge theories   as well as $\mathsf{SPT}$ phases \cite{spt6,spt15,aada_ryu1,aada_ryu2,aada_wang}.    
The presence of interlayer topological interactions leads to \emph{twisted} $G_g$ gauge theories.   Since these new topological terms are explicitly not gauge invariant (even in a  closed manifold) alone, the definitions of usual gauge transformations on $\{b^I\}$ must be properly modified [to appear in Eq.~(\ref{eq:modified_gauge_a1a2da2})]. 
To be a legitimate $\mathsf{GT}$ action, $\{q^{IJK}\}$ and $\{t^{IJKL}\}$ are expected to be   quantized and compact (i.e., periodic), 
  which eventually leads to finite number of distinct $\mathsf{GT}$s \emph{before} global symmetry is imposed.    All of them are classified by the fourth group cohomology with U(1) coefficient:
$ \mathcal{H}^{4}(\mathbb{Z}_{N_1}\times \mathbb{Z}_{N_2}\cdots,\mathrm{U(1)})= \prod_{I<J}(\mathbb{Z}_{N_{IJ}})^2\!\times\!\prod_{I<J<K}(\mathbb{Z}_{N_{IJK}})^2\times\!\prod_{I<J<K<L}\mathbb{Z}_{N_{IJKL}}$,  
where $N_{IJ,...}$ is the greatest common divisor of $N_I, N_J,\cdots$.  Technical details are shown in Sec.~\ref{cal_1}.
  
  In addition, one may always add arbitrary number of ``trivial layers'' into the action $S$ in Eq.~(\ref{eq:action_of_pure_gauge}): 
  \begin{align}
  S\!\rightarrow& S+i \frac{1}{2\pi}\int \!\!b^{n+1}\!\wedge \!da^{n+1}+i  \frac{1}{2\pi} \!\int \!\!b^{n+2}\!\wedge\! da^{n+2}+\cdots\,  .  \label{eq:trivial_layer_action}
  \end{align}
These trivial layers do not introduce additional gauge structures. However,  as we will see,  adding trivial layers will be very useful and sometimes necessary when global symmetry $G_s$ is imposed. 
 
 \subsection{Symmetry assignment}\label{sec:impose_symmetry}
Now, let us consider how to impose global symmetry group $G_s=\Z_{K_1}\times \Z_{K_2}\times\cdots\times\Z_{K_m}$. In topological quantum field theory, there is  a 1-form topological current $J^I$ for each $I$:
$*J^I=\frac{1}{2\pi}db^I\,,
$ where $*$ denotes the Hodge dual operation. It is conserved automatically since $d^2=0$.  The fact that the total particle number is  integral is nicely guaranteed by Dirac quantization condition (\ref{eq:dirac_b}).  
Therefore, a natural definition of global symmetry is to enforce that the symmetry charge is carried by this topological current. This is the so-called hydrodynamical approach that was    applied successfully in the fractional quantum Hall effect with the multi-component Chern-Simons theory description \cite{wenbook}. This is also a key step of the topological quantum field theory description of SPTs \cite{spt6}. 

In order to identify global symmetry,  a background gauge field $A^{i}$ is turned on. Mathematically,  a minimal coupling term between background gauge fields and topological currents is introduced into the action (\ref{eq:trivial_layer_action}): 
$S_{sym.}\!\!=\!i\! \sum^n_I\!\sum^m_i L^{Ii} \!\!\int \!\!J^I\wedge \!* \!A^{i}$\,,  where $L^{Ii}$ is an $n\times m$ integer matrix. By noting that the total symmetry group $G_s=\Z_{K_1}\times\Z_{K_2}\times\cdots$,  the background 1-form U(1) gauge field $A^{i}$ is subject to the following constraints:
\begin{align}
\frac{K_i}{2\pi}\int_{\mathcal{M}^1}A^{i}\in\Z\, \text{ for }\Z_{K_i}\text{ symmetry subgroup}\,,\label{eq:holo}
\end{align} 
where $\mathcal{M}^1$ denotes a closed spacetime loop.  
%As a result, the total symmetry charge is conserved mod $K_i$ for subgroup $\Z_{K_i}$. Indeed, by definition, the conservation of total $\Z_{K_i}$ symmetry charge allows a change of $K_i$, in contrast to U(1) symmetry where total symmetry charge must be rigorously conserved. 
As mentioned in Sec.~\ref{sec:pure_gauge_theory}, trivial layers in Eq.~(\ref{eq:trivial_layer_action}) may be taken into consideration once symmetry is imposed. Therefore, the index $I$ in $S_{sym.}$ is  allowed to be larger than $n$. 
Once the topological current carries symmetry charge,  a new set of stringent constraints on the coefficients $\{q^{IJK}\}$ and $\{t^{IJKL}\}$ will be imposed  such that  global symmetry is compatible with gauge invariance principle, the quantization and periodicity of $\{q^{IJK}\}$ and $\{t^{IJKL}\}$ may be changed dramatically after global symmetry is imposed.  It means that, one  $\mathsf{GT}$ may generate  many distinct $\mathsf{SEG}$ descendants  after   symmetry is imposed, which manifestly shows   patterns of symmetry enrichment (see Fig.~\ref{figure_tree}). If symmetry is not imposed, those distinct $\mathsf{SEG}$s become indistinguishable and reduce back to the same parent $\mathsf{GT}$.

\subsection{Summary of the 5-step general procedure}\label{sec:general_procedures}
Based on the preparation done in Sec.~\ref{sec:pure_gauge_theory} and \ref{sec:impose_symmetry}, in this part, we summarize the general procedure for obtaining $\mathsf{SEG}$s and connecting them to their parent $\mathsf{GT}$s. There are five main steps. 

\underline{\textit{Step-1}}. Add trivial layers (i.e., type-II in Fig.~\ref{figure_steps}). Mathematically, trivial layers are described by Eq.~(\ref{eq:trivial_layer_action}).

\underline{\textit{Step-2}}. Assign symmetry via the minimal coupling terms ($\sim J\wedge * A$).  Symmetry assignment can be either made purely inside type-I or purely inside type-II or both \cite{footnote_symmetry}.

\underline{\textit{Step-3}}.  Add all possible topological interactions among layers via the topological terms with coefficients $\{q^{IJK}\}$ and $\{t^{IJKL}\}$  in Eq.~(\ref{eq:action_of_pure_gauge}) and the indices $I,J,K,\cdots$ are extended to all layers including trivial layers.  In Fig.~\ref{figure_steps}, only two-layer interactions (denoted by   dashed lines) are drawn for simplicity. However, generic three-layer and four-layer interactions should also be taken into considerations.

\underline{\textit{Step-4}}. Consider all consistent conditions and determine the quantization and periodicity of coefficients of topological interactions. These consistent conditions are \textit{(i)} Dirac quantization conditions; \textit{(ii)} ``small'' gauge transformations; \textit{(iii)} ``large'' gauge transformations; \textit{(iv)} shift operation of coefficients that leads to coefficient periodicity; \textit{(v)}. total symmetry charge for $\Z_{K_i}$ subgroup is conserved mod $K_i$. Once the above four steps are done, the path-integral expressions and symmetry assignment for $\mathsf{SEG}$s are obtained. Definitions and quantitative studies of these consistent conditions will be provided in Sec.~\ref{cal_1},\ref{appendix_sub_n1n2_k1k2_1},\ref{appendix_sub_n1n2_k1k2_2}, and Appendix~\ref{sec:tech1}.

\underline{\textit{Step-5}}. Regroup all $\mathsf{SEG}$s obtained above into  distinct $\mathsf{GT}$s in Fig.~\ref{figure_tree}. For example, in Fig.~\ref{figure_tree}, $\mathsf{SEG}_{1,\cdots,5}$  are $\mathsf{SEG}$ descendants of  $\mathsf{GT}_1$, while, $\mathsf{SEG}_6$  is a $\mathsf{SEG}$ descendant of  $\mathsf{GT}_2$.  If gauge group is $G_g=\Z_N$ that will be calculated in Sec.~\ref{sec:one}, this step can be skipped for the reason that there is   only one $\Z_N$ $\mathsf{GT}$, i.e., the untwisted $\mathsf{GT}$.    If gauge group contains more than one $\Z_N$s, e.g., $G_g=\Z_{N_1}\times\Z_{N_2}$, usually gauge theories have twisted versions. Under the circumstances,  the role of Step-5 becomes critical. We will discuss pertinent details in Sec.~\ref{sec:two}.

  \subsection{General calculation on $G_g=\Z_{N_1}\times \Z_{N_2}$ with no symmetry}\label{cal_1}
  In the following, we present some useful calculation details on gauge theories with $G_g=\Z_{N_1}\times \Z_{N_2}$ and demonstrate, especially, what the consistent conditions listed in \text{Step-4} of Sec.~\ref{sec:general_procedures} are, at quantitative level.
%   Furthermore, by simply setting $N_2=1$, the gauge theories directly reduce to those with $G_g=\Z_{N_1}$.
    Several mathematical notations are introduced and will be frequently used in the remaining parts of this paper.   All other calculation details are present in Appendix~\ref{sec:tech1}.
  
  Consider the following two-layer BF theories with inter-layer topological couplings in the form of ``$aada$'':
\begin{align}
 S=&\sum_{I=1}^2\frac{i N_I}{2\pi} \int b^I \wedge d a^I+i \frac{q}{4\pi^2} \int a^1\wedge a^2 \wedge da^2  \nonumber\\
& +i \frac{\bar q}{4\pi^2} \int a^2\wedge a^1 \wedge da^1\,,\label{eq:aada_general_q}
\end{align}
where $q\equiv q^{122}$ and $\bar q\equiv q^{211}$. 
Since $a^1a^2da^2$ and $a^2a^1da^1$ are linearly independent, we may study them separately. First consider $\bar q=0$. The action  is invariant under the following gauge transformations parametrized by scalars $\{\chi^I\}$ and vectors $\{V^I\}$:
\begin{align}
a^I&\longrightarrow a^I +d\chi^I\,,\\
b^I&\longrightarrow b^I+dV^I-\frac{ q}{2\pi N^I} \epsilon^{IJ3} \chi^J \wedge da^2\,,\label{eq:modified_gauge_a1a2da2}
\end{align}
where $\epsilon^{123}=-\epsilon^{213}=1$.  
It is clear that   the usual gauge transformations of  $b^I$ \cite{horowitz} are modified through adding a $q$-dependent term in Eq.~(\ref{eq:modified_gauge_a1a2da2}). As usual,  the gauge parameters $\chi^I$ and $V^I$ satisfy the following conditions:
\begin{align}
\frac{1}{2\pi}\int_{\mathcal{M}^1} d\chi^I \in\Z, \quad \frac{1}{2\pi}\int_{\mathcal{M}^2} dV^I \in\Z \,.\label{eq:large}
\end{align}
Once the integers on the r.h.s. are nonzero, the associated gauge transformations are said to be ``large''.  Let us investigate  the integral $\frac{1}{2\pi}\int_{\mathcal{M}^3}db^I$.

 Under the above modified gauge transformations (\ref{eq:modified_gauge_a1a2da2}), the integral will be changed by the amount below (for $I=1$, $\mathcal{M}^3=\mathcal{M}^1\times\mathcal{M}^2$ is considered):
\begin{align}
 \frac{1}{2\pi}\int_{\mathcal{M}^3}db^1\longrightarrow& \frac{1}{2\pi}\int_{\mathcal{M}^3}db^1 - \frac{ q}{4\pi^2 N_1} \int _{S^1}d\chi^2 \int_{M^2} da^2\nonumber\\
  =& \frac{1}{2\pi}\int_{\mathcal{M}^3}db^1 - \frac{ q}{4\pi^2 N_1}  \times  2\pi \ell \times 2\pi \ell' \,,\label{eq:key_1}
\end{align}
where $\ell\,,\,\ell' \in  \Z$, and, the Dirac quantization condition (\ref{eq:dirac_a}) and gauge parameter condition (\ref{eq:large}) are applied. In order to be consistent with the Dirac quantization condition (\ref{eq:dirac_b}), the change amount must be integral, namely, $q$ must be divisible by $N_1$. Similarly,   $ q$ is also divisible by $N_2 $ due to:
\begin{align}
 \frac{1}{2\pi}\int_{\mathcal{M}^3}db^2\longrightarrow& \frac{1}{2\pi}\int_{\mathcal{M}^3}db^2 + \frac{ q}{4\pi^2 N_2} \int _{S^1}d\chi^1 \int_{M^2} da^2 \nonumber\\
=&\frac{1}{2\pi}\int_{\mathcal{M}^3}db^2 + \frac{ q}{4\pi^2 N_2}  \times  2\pi \ell'' \times 2\pi \ell''' ,\label{eq:key_2}
\end{align}
where $  \ell''\,,\,\ell''' \in  \Z$. 
 Hence, $q=\frac{kN_1N_2}{ N_{12}}, k \in \Z$, where the symbol ``$N_{12}$'' denotes   the greatest common divisor of $N_1$ and $N_2$.
 
Below, we will show that $k$ has a periodicity $N_{12}$ and thereby $q$ is compactified: $q\sim q+{N_1N_2}$. Let us consider the following redundancy due to shift operations:
 \begin{align}
\frac{1}{2\pi}\int  db^1 \longrightarrow &\frac{1}{2\pi} \int db^1-\frac{ N_2 \tilde{K}_1}{4\pi^2 N_{12}} \int a^2\wedge da^2 \,, \label{eq:key_3}\\
\frac{1}{2\pi}\int d b^2\longrightarrow  &\frac{1}{2\pi}\int db^2+\frac{N_1  \tilde{K}_2}{4\pi^2 N_{12}} \int a^1\wedge da^2 \,,\label{eq:key_4}\\
 k\longrightarrow\,& \,k+\tilde{K}_1+\tilde{K}_2\,.
 \end{align}
 Under the above shift operation, the total action (\ref{eq:aada_general_q}) is invariant. 
Again, in order to be consistent with  Dirac quantization (\ref{eq:dirac_b}), the change amount of the integral $\frac{1}{2\pi}\int_{\mathcal{M}^3}db^I$ should be integral, namely:
 \begin{align}
  \frac{N_2 \tilde{K}_1}{4\pi^2 N_{12}} \int_{\mathcal{M}^3}   a^2\wedge da^2&\in \Z\,,\,\\
   \frac{N_1 \tilde{K}_2}{4\pi^2 N_{12}} \int_{\mathcal{M}^3}  a^1\wedge d a^2&\in \Z\,.
\end{align} 
We may apply the Dirac quantization condition (\ref{eq:dirac_a}) and the quantized Wilson loop $ \frac{N_I}{2\pi} \int_{\mathcal{M}^1} a^I \in\Z$ that is obtained via  equations of motion of $b^I$.  As a result, two constraints are achieved: $\tilde{K}_1/N_{12}\in\Z$, $\tilde{K}_2/N_{12}\in\Z$. By using Bezout's lemma, the minimal periodicity of $k$  is given by the greatest common divisor (GCD) of $N_{12}$ and $N_{12}$, which is still $N_{12}$.  As a result, we obtain the conditions on $q$ \emph{if} symmetry is not taken into consideration.
\begin{align}
q=k\frac{N_1N_2}{N_{12}} \text{  mod  } N_1N_2\,,~~~ k\in\Z_{N_{12}}.\label{eq:key_quantized_no_symmetry}
\end{align}
 Similarly, for  $\frac{\bar q}{4\pi^2} a^2\wedge a^1 \wedge da^1$ term, we also have   the same quantization and the same periodicity:
 \begin{align}
\bar q=k\frac{N_1N_2}{N_{12}} \text{  mod  } N_1N_2\,,~~~ k\in\Z_{N_{12}}.\label{eq:key_quantized_no_symmetry_plus}
\end{align}
In conclusion, we have $(\Z_{N_{12}})^2$ different kinds of gauge theories with $G_g=\Z_{N_{1}}\times \Z_{N_2}$.

 \subsection{General calculation on $G_g=\Z_{N_1}\times \Z_{N_2}$ with $G_s=\Z_{K_1} \times \Z_{K_2}$-(I)}\label{appendix_sub_n1n2_k1k2_1}

  To impose the symmetry, we add the following coupling term into $S$ in Eq.~(\ref{eq:aada_general_q}) (again, we consider $\bar q=0$ only): 
 \begin{align}
 \sum_{i}^2\frac{i}{2\pi} \int   A^{i}\wedge  db^{i}\,,\label{eq:symmetry_coupling_aada}
\end{align}  
where $A^i$ is subject to the constraints in Eq~(\ref{eq:holo}). This coupling term simply means that the first layer carries $\Z_{K_1}$ symmetry while the second layer carries $\Z_{K_2}$ symmetry.  The total symmetry group $G_s=\Z_{K_1}\times\Z_{K_2}$.
 
 Our goal is to determine all legitimate values of $q$ in the presence of global symmetry. And we expect that the period of $q$ is in general  larger than the original gauge theory with no symmetry, which leads to a set of $\mathsf{SEG}$s.  
{The key observation is that the change amounts of the integral $\frac{1}{2\pi}\int_{\mathcal{M}^3}db^I$ in both gauge transformations and shift operations should not only be integral [in order to be consistent with the Dirac quantization condition (\ref{eq:dirac_b})] but also be multiple of $K_i$ such that the coupling term (\ref{eq:symmetry_coupling_aada}) is gauge invariant \text{modular} $2\pi$.} Physically, it can be understood via the definition of the integral. This integral is nothing but the total symmetry charge of the associated symmetry group. Since the total symmetry charge of $\Z_{K_i}$ is allowed to be changed by $K_i$ while still respecting symmetry. This is a peculiar feature of cyclic symmetry group, compared to  U(1) symmetry. 
 
 More quantitatively,  with symmetry taken into account, from Eqs.~(\ref{eq:key_1},~\ref{eq:key_2}), we may obtain the quantization of $q$: $q=\frac{kN_1N_2K_1K_2}{\mathrm{GCD}(N_1K_1,N_2K_2)}$  with $k\in\Z$ such that the change amounts are multiple of $K_i$. Then, with these new quantized values, the shift operations (\ref{eq:key_3},\ref{eq:key_4}) are changed to:
 \begin{align}
\frac{1}{2\pi} \int db^1 \longrightarrow &\frac{1}{2\pi} \int db^1-\frac{\tilde{K}_1  N_2K_1K_2 \int a^2\wedge da^2}{4\pi^2\mathrm{GCD}(N_1K_1,N_2K_2)}  \,, \label{eq:key_5}\\
\frac{1}{2\pi}\int d b^2\longrightarrow  &\frac{1}{2\pi}\int db^2+\frac{\tilde{K}_2 N_1K_1K_2 \int a^1\wedge da^2}{4\pi^2\mathrm{GCD}(N_1K_1,N_2K_2)}  \,.\label{eq:key_6}
 \end{align}
The change amounts should be quantized at $K^1$ in Eq.~(\ref{eq:key_5}) and $K^2$ in Eq.~(\ref{eq:key_6}), respectively, such that symmetry is kept. 
 We may apply the Dirac quantization condition (\ref{eq:dirac_a}) and the quantized Wilson loop $ \frac{N_IK_I}{2\pi} \int_{\mathcal{M}^1} a^I \in\Z$ that is obtained via  equations of motion of $b^I$ in the presence of $A^I$ background.  As a result, two necessary and sufficient constraints are achieved: $\frac{\tilde{K}_1}{\mathrm{GCD}(N_1K_1,N_2K_2)}\in\Z$, $\frac{\tilde{K}_2}{\mathrm{GCD}(N_1K_1,N_2K_2)}\in\Z$. By using Bezout's lemma, the minimal periodicity of $k$  is given by   $\mathrm{GCD}$ of  $\mathrm{GCD}(N_1K_1,N_2K_2)$ and $\mathrm{GCD}(N_1K_1,N_2K_2) $,  which is still $\mathrm{GCD}(N_1K_1,N_2K_2)$. Therefore, once symmetry is imposed, $q$ is changed from Eq.~(\ref{eq:key_quantized_no_symmetry}) to:
 \begin{align} 
q=&k\frac{N_1N_2K_1K_2}{\mathrm{GCD}(N_1K_1,N_2K_2)} \text{  mod  }  N_1N_2K_1K_2\,,\nonumber\\
& \text{with } k\in\Z_{\mathrm{GCD}(N_1K_1,N_2K_2)}\label{eq:useful_1}
\end{align}
which gives $\mathrm{GCD}(N_1K_1,N_2K_2)$ $\mathsf{SEG}$s. In other words, the allowed values of $q$ are enriched by symmetry. 
For $\bar q$ term, the conditions are completely the same as $q$, which leads to another  $\mathrm{GCD}(N_1K_1,N_2K_2)$ $\mathsf{SEG}$s.
 \begin{align} 
\bar q=&k\frac{N_1N_2K_1K_2}{\mathrm{GCD}(N_1K_1,N_2K_2)} \text{  mod  }  N_1N_2K_1K_2\,,\nonumber\\
& \text{with } k\in\Z_{\mathrm{GCD}(N_1K_1,N_2K_2)}\,.\label{eq:useful_1_2_bar_q}
\end{align}

In short, before imposing symmetry, according to Eqs.~(\ref{eq:key_quantized_no_symmetry},\ref{eq:key_quantized_no_symmetry_plus}),  there are $(N_{12})^2$ distinct $\mathsf{GT}$s with gauge group $G_g=\Z_{N_1}\times \Z_{N_2}$. After imposing symmetry group $G_s=\Z_{K_1}\times\Z_{K_2}$, according to Eqs.~(\ref{eq:useful_1},\ref{eq:useful_1_2_bar_q}), there are $\left[\mathrm{GCD}(N_1K_1,N_2K_2)\right]^2$ distinct $\mathsf{SEG}$s \emph{if} the symmetry assignment is given by Eq.~(\ref{eq:symmetry_coupling_aada}). Likewise, one can consider that $\Z_{K_1}$ and $\Z_{K_2}$ symmetry charges are carried by the second layer and the first layer respectively, i.e., Eq.~(\ref{eq:symmetry_coupling_aada}) is changed to:
\begin{align}
\frac{i}{2\pi}\int (A^1\wedge d b^2+A^2\wedge d b^1)\,.
\end{align} Then, there will be  $\left[\mathrm{GCD}(N_1K_2,N_2K_1)\right]^2$ new $\mathsf{SEG}$s.

 \subsection{General calculation on $G_g=\Z_{N_1}\times \Z_{N_2}$ with  $G_s=\Z_{K_1} \times \Z_{K_2}$-(II)}\label{appendix_sub_n1n2_k1k2_2}

In the following, we alter the definition of symmetry assignment and still consider $a^1a^2da^2$ first. The coupling term in Eq.~(\ref{eq:symmetry_coupling_aada}) is now changed to:
 \begin{align}
 \frac{i}{2\pi} \int  (A^1+A^2)\wedge db^1 \label{eq:symmetry_coupling_aada_same_layer}
  \end{align} 
 which means that both $\Z_{N_1}$ and $\Z_{N_2}$ symmetry charges are carried by the first layer.  We will show that ($\mathrm{LCM}$ stands for ``least common multiple''):
  \begin{align} 
q= &  k \,\mathrm{LCM}(N_1K_1,N_1K_2,N_2) \text{ mod }   {N_1N_2\, \mathrm{LCM}(K_1,K_2)}\,\,, \nonumber\\
&\text{ with } k\in\Z_{  \frac{N_1N_2\, \mathrm{LCM}(K_1,K_2)}{\mathrm{LCM}(N_1K_1,N_1K_2,N_2)}}\,\label{eq:q_12_1}
\end{align} 
meaning that the total number of $\mathsf{SEG}$s are  $\frac{N_1N_2\, \mathrm{LCM}(K_1,K_2)}{\mathrm{LCM}(N_1K_1,N_1K_2,N_2)}$ if (i) both symmetry charges are carried by the first layer shown in Eq.~(\ref{eq:symmetry_coupling_aada_same_layer}) and  (ii) $a^1a^2da^2$ is considered (i.e., $\bar q=0$).  As a side note, by exchanging $1\leftrightarrow 2$, the above result directly implies that   the total number of $\mathsf{SEG}$s are  $\frac{N_1N_2\, \mathrm{LCM}(K_1,K_2)}{\mathrm{LCM}(N_2K_1,N_2K_2,N_1)}$ if (i) both symmetry charges are carried by the second layer [replacing $b^1$  in Eq.~(\ref{eq:symmetry_coupling_aada_same_layer}) by $b^2$] and  (ii) $a^2a^1da^1$ is considered (i.e., $q=0$):
  \begin{align} 
\bar{q}= &  k \,\mathrm{LCM}(N_2K_1,N_2K_2,N_1) \text{ mod }   {N_1N_2\, \mathrm{LCM}(K_1,K_2)}\,,\nonumber\\
&\text{  with } k\in\Z_{  \frac{N_1N_2\, \mathrm{LCM}(K_1,K_2)}{\mathrm{LCM}(N_2K_1,N_2K_2,N_1)}}\,.
\end{align}

 Let us present several key steps towards  Eq.~(\ref{eq:q_12_1}) below. The change amount in Eq.~(\ref{eq:key_1}) should be divisible \emph{simultaneously} by $K_1$ and $K_2$ such that symmetry is kept. Meanwhile, the change amount in Eq.~(\ref{eq:key_2}) should be integral in order to be consistent with Dirac quantization condition (\ref{eq:dirac_b}). Therefore, $q$ should be quantized as: $q=k \,\mathrm{LCM}(N_1K_1,N_1K_2,N_2)$ with $k\in\Z$. Then, with these new quantized values, the shift operations (\ref{eq:key_3},\ref{eq:key_4}) are changed to:
 \begin{align}
\frac{1}{2\pi}\int db^1 \longrightarrow &\frac{1}{2\pi} db^1+ \frac{1}{4\pi^2 N_1}\tilde{K}_1  \,\mathrm{LCM}(N_1K_1,N_1K_2,N_2) \nonumber\\
&~~~~~~~ \int a^2\wedge da^2 \,, \label{eq:key_55}\\
\frac{1}{2\pi}\int d b^2\longrightarrow  &\frac{1}{2\pi}db^2-  \frac{1}{4\pi^2 N_2} \tilde{K}_2 \,\mathrm{LCM}(N_1K_1,N_1K_2,N_2)\nonumber\\
&~~~~~~~ \int a^1\wedge da^2 \,.\label{eq:key_66}
 \end{align}
Again, the change amount in Eq.~(\ref{eq:key_55})  should be divisible simultaneously by   $K_1$ and $K_2$ such that symmetry is kept. The change amount in Eq.~(\ref{eq:key_66})  should be integral such that Dirac quantization condition (\ref{eq:dirac_b}) is satisfied.  Before evaluating the integral, the Wilson loop of $a^1$ may be obtained via equation of motion of $b^1$:
\begin{align}
\frac{N_1K_1K_2}{2\pi\,\mathrm{GCD}(K_1,K_2)   }\int_{\mathcal{M}^1}a^1\in\Z \,,
\end{align}
where Eq.~(\ref{eq:holo}) and Bezout's lemma are applied. The Wilson loop of $a^2$ may be obtained via equation of motion of $b^2$:
\begin{align}
\frac{N_2}{2\pi}\int_{\mathcal{M}^1}a^2\in\Z \,.
\end{align}
With this preparation, we may calculate the change amounts in Eqs.~(\ref{eq:key_55},\ref{eq:key_66}) and obtain the conditions on $\tilde{K}_1$ and $\tilde{K}_2$:
\begin{align}
&\frac{\mathrm{LCM}(N_1K_1,N_1K_2,N_2)}{N_1N_2\, \mathrm{LCM}(K_1,K_2)}\tilde{K}_1\in\Z\,,\\
&\frac{\mathrm{LCM}(N_1K_1,N_1K_2,N_2)}{N_1N_2\, \mathrm{LCM}(K_1,K_2)}\tilde{K}_2\in\Z\,.
\end{align}
 Therefore, by using Bezout's lemma, the minimal periodicity of $k$ can be fixed and $k$ is thus compactified: $k\in\Z_{\frac{N_1N_2\, \mathrm{LCM}(K_1,K_2)}{\mathrm{LCM}(N_1K_1,N_1K_2,N_2)}}$.
 
  Following the similar procedure, we may obtain the results for the remaining two cases: 
  (i). $a^2a^1da^1$ (labeled by $\bar q$) and both symmetry charges are in the first layer;  (ii). $a^1a^2da^2$ (labeled by $ q$) and both symmetry charges are in the second layer.  
  For (a), $\bar q$ is given by:
   \begin{align} 
\bar q= &  k \,\mathrm{LCM}(N_1K_1,N_1K_2,N_2) \text{ mod }   {N_1N_2\, \mathrm{LCM}(K_1,K_2)}\,\,,\nonumber\\
&\text{ with } k\in\Z_{  \frac{N_1N_2\, \mathrm{LCM}(K_1,K_2)}{\mathrm{LCM}(N_1K_1,N_1K_2,N_2)}}\,.
\end{align} 
For (b), $q$ is given by:
  \begin{align} 
q= &  k \,\mathrm{LCM}(N_2K_1,N_2K_2,N_1) \text{ mod }   {N_1N_2\, \mathrm{LCM}(K_1,K_2)}\,\,, \nonumber\\
&\text{ with } k\in\Z_{  \frac{N_1N_2\, \mathrm{LCM}(K_1,K_2)}{\mathrm{LCM}(N_2K_1,N_2K_2,N_1)}}\,.
\end{align}

 \section{Typical examples of symmetry-enriched gauge theories}\label{sec:examples}

 In this section, through a few concrete examples, we apply the general procedure shown in Sec.~\ref{sec:general_procedures} and construct $\mathsf{SEG}$s that satisfy the definition and conditions listed in Sec.~\ref{sec:introduction_main}.  Useful technical  details are present in Sec.~\ref{cal_1},\ref{appendix_sub_n1n2_k1k2_1},\ref{appendix_sub_n1n2_k1k2_2} and Appendix~\ref{sec:tech1}. More examples are collected in Appendix~\ref{appendix_examples}.

    \subsection{ $\mathsf{SEG}(\Z_2,\Z_K)$}\label{sec:one}

We begin with $G_g=\Z_N$ and $G_s=\Z_K$.  The common features of this class are that: \textit{(i)} there is only one gauge theory before imposing global symmetry; \textit{(ii)} there are two complementary choices of symmetry assignment \cite{footnote_symmetry}, namely, $\Z_K$ is either in the first layer or in the second layer (trivial layer).  
More concretely, before imposing global symmetry, there is only one $\Z_N$ gauge theory since all additional topological terms like $aada, aaaa$ vanish identically. Despite that, we still formally explicitly add $a^1a^2da^2$ and $a^2a^1da^1$ in all tables in order to see whether or not these topological terms will eventually have chance to be nonvanishing after symmetry is taken into consideration.  Since we only have one cyclic symmetry subgroup, i.e., $G_s=\Z_K$,  inclusion of two layers (the second one is a trivial layer in a sense that the level of $b^2da^2$ term is 1) is enough in the current simple cases.

\begin{table}[t]\small
\caption{$\mathsf{SEG}$($\Z_2 $, $ \Z_K$). Both gauge group and symmetry group are in the same layer (the first layer). There is no nontrivial symmetry enrichment but a trivial stacking of symmetry and gauge theory.}
\label{table:z2_zk_1}
\begin{tabular}{|c|c|c|c|}
\hline
\hline
                   \multirow{4}{*}{    \begin{minipage}   {0.65in}Symmetry assignment\end{minipage}  }  & \multicolumn{3}{c|}{ \multirow{4}{*}{ \includegraphics[width=5cm]{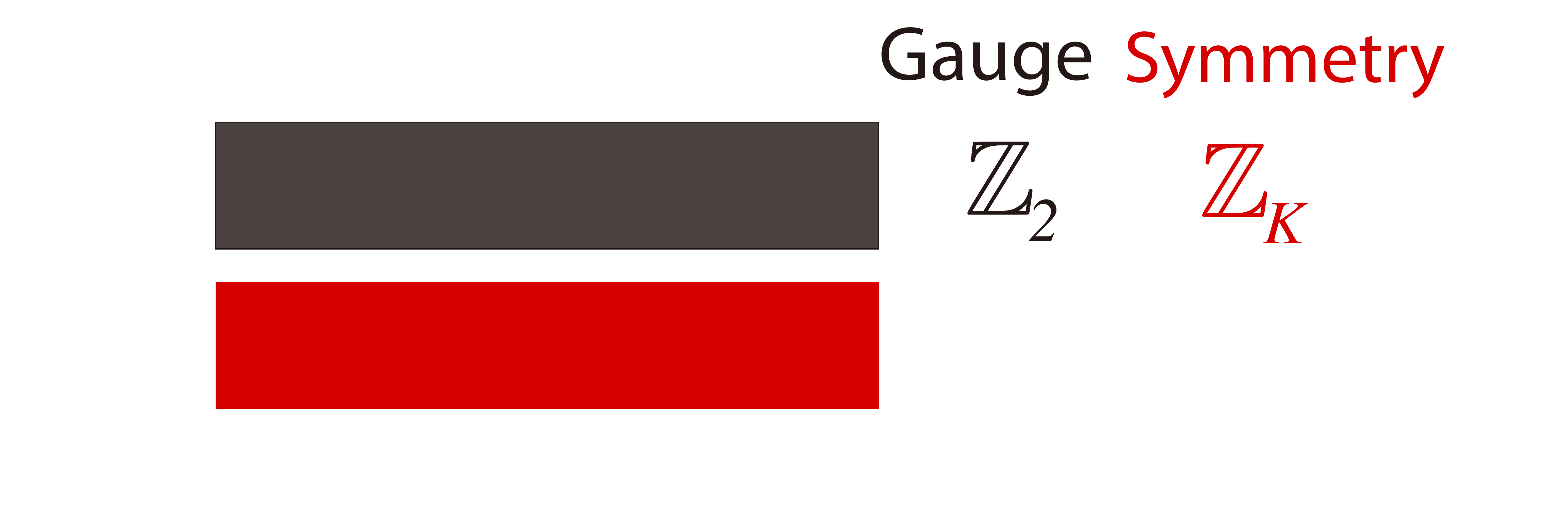}  }}     \\
                    & \multicolumn{3}{l|}{  }   \\ 
                    & \multicolumn{3}{l|}{  }   \\
                    & \multicolumn{3}{l|}{  }   \\ 
                   \hline
\multirow{2}{*}{$\mathsf{GT}$}   & $ q/4 \pi^2 a^1a^2da^2$ & $ \bar q /4 \pi^2 a^2a^1da^1$  &   \\ \cline{2-4} 
                      & 0 mod 2                                  & 0 mod 2                              &   \\ \hline
\multirow{1}{*}{$\mathsf{SEG}$} &               \multirow{1}{*}{0 mod $2K$}                  &         \multirow{1}{*}{ 0 mod $2K$}               &\multirow{4}{*}{}1     \\ 
%                      & 0 mod 4                                  & 0 mod 4                  &(0,2)                \\ 
%                       & 2 mod 4                                  & 2 mod 4                  &(2,0)                \\ 
%                        &                               &              &(2,2)                \\ \cline{2-4} 
                       \hline\hline
\end{tabular}
\end{table}
\begin{table} [h]\small
\caption{$\mathsf{SEG}$($\Z_2 $, $ \Z_K$). Gauge group and symmetry group are in different layers. $K\in\Z_{\rm odd}$ ($\Z_{\rm even}$) for  first (second) sub-table.}
\label{table:z2_zk_2}
\begin{tabular}{|c|c|c|c|}
\hline
\hline
                   \multirow{4}{*}{   \begin{minipage}   {0.65in}Symmetry assignment\end{minipage}    }  & \multicolumn{3}{c|}{ \multirow{4}{*}{ \includegraphics[width=4.5cm]{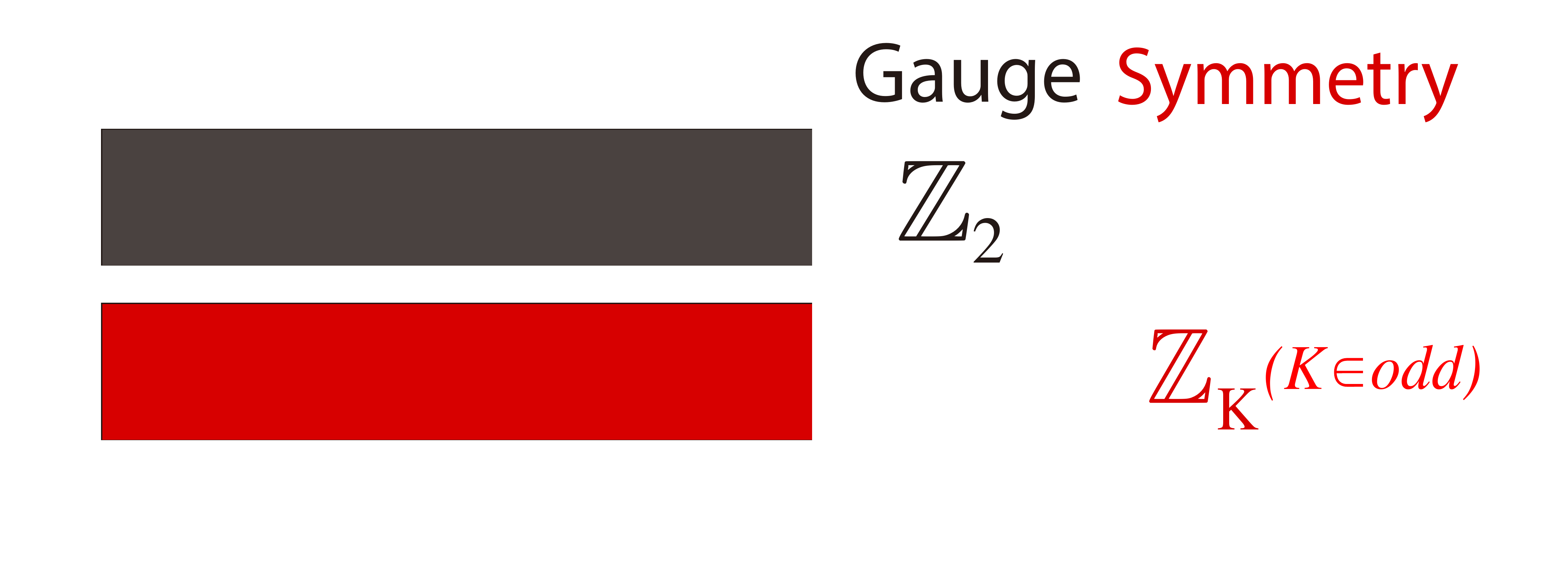}  }}     \\
                    & \multicolumn{3}{l|}{  }   \\ 
                    & \multicolumn{3}{l|}{  }   \\
                    & \multicolumn{3}{l|}{  }   \\ 
                   \hline
\multirow{2}{*}{$\mathsf{GT}$}   & $ q/4 \pi^2 a^1a^2da^2$ & $ \bar q /4 \pi^2 a^2a^1da^1$  &   \\ \cline{2-4} 
                      & 0 mod 2                                  & 0 mod 2                              &    \\ \hline
\multirow{1}{*}{$\mathsf{SEG}$} &               \multirow{1}{*}{0 mod $2K$}                  &         \multirow{1}{*}{ 0 mod $2K$}               &\multirow{4}{*}{}~~~~1~~~~  \\ 
    %     &                                &                    &              \\ 
%                       & 2 mod 4                                  & 2 mod 4                  &(2,0)                \\ 
%                        &                               &              &(2,2)                \\ \cline{2-4} 
                       \hline 
\end{tabular}  
\begin{tabular}{|c|c|c|c|}

\hline
                   \multirow{4}{*}{ \begin{minipage}   {0.65in}Symmetry assignment\end{minipage}    }  & \multicolumn{3}{c|}{ \multirow{4}{*}{ \includegraphics[width=5cm]{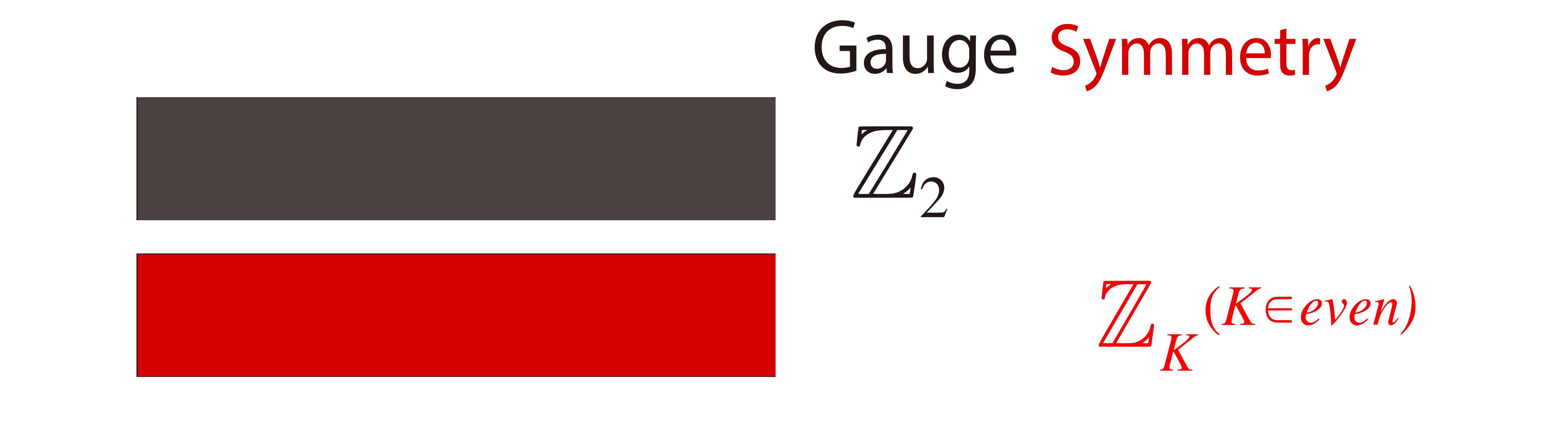}  }}     \\
                    & \multicolumn{3}{l|}{  }   \\ 
                    & \multicolumn{3}{l|}{  }   \\
                    & \multicolumn{3}{l|}{  }   \\ 
                   \hline
\multirow{2}{*}{$\mathsf{GT}$}   & $ q/4 \pi^2 a^1a^2da^2$ & $ \bar q /4 \pi^2 a^2a^1da^1$  &   \\ \cline{2-4} 
                      & 0 mod 2                                  & 0 mod 2                              &   \\ \hline
\multirow{2}{*}{$\mathsf{SEG}$} &               \multirow{2}{*}{}       0 mod $2K$                 &         \multirow{2}{*}{}          0 mod $2K$           &\multirow{2}{*}{}    \\ 
                      &                     $K$ mod $2K$        & $K$ mod $2K$           &      $2^2$         \\ 
 \cline{2-4} 
                       \hline\hline
\end{tabular}
 \end{table}
 
 We choose $N=K=2$ which was studied thoroughly in Ref.~\cite{3dset_chen} via a completely   different approach.  The results are collected in Tables \ref{table:z2_zk_1} and \ref{table:z2_zk_2} ($K=2$). 
  In Table \ref{table:z2_zk_1}, the symmetry charge is carried by the first layer. Before imposing symmetry, we find that both $q$  and $\bar q$  are $0\text{  mod  }2$, indicating that  topological interactions between layers are irrelevant. Mathematically, this conclusion can be achieved  from Eqs.~(\ref{eq:key_quantized_no_symmetry},\ref{eq:key_quantized_no_symmetry_plus}) by simply setting $N_1=2,N_2=1$.  Physically, it means that there is only one $\Z_2$ $\mathsf{GT}$ which is described by  the BF term with level-2: $i\frac{2}{2\pi}\int b\wedge da$.   
  After symmetry is imposed, both $q$ and $\bar q$ are $0\text{  mod  }4$.  This conclusion can be easily obtained by setting $N_1=2,K_1=2,N_2=K_2=1$ in Eq.~(\ref{eq:useful_1}). Physically,  after imposing symmetry, for each topological interaction, there is still only one choice of the coefficient but which is  always connected to zero   via a   periodic shift. As a result, the total number of $\mathsf{SEG}$s from this table is just one although the periodicity of both $q$ and $\bar q$ is enhanced by symmetry.

   In Table \ref{table:z2_zk_2} ($K=2$), the symmetry charge is carried by the second layer that is a trivial layer. In this case, we find that there are 2 distinct choices for both $q$  and $\bar q$: either $0\text{  mod  }4$ or $2\text{   mod  }4$.  Quantitatively, this result can be obtained by simply setting $N_1=2,N_2=1,K_1=1,K_2=2$ in Eqs.~(\ref{eq:useful_1},\ref{eq:useful_1_2_bar_q}).  
   As a result, there are in total $2^2$ $\mathsf{SEG}$s from this table.  Among them, the $\mathsf{SEG}$ with $q=\bar q=2\text{ mod }4$ can be simply regarded as stacking of symmetry enrichments from $(q,\bar q)$=$(2 \text{ mod }4, 0\text{ mod }4)$ and $(q,\bar q)$=$(0\text{ mod }4,2\text{ mod }4)$. In other words, both $a^1a^2da^2$ and $a^2a^1da^1$ topological interactions are present in this $\mathsf{SEG}$.

In summary, there are $1+2^2=5$ $\mathsf{SEG}$s with $G_g=\Z_2$ and $G_s=\Z_2$. One of them, labeled by $(2,2)$ in Table \ref{table:z2_zk_2} can be regarded as stacking of symmetry enrichment patterns of $(0,2)$ and $(2,0)$.
For generic even $K$ in Tables \ref{table:z2_zk_1} and \ref{table:z2_zk_2}, there are in total five $\mathsf{SEG}$s, just like $K=2$ case. For odd $K$, there are two $\mathsf{SEG}$s only. One is from Table \ref{table:z2_zk_1} where symmetry group is in the same layer as   gauge group. The other one is from Table \ref{table:z2_zk_2} where   gauge group and symmetry group are in different layers.    
%%%%%%%%%%%%%%%%%%

   \subsection{$\mathsf{SEG}(\Z_2\times\Z_2,\Z_2)$}\label{sec:two}
\begin{table*}[t]
\caption{$\mathsf{SEG}$$(\Z_2 \times  \Z_2, \Z_2)$. Three different ways of symmetry assignment are considered. Interestingly, all SEGs in first and second ways of symmetry assignment come from the untwisted $\Z_2\times\Z_2$ $\mathsf{GT}$ only. ``N/A'' means that SEGs do not exist. Those states necessarily either break symmetry or violate gauge invariance principle. For the former, the ground states should be discrete-symmetry-breaking phases. The latter may exist on the boundary of some (4+1)D systems.}\label{table:z2z2_z2_1}
\begin{tabular}{|c|c|c|c|c|c|c|c|c|c|c|c|}
\hline\hline
   \multirow{5}{*}{ \begin{minipage}   {0.7in}Symmetry assignment\end{minipage} } & \multicolumn{5}{c|}{\multirow{6}{*}{\includegraphics[width=4cm]{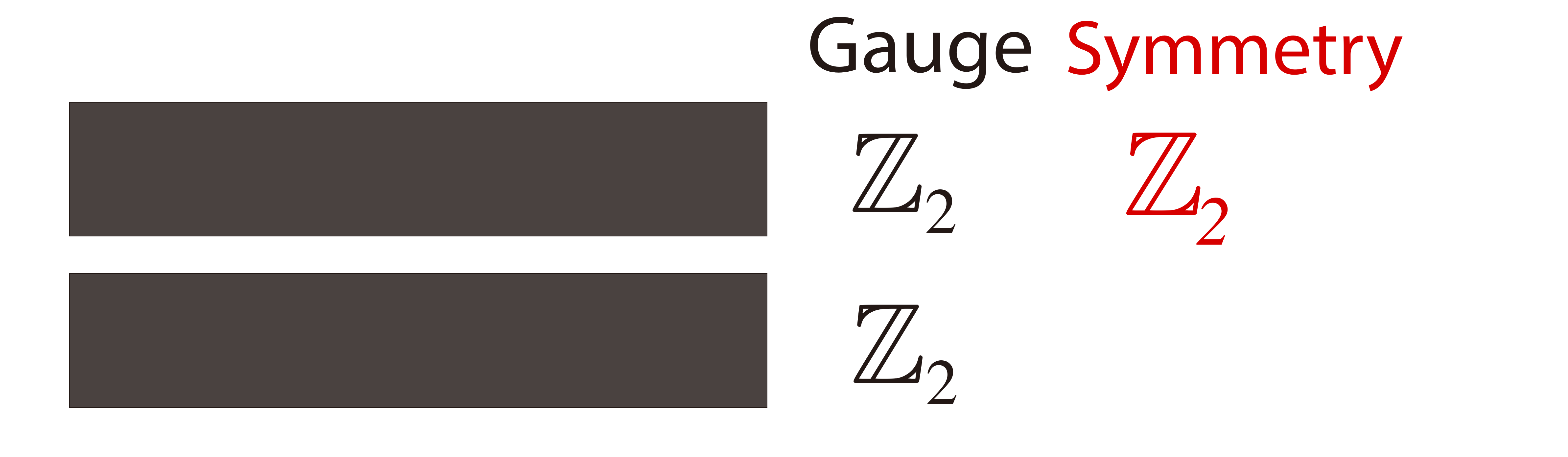}}}                                                                                                                                                                                                                                                                         \\
               & \multicolumn{5}{l|}{  }   \\
               & \multicolumn{5}{l|}{  }   \\ 
               & \multicolumn{5}{l|}{  }   \\
               & \multicolumn{5}{l|}{  }   \\  \hline
\multirow{2}{*}{\begin{minipage}   {0.6in}$\mathsf{GT}$\end{minipage}} &\multicolumn{2}{c|}{$\frac{q}{4\pi^2} a^1a^2da^2 $}&\multicolumn{2}{c|}{$ \frac{\bar q}{4\pi^2}a^2a^1da^1$}&\\
\cline{2-6 }&0 mod 4&2 mod 4&0 mod 4&2 mod 4&\\
\hline
\multirow{2}{*}{$\mathsf{SEG}$} &0 mod 8&\multirow{2}{*}{N/A}&0 mod 8&\multirow{2}{*}{N/A}& \\
\cline{2-2} \cline{4-4}  & 4 mod 8& & 4 mod 8&& $~2^2~$\\
\hline
\end{tabular}
\begin{tabular}{|c|c|c|c|c|c|c|c|c|c|c|c|}
\hline
\hline
   \multirow{5}{*}{ \begin{minipage}   {0.7in}Symmetry assignment\end{minipage} } & \multicolumn{5}{c|}{\multirow{6}{*}{\includegraphics[width=5cm]{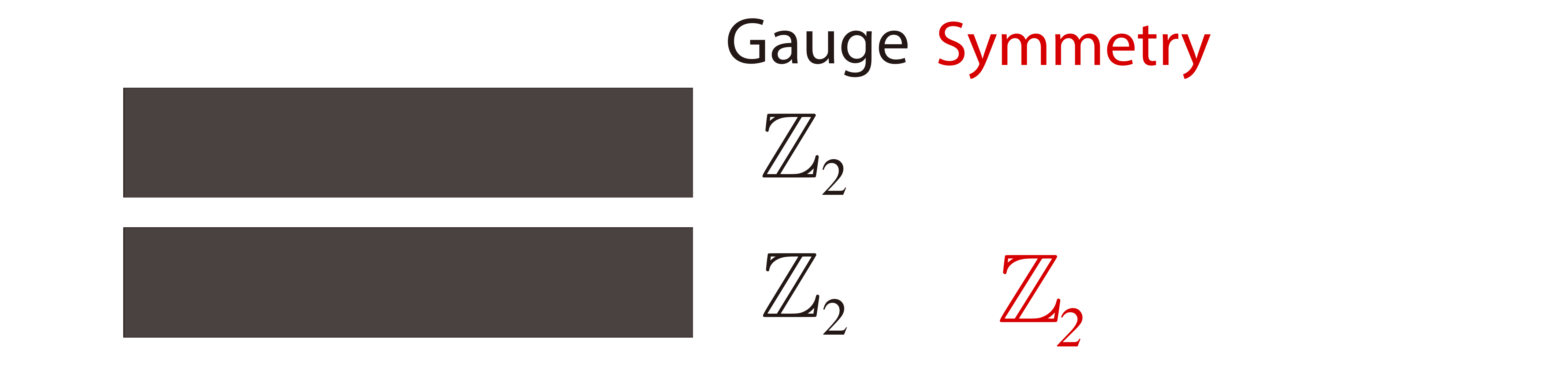}}}                                                                                                                                                                                                                                                                         \\
               & \multicolumn{5}{l|}{  }   \\
               & \multicolumn{5}{l|}{  }   \\ 
               & \multicolumn{5}{l|}{  }   \\
               & \multicolumn{5}{l|}{  }   \\  \hline
\multirow{2}{*}{\begin{minipage}   {0.5in}$\mathsf{GT}$\end{minipage}   } &\multicolumn{2}{c|}{$ \frac{q}{4\pi^2} a^1a^2da^2 $}&\multicolumn{2}{c|}{$ \frac{\bar q}{4\pi^2} a^2a^1da^1$}&\\
\cline{2-6 }&0 mod 4&2 mod 4&0 mod 4&2 mod 4&\\
\hline
\multirow{2}{*}{$\mathsf{SEG}$} &0 mod 8&\multirow{2}{*}{N/A}&0 mod 8&\multirow{2}{*}{N/A}& \\
\cline{2-2} \cline{4-4}  & 4 mod 8& & 4 mod 8&& $~2^2~$\\
\hline
\end{tabular}
\begin{tabular}{|c|c|c|c|c|c|c|c|c|c|c|c|}
\hline
   \multirow{5}{*}{ \begin{minipage}   {0.9in}Symmetry assignment\end{minipage} } & \multicolumn{11}{c|}{\multirow{6}{*}{\includegraphics[width=4cm]{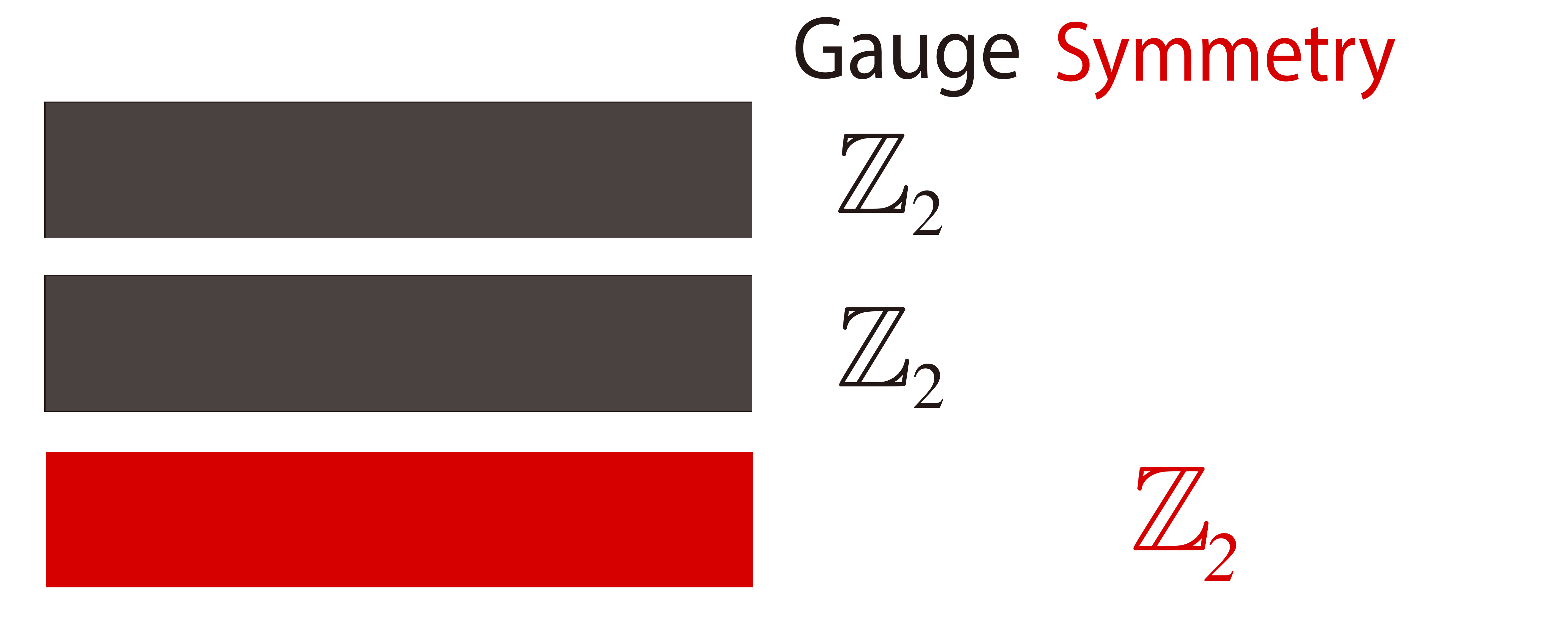}}}                                                                                                                                                                                                                                                                         \\
                            & \multicolumn{11}{l|}{  }   \\
               & \multicolumn{11}{l|}{  }   \\ 
               & \multicolumn{11}{l|}{  }   \\
               & \multicolumn{11}{l|}{  }   \\  \hline
\multirow{2}{*}{\begin{minipage}   {0.5in}$\mathsf{GT}$\end{minipage} } &\multicolumn{2}{c|}{$ a^1a^2da^2 $}&\multicolumn{2}{c|}{$ a^2a^1da^1$}& $ a^1a^3da^3$ & $ a^3a^1da^1$ &$ a^2a^3da^3$&$ a^3a^2da^2$& $ a^1a^2da^3$ &$a^2a^3da^1$ &\\
\cline{2-12 }&0 mod 4&2 mod 4&0 mod 4&2 mod 4&0 mod 2&0 mod 2&0 mod 2&0 mod 2&0 mod 2&0 mod 2&\\
\hline
\multirow{2}{*}{$\mathsf{SEG}$} &\multirow{2}{*}{0 mod 4}&\multirow{2}{*}{2 mod 4}&\multirow{2}{*}{0 mod 4}&\multirow{2}{*}{2 mod 4}&0 mod 4&0 mod 4&0 mod 4&0 mod 4&0 mod 4&0 mod 4&\quad ~\\
\cline{6-11} &&&&&2 mod 4&2 mod 4&2 mod 4&2 mod 4&2 mod 4&2 mod 4 & $2^8$\\
\hline
\hline
\end{tabular}
\end{table*}

The calculation in Sec.~\ref{sec:one} only involves one gauge group. Therefore, before imposing symmetry group, there is only one gauge theory, i.e., the untwisted one.  In the following, we calculate SEGs with $G_g=\Z_2\times\Z_2$ and $G_s=\Z_2$. Before imposing symmetry, there are already four topologically distinct $\mathsf{GT}$s labeled by $(q,\bar q)=(0 \text{ mod }4,0\text{ mod }4)$, $(0\text{ mod }4,2\text{ mod }4)$, $(2\text{ mod }4,0\text{ mod }4)$, and $(2\text{ mod }4, 2\text{ mod }4)$, which can be derived from Eqs.~(\ref{eq:key_quantized_no_symmetry},\ref{eq:key_quantized_no_symmetry_plus}) by setting $N_1=N_2=2$.  Under this circumstances,  \textit{Step-5} in Sec.~\ref{sec:general_procedures} cannot be skipped.   All SEGs are listed in Table~\ref{table:z2z2_z2_1}, where three different ways of symmetry assignment are considered. 

Taking the first symmetry assignment ($\Z_2$ symmetry is assigned to the first layer, see the first subtable of Table~\ref{table:z2z2_z2_1}) as an example, there are two choices of $q$ after symmetry is imposed: either $0\text{  mod  }8$ or $4\text{   mod  }8$. This result can be easily obtained by setting $N_1=2,K_1=2,N_2=2, K_2=1$ in Eq.~(\ref{eq:useful_1}).  Similarly, there are also two choices of $\bar q$. Therefore, totally there are $2^2$ $\mathsf{SEG}$s from the first subtable of Table~\ref{table:z2z2_z2_1}.  
However, one may wonder what is the parent gauge theory ($\mathsf{GT}$) for each choice. This line of thinking is the goal of Step-5 in Sec.~\ref{sec:general_procedures}. 
 Interestingly, both choices of $q$ mathematically  belong to the sequence ``$0 \text{ mod }4$''. In other words, $0\text{ mod }8$ and $4\text{ mod }8$, both of which belong to the sequence $0\text{ mod }4$ and thus are indistinguishable before imposing symmetry, become distinguishable after symmetry is imposed. This is nothing but a consequence of symmetry enrichment. 
 
 Meanwhile, both choices do not match the sequence ``$2\text{   mod  }4$'' at all, which is indicated by the mark ``N/A'' in the table. Similar analysis can be applied to $a^2a^1da^1$. 
 This phenomenon tells us that,   $\Z_2\times\Z_2$ $\mathsf{GT}$ labeled by $(q,\bar q)=(2\text{ mod }4, 2\text{ mod }4)$ cannot generate $\mathsf{SEG}$ descendants \emph{if symmetry is  assigned to either the first layer (the first subtable of Table~\ref{table:z2z2_z2_1}) or the second layer  (the second subtable of Table~\ref{table:z2z2_z2_1})}. Both layers are of type-I in Fig.~\ref{figure_steps}.  One may wonder what will happen if we still enforce $G_s$ on this twisted $\mathsf{GT}$ in such kinds of symmetry assignment.  Can the gauge group and symmetry group be compatible  with each other simultaneously? To answer these questions, recalling   the general procedure shown in Sec.~\ref{sec:general_procedures}, there are several conditions (symmetry requirement and gauge invariance) listed in Step-4 that determine  $\mathsf{SEG}(G_g,G_s)$.  Therefore, if there is a $\mathsf{SEG}$ replacing the mark ``N/A'', it \emph{either} breaks symmetry  \emph{or} preserves symmetry but violates gauge invariance principle.  The latter case is an anomalous $\mathsf{SEG}$ and possibly realizable on the boundary of some (4+1)D system.
 
In the third subtable of Table~\ref{table:z2z2_z2_1}, symmetry is assigned to the third layer, i.e., the type-II layer in Fig.~\ref{figure_steps}. It is clear that there are 8 linearly independent topological interaction terms that can be applied \cite{footnote_8_topo}.  In this symmetry assignment, each topological interaction term has two choices of its coefficient: either $0\text{  mod } 4$ or $2\text{ mod } 4$ (for $a^1a^2da^3$ and $a^2a^3da^1$, the result can be obtained from the general calculation in  Appendix~\ref{section_useful_1} and \ref{section_useful_2}). Therefore, totally, there are $2^8$ $\mathsf{SEG}$s. Interestingly, for those four $\mathsf{SEG}$s with topological interactions  $a^1a^2da^2$ and $a^2a^1da^1$ only, they can be simply regarded as stacking of a twisted $\Z_2\times\Z_2$ gauge theory and a direct product state with $\Z_2$ symmetry.

 \subsection{$\mathsf{SEG}(\Z_2\times\Z_2,\mathrm{U(1)})$}\label{sec:seg_with_u1} 
  
 In this part,  we discuss the gauge theory $\Z_2\times \Z_2$ enriched by the continuous symmetry $\mathrm{U(1)}$. The result can be obtained by following the general calculation in Appendix~\ref{appendix_sub_zn1zn2_k1k2u1_123}, \ref{appendix_subsection_zn1zn2_zku1_1}, and \ref{appendix_subsection_zn1zn2_zku1_2}.  Similar to the case of $\mathsf{SEG}(\Z_2\times \Z_2, \Z_2)$, we consider 3 ways to assign the symmetry, as shown in Table~\ref{table:z2z2_U(1)}.  
 Considering the first symmetry assignment  ($\mathrm{U(1)}$ is assigned at the first layer),  we  find that there is only one $\mathsf{SEG}(\Z_2\times \Z_2, \mathrm{U(1)})$ for both interaction terms with $q=\bar q=0$. In other words, this $\mathsf{SEG}$ is a descendant of  the untwisted $\Z_2\times\Z_2$ $\mathsf{GT}$ with $q=\bar q=0$, while all other three twisted $\mathsf{GT}$s do not have $\mathsf{SEG}$ descendants \emph{in this symmetry assignment}. Similarly, for the second symmetry assignment, there is also only one $\mathsf{SEG}$ and it is also a descendant of  the untwisted $\mathsf{GT}$. 
 
 However, there is one subtle feature that is absent for discrete symmetry group. $q=\bar q=0$ means that $q$ and $\bar q$ are absolutly zero with no periodicity (or periodicity=0 formally) after symmetry is imposed. We note that periodicity is always nonzero in all previous examples with discrete symmetry group.   It means that if we start with an untwisted $\mathsf{GT}$ but with $q=4$, the resulting gauge theory after imposing $\mathrm{U(1)}$ symmetry either breaks symmetry or violates gauge invariance principle.  For the latter case, the theory can be regarded as an anomalous $\mathsf{SEG}$ which is possibly realizable on the boundary of certain (4+1)D systems.

 Now we consider the third symmetry assignment (the last row in Table~\ref{table:z2z2_U(1)}) which is much more complex. There are 8  linearly independent  topological interaction terms of  $aada$ type \cite{footnote_8_topo}. We find that there are $2^3$   $\mathsf{SEG}(\Z_2\times \Z_2,\mathrm{U(1)})$. Each coefficient of $a^1a^2da^2$, $a^2a^1da^1$ and $a^1a^2da^3$ has two choices while the coefficient of  other $aada$ interaction terms vanish identically after periodicity shift, which leads to $2^3$ $\mathsf{SEG}$s.    
For $\mathsf{SEG}$s where only $a^1a^2da^2$  $a^2a^1da^1$ are considered (other topological interaction terms vanish), they can be simply regarded as the stacking of a twisted $\Z_2\times\Z_2$ gauge theory and a direct product state with $\mathrm{U(1)}$ symmetry. For $\mathsf{SEG}$s with   at least  $a^1a^2da^3$ topological interaction term, they are more interesting ones since they induce the nontrivial couplings between type-I layers and type-II layers as shown in Fig.~\ref{figure_steps}.

\begin{table*}[t] 
\centering
\caption{$\mathsf{SEG}(\Z_2\times \Z_2, \mathrm{U(1)})$ When the symmetry is assigned at the first and second layer, there is only one $\mathsf{SEG}$:$(q,\bar{q})$=(0,0). The notation N/A denotes that there is no $\mathsf{SEG}$ descendant  for \textit{the} specific symmetry assignment. $0$ means that $q$ or $\bar{q}$ exactly takes zero.}\label{table:z2z2_U(1)}
\begin{tabular}{|c|c|c|c|c|c|}
\hline
\hline
\multirow{4}{*}{ \begin{minipage}   {0.7in}Symmetry assignment\end{minipage} } &\multicolumn{5}{c|}{\multirow{4}{*}{\includegraphics[width=1.5in]{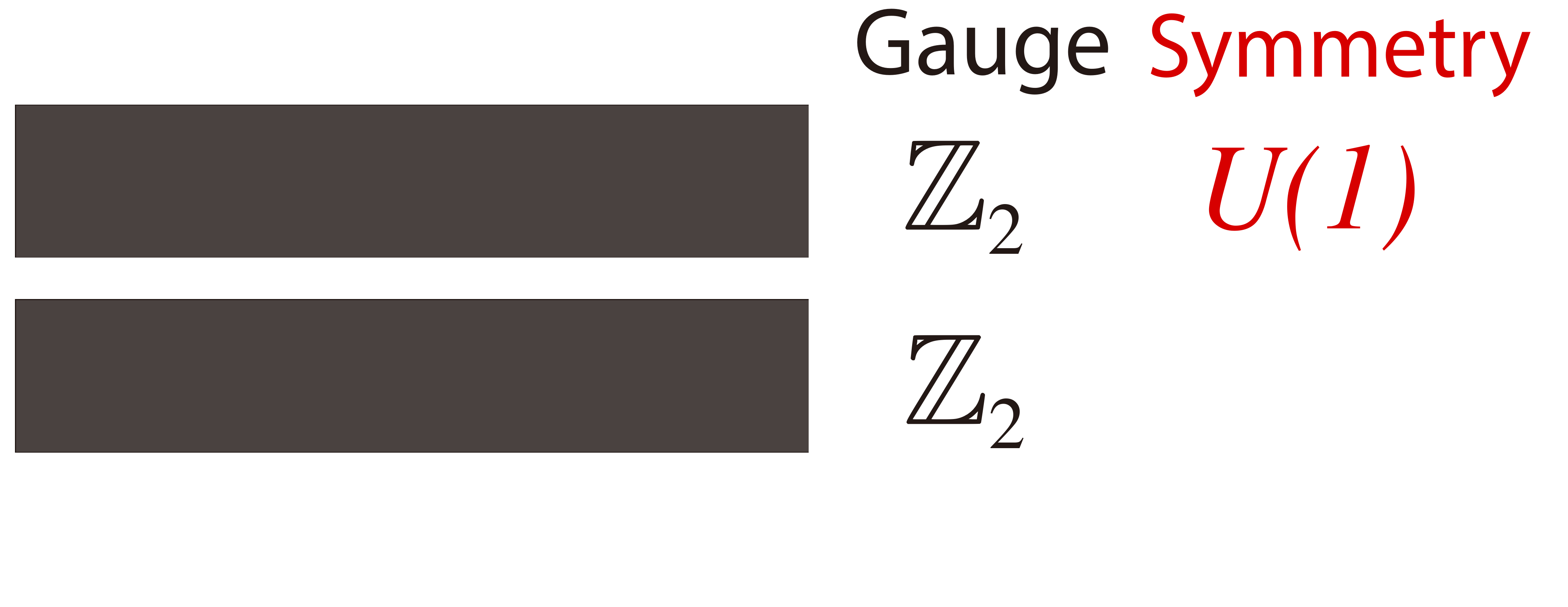}}} \\
   & \multicolumn{5}{l|}{  }   \\
              & \multicolumn{5}{l|}{  }   \\ 
               & \multicolumn{5}{l|}{  }   \\
\hline
\multirow{2}{*}{$\mathsf{GT} $} &\multicolumn{2}{c|}{ $q/4\pi^2 a^1a^2da^2$} &\multicolumn{2}{c|}{$\bar{q}/4\pi^2 a^2a^2da^1$}& \\
\cline{2-6}
&  0 mod 4 & 2 mod 4& 0 mod 4 &2 mod 4& $~~~~~~~$\\
\hline 
$\mathsf{SEG}$ & $0$& N/A & $0$ &N/A &1\\
\hline
\end{tabular}
\begin{tabular}{|c|c|c|c|c|c|}
\hline
\hline
\multirow{4}{*}{ \begin{minipage}   {0.7in}Symmetry assignment\end{minipage} } &\multicolumn{5}{c|}{\multirow{4}{*}{\includegraphics[width=1.52in]{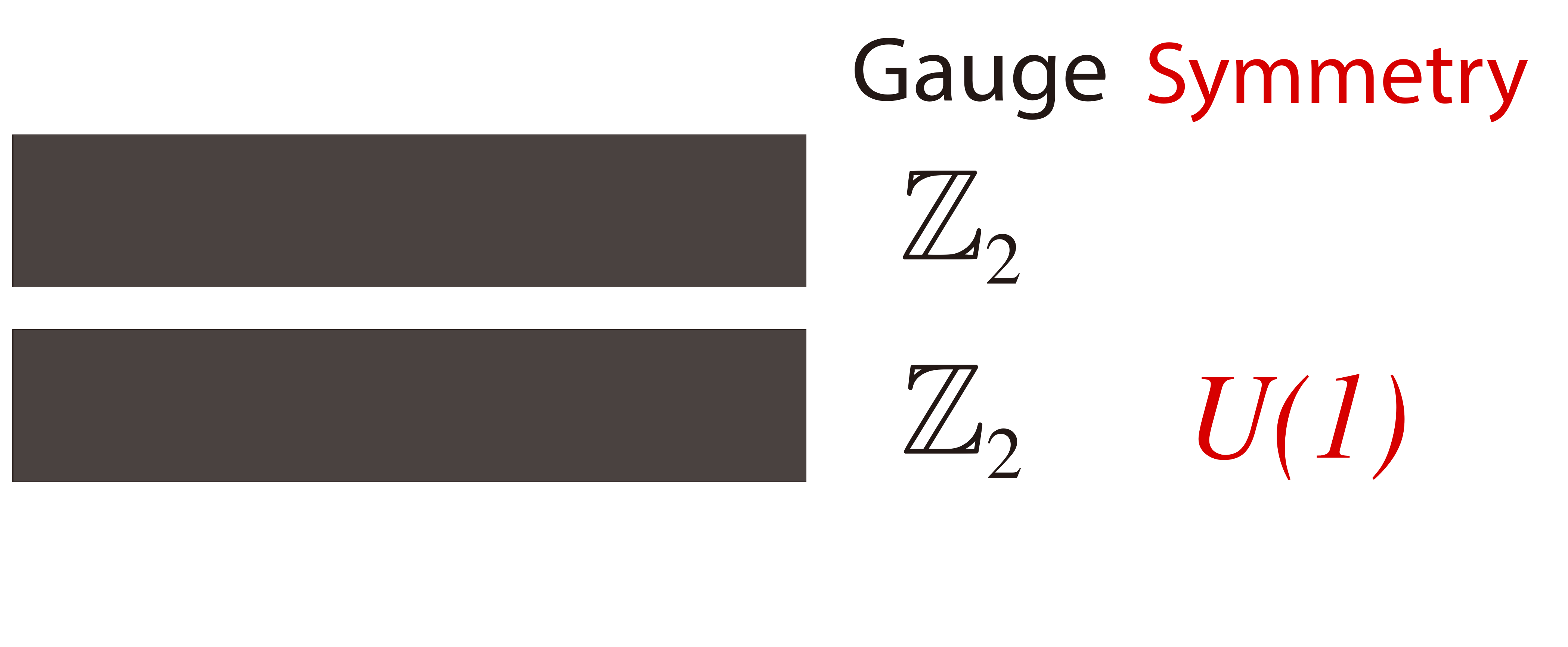}}} \\
   & \multicolumn{5}{l|}{  }   \\
              & \multicolumn{5}{l|}{  }   \\ 
               & \multicolumn{5}{l|}{  }   \\
\hline
\multirow{2}{*}{$\mathsf{GT} $} &\multicolumn{2}{c|}{ $q/4\pi^2 a^1a^2da^2$} &\multicolumn{2}{c|}{$\bar{q}/4\pi^2 a^2a^2da^1$}& \\
\cline{2-6}
&  0 mod 4 & 2 mod 4& 0 mod 4 &2 mod 4&$~~~~~~$ \\
\hline 
$\mathsf{SEG}$ & $0$& N/A & $0$ &N/A &1\\
\hline
\end{tabular}
\begin{tabular}{|c|c|c|c|c|c|c|c|c|c|c|c|}
\hline
   \multirow{5}{*}{ \begin{minipage}   {0.9in}Symmetry assignment\end{minipage} } & \multicolumn{11}{c|}{\multirow{6}{*}{\includegraphics[width=4cm]{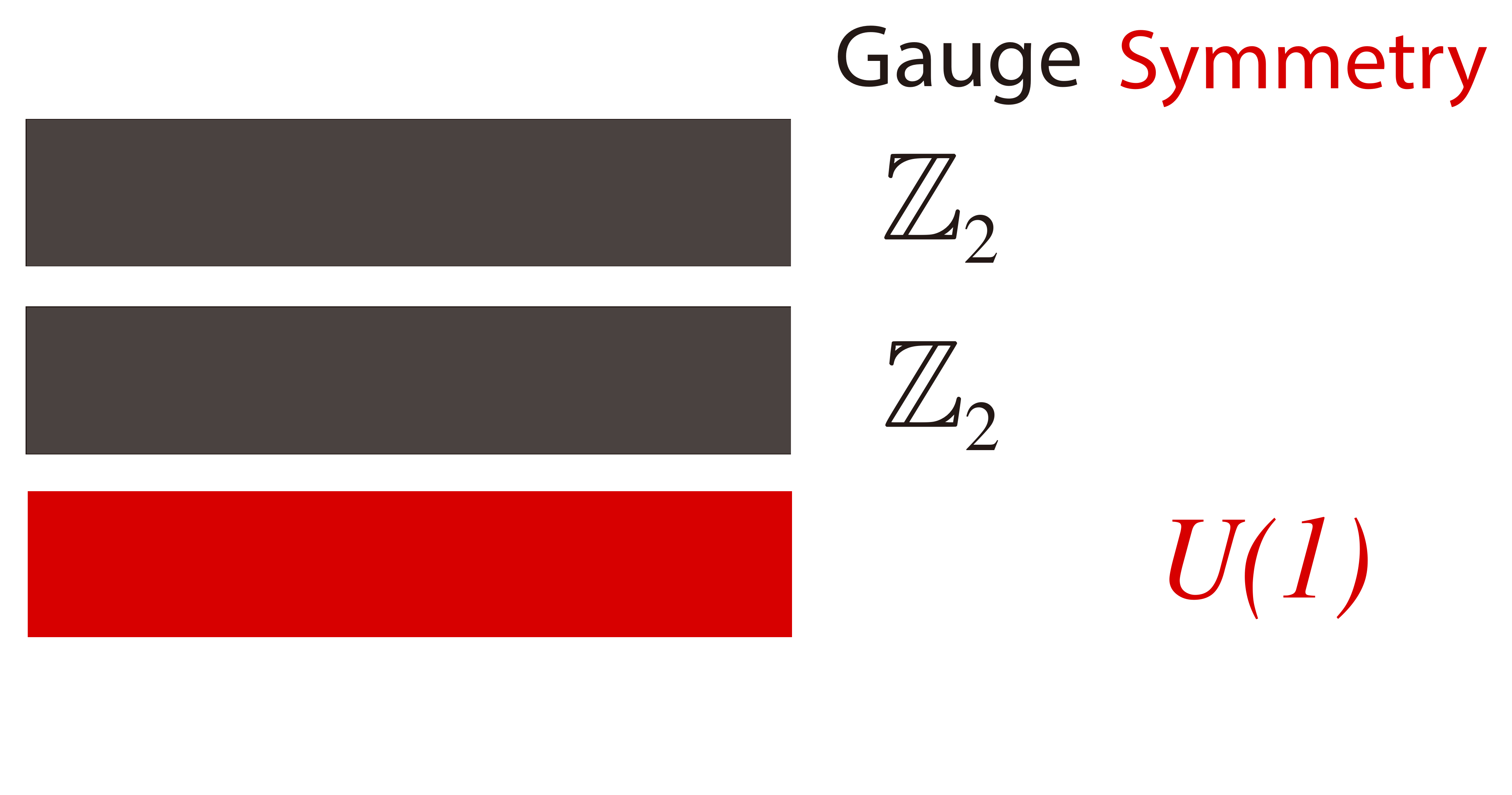}}}                                                                                                                                                                                                                                                                         \\
                            & \multicolumn{11}{l|}{  }   \\
               & \multicolumn{11}{l|}{  }   \\ 
               & \multicolumn{11}{l|}{  }   \\
               & \multicolumn{11}{l|}{  }   \\  \hline
\multirow{2}{*}{\begin{minipage}   {0.5in}$\mathsf{GT}$\end{minipage} } &\multicolumn{2}{c|}{$ a^1a^2da^2 $}&\multicolumn{2}{c|}{$ a^2a^1da^1$}& $ a^1a^3da^3$ & $ a^3a^1da^1$ &$ a^2a^3da^3$&$ a^3a^2da^2$& $ a^1a^2da^3$ &$a^2a^3da^1$ &\\
\cline{2-12 }&0 mod 4&2 mod 4&0 mod 4&2 mod 4&0 mod 2&0 mod 2&0 mod 2&0 mod 2&0 mod 2&0 mod 2&\\
\hline
\multirow{2}{*}{$\mathsf{SEG}$} &\multirow{2}{*}{0 mod 4}&\multirow{2}{*}{2 mod 4}&\multirow{2}{*}{0 mod 4}&\multirow{2}{*}{2 mod 4}&\multirow{2}{*}{$0$}&\multirow{2}{*}{$0$}&\multirow{2}{*}{$0$}&\multirow{2}{*}{$0$}&0 mod 4&\multirow{2}{*}{$0$}&     \\
 &&&&&&&&&2 mod 4& &  ~ $2^3$ ~~\\
\hline
\hline
\end{tabular}
\end{table*}

 \section{Probing $\mathsf{SET}$ orders}\label{sec:fractionalization_3loop}
 
In Sec.~\ref{sec:examples},  we have constructed anomaly-free $\mathsf{SEG}$s in a few concrete examples.  In this section, we probe   $\mathsf{SET}$  orders possessed by the ground states of $\mathsf{SEG}$s. Then, the map from $\mathsf{SEG}$s to $\mathsf{SET}$s in Fig.~\ref{figure_tree} is achieved.  In order to identify $\mathsf{SET}$ order in a given $\mathsf{SEG}$, one should know the topological orders and symmetry-enriched properties.  

Given a gauge group $G_g$, the total number of topological orders is generically smaller than that of $\mathsf{GT}$s that are classified by $\mathcal{H}^4(G_g, \mathrm{U(1)})$. Intuitively, the labelings of gauge fluxes / gauge charges probably have redundancy from the aspect of  topological orders. For example, if $G_g=\Z_2\times\Z_2$, there are four $\mathsf{GT}$s. However, at least $\mathsf{GT}$ with $q=2\text{ mod } 4$ and $\bar q=0 \text{ mod }4$ and $\mathsf{GT}$ with $\bar q=2\text{ mod } 4$ and $ q=0 \text{ mod }4$ share the same topological order since both are just connected to each other via exchanging superscripts $1$ and $2$. 
 
 For the sake of simplicity, in this section, we will only consider $G_g=\Z_N$ such that both $\mathsf{GT}$ and topological order are unique. In these cases,  we find that: \emph{(i)} quasiparticles that carry unit gauge charge of the gauge group $G_g$ may carry fractionalized symmetry charge of the symmetry group $G_s$, which is classified by the second group cohomology with $G_g$ coefficient: $\mathcal{H}^2(G_s,G_g)$; \emph{(ii)} there is an interesting mixed version of three-loop braiding statistics among symmetry fluxes and gauge fluxes.   Both features are  gauge-invariant and topological, which can be detected in experiments.

  \subsection{$\mathsf{SET}$ orders in $\mathsf{SEG}(\Z_2,\Z_K)$ with $K\in\Z_{\rm even}$}\label{sec:twist_z2z2_subsection}
  In this part, we probe $\mathsf{SET}$ orders with $\Z_2$ gauge group and $\Z_2$ symmetry group in the five $\mathsf{SEG}$s listed in Table~\ref{table:z2_zk_1} and Table~\ref{table:z2_zk_2}. General even $K$ is straightforward. When the gauge group $G_g$ only includes one $\Z_N$ subgroup, e.g., $G_g=\Z_2$, there is only one $\mathsf{GT}$, i.e., the untwisted one. The topological order of the $\mathsf{GT}$  is dubbed ``$\Z_N$ topological order'', characterized by the charge-loop braiding statistics data, i.e., the $e^{i 2\pi/N}$ phase accumulated by a unit gauge charge moving around a unit gauge flux. For $N=2$, the phase is just $e^{i\pi}=-1$.  Due to this simplification,  in order to characterize $\mathsf{SET}$ orders in these five $\mathsf{SEG}$s,  the only remaining task is to diagnose the symmetry-enriched properties. From the following analysis, we  obtain five distinct $\mathsf{SET}$ orders with $\Z_2$ topological order and $\Z_2$ global symmetry.

 \begin{figure}[t]
\centering
\includegraphics[width=8.5cm]{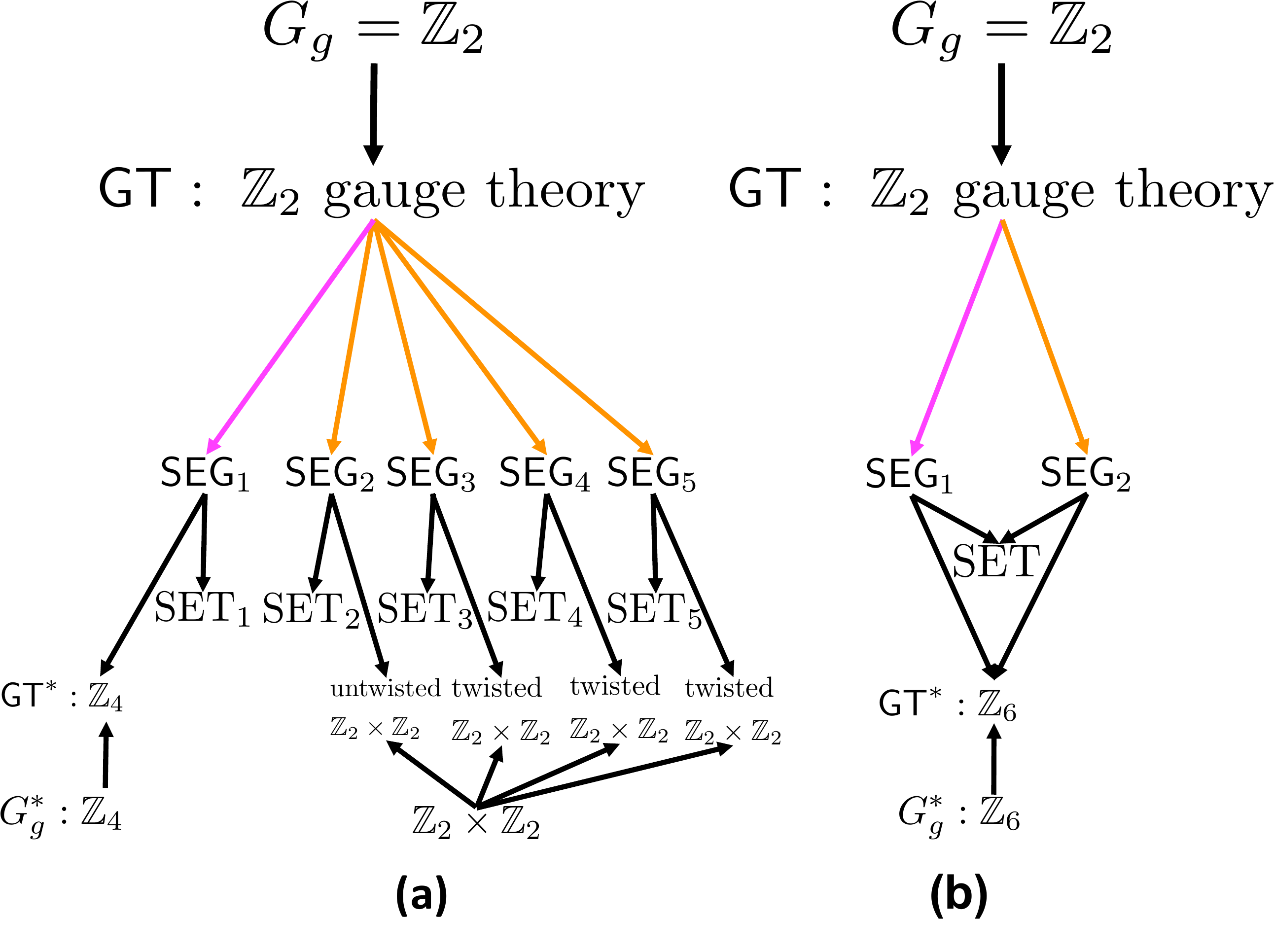}
\caption{(Color online) Two concrete examples of webs of gauge theories shown in  Fig.~\ref{figure_tree}. $\mathsf{SEG}(\Z_2,\Z_2)$ and $\mathsf{SEG}(\Z_2,\Z_3)$ are shown in (a) and (b), respectively. $\mathsf{SEG}_1$ in (a) can be found in Table~\ref{table:z2_zk_1}. $\mathsf{SEG}_{2,\cdots,5}$ in (a) can be found in Table~\ref{table:z2_zk_2}. $\mathsf{SEG}_1$ in (b) can be found in Table~\ref{table:z2_zk_1} by setting $K=3$. $\mathsf{SEG}_2$ in (b) can be found in the first subtable of Table~\ref{table:z2_zk_2} by setting $K=3$.}
\label{figure_example}
\end{figure}

  \subsubsection{  $\mathsf{SEG}(\Z_2,\Z_K)$ with $K\in\Z_{\rm even}$  in Table~\ref{table:z2_zk_1}}\label{sec:twist_z2z2_1}
 For \textit{the} $\mathsf{SEG}$ in Table~\ref{table:z2_zk_1}, we may consider the following action in the presence of excitation terms ($K=2$ as an example):
  \begin{align}
     S\!=i\frac{2}{2\pi}\!\!\int \! b\wedge da +i\frac{1}{2\pi}\! \int \!b\wedge dA  +i\!\int \!b\wedge *\Sigma +i\!\int \!a\wedge *j\,,\label{eq:action_z2_z2_1_twist}
\end{align}
where the 2-form tensor $\Sigma$ represents the unit loop excitation current (world-sheet) of the $\Z_2$ gauge theory.  The 1-form vector $j$ represents the unit gauge particle current (world-line) of the $\Z_2$ gauge theory. Since only one layer is considered in this case, the superscripts of $b^1,a^1$ are removed. The background gauge field $A$ is   constrained by Eq.~(\ref{eq:holo}) with $K_1=2$.  Next, integrating out $b$ field leads to:
$\frac{2}{2\pi}da=-*\Sigma-\frac{1}{2\pi}dA\,.
$ 
Then, $a$ can be formally solved by adding $*d*$ in both sides: $a=-\pi\frac{*d}{\hat{\Delta} }\Sigma-\frac{1}{2}A\,$, where the Laplacian operator $\hat{\Delta}\equiv *d*d$.  Plugging this expression into the last term of Eq.~(\ref{eq:action_z2_z2_1_twist}), we obtain the following effective action about excitations in the presence of symmetry twist: 
$-i\frac{1}{2}\int A\wedge *j+i \pi \int j\wedge d^{-1} \Sigma\,
$. In this effective action, the second term    characterizes the   $\Z_2$ topological order with charge-loop braiding phase $e^{i\pi}=-1$. Mathematically, this is a Hopf term and represents the long-range Aharonov-Bohm statistical interaction between gauge fluxes (i.e., the loop excitations) and particles. The operator $d^{-1}=\frac{d}{\hat{\Delta} }$ is a formal notation defined as the operator inverse of $d$, whose exact form can be understood in momentum space by Fourier transformations. The first term of this effective action encodes the symmetry-enriched  properties of the $\mathsf{SEG}$. It indicates that the unit gauge charge carries $1/2$ symmetry charge of symmetry group $G_s=\Z_2$, which corresponds to the second group cohomology classification $\mathcal{H}^2(G_s,G_g)=\Z_2$ (see Appendix~\ref{appendix_h2z3z2} for details).

\textbf{In summary}, for the $\mathsf{SEG}$ given by Table~\ref{table:z2_zk_1},  the $\Z_2$ gauge charged bosons carry half quantized $\Z_2$ symmetry charge.  This is the first $\mathsf{SET}$ order we identify. 

  \subsubsection{$\mathsf{SEG}(\Z_2,\Z_K)$ with $K\in\Z_{\rm even}$   in Table~\ref{table:z2_zk_2}}

For Table~\ref{table:z2_zk_2}, we first consider the $q$-topological interaction term. The action in the presence of $A$ is given by ($K=2$ as an example):
  \begin{align}
     &S=i\frac{2}{2\pi}\int  b^1\wedge da^1+i\frac{1}{2\pi}\int  b^2\wedge da^2 +i\frac{1}{2\pi} \int b^2\wedge dA \nonumber\\
    &\! \!+\!i\frac{q}{4\pi^2}\!\!\int \!a^1\!\wedge a^2\!\wedge \! d a^2\!+\!i\!\!\int\! b^1\!\wedge *\Sigma^1\! +\!i\sum_I^2\! \!\int \! a^I\!\wedge \!*j^I\!,\label{eq:action_z2_z2_2_twist}
\end{align}
where $\Sigma^1$ and $\{j^I\}$ are loop excitation currents and particle excitation currents of the $I$th layer respectively. $\Sigma^2$ is not considered for the reason that the second layer is trivial and   $\Sigma^2$ carries $0\text{  mod  }2\pi$ fluxes which are not detectable. One may first integrate out $\{b^I\}$, which enforces that the path-integral configurations of $\{a^I\}$ are completely fixed by excitations and the background gauge field:
$a^1=-\frac{2\pi}{2}*d^{-1}\Sigma^1 \,,\,a^2=-\frac{2\pi}{2}*d^{-1}\sigma\,
$. Here, the symbol $d^{-1}$ has been defined in Sec.~\ref{sec:twist_z2z2_1}. The new 2-form variable $\sigma$ is defined through: 
$\sigma=*\frac{2}{2\pi}dA\,
$ which represents the number density / current of the $\pi$-symmetry  twist induced by the background gauge field. Plugging the expressions of $\{a^I\}$  into $\{j^I\}$-dependent terms in Eq.~(\ref{eq:action_z2_z2_2_twist}), we obtain the following effective action terms:
$i\pi\int j^1\wedge d^{-1}\Sigma^1+i\int j^2\wedge *A\,$, 
where the first Hopf term indicates that the first layer has a $\Z_2$ topological order. The second term indicates that the quasiparticles in the second layer carry integer symmetry charge. In other words, there doesn't exist symmetry fractionalization.

\begin{figure}[t]
\centering
\includegraphics[width=7.5cm]{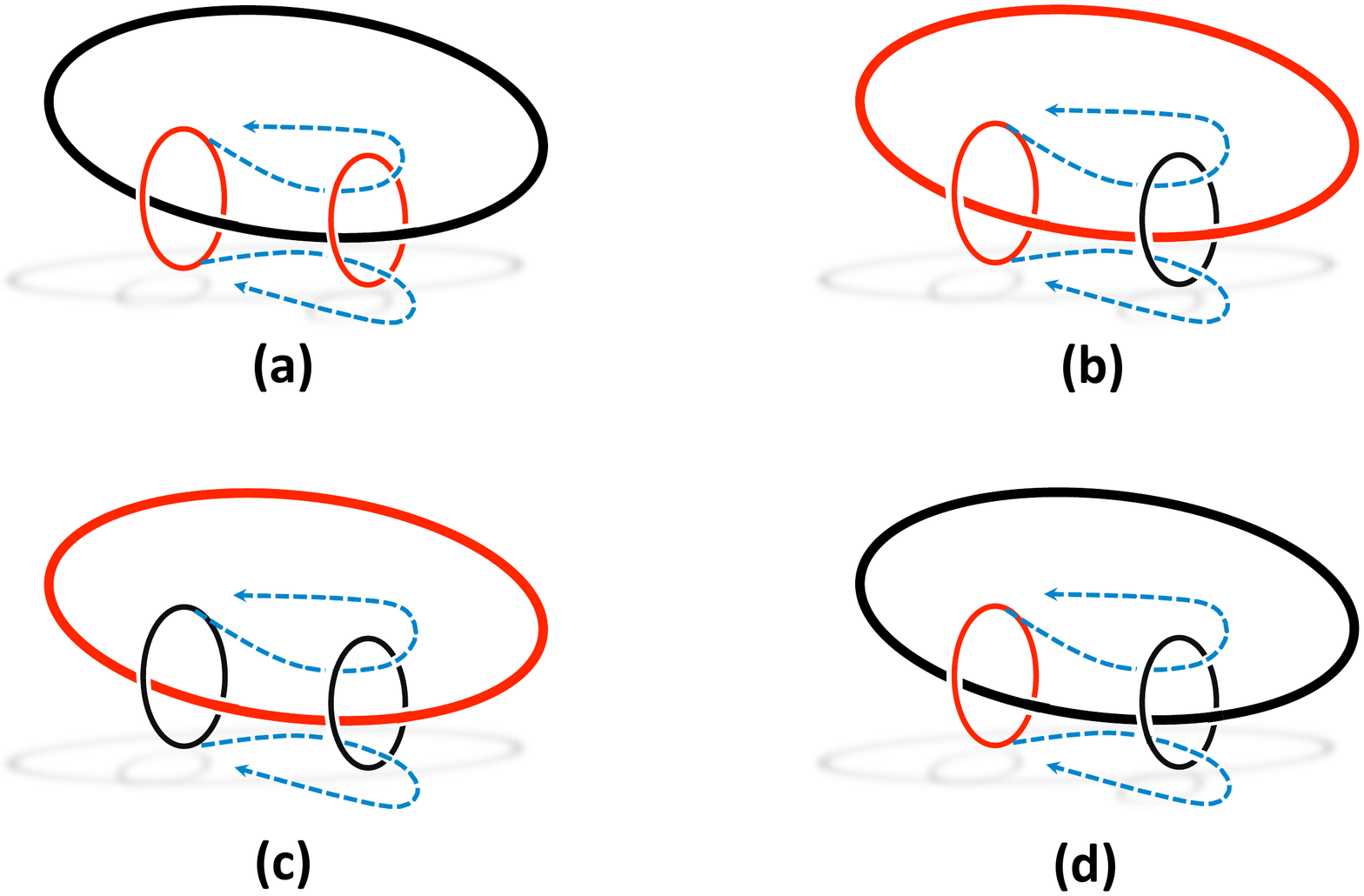}
  \caption{(Color online) A mixed version of three-loop braiding process among gauge fluxes and symmetry fluxes in $\mathsf{SEG}(\Z_2,\Z_2)$ of Table~\ref{table:z2_zk_2}. Loops in red and black denote symmetry flux loop ($\sigma$) and gauge flux loop ($\Sigma^1$), respectively. The dashed curves shows the trajectory of one loop that moves around another loop, both of which are linked to the base loop. } 
\label{figure_3loop}
\end{figure}

Despite that, we will show that there is interesting \emph{mixed} three-loop statistics among symmetry fluxes ($\sigma$) and gauge fluxes ($\Sigma^1$). For this purpose,  plugging the expressions of $\{a^I\}$   into the $q$-dependent term in Eq.~(\ref{eq:action_z2_z2_2_twist}), we obtain:
$ -i\frac{q\pi}{4}\int (*d^{-1}\Sigma^1)\wedge (*d^{-1}\sigma)\wedge (* \sigma)=-i\int (*d^{-1}\Sigma^1)\wedge \frac{\pi q}{4} (*d^{-1}\sigma)\wedge (* \sigma)
$ which is the topological invariant    that characterizes the  \emph{mixed}  three-loop statistics among symmetry fluxes and gauge fluxes and provides important symmetry-enriched properties of $\mathsf{SEG}$s. This mixed version of three-loop statistics  enriches our previous understandings on three-loop statistics among gauge fluxes \cite{spt18,spt19,spt20,jiang_ran04,wang_wen_3loop}.  Pictorially, the topological invariant corresponds to the three-loop process shown in Fig.~\ref{figure_3loop}(a) where the gauge flux $\Sigma^1$ is a  {base loop} (a term coined by Wang and Levin \cite{spt18}). The entire process leads to Berry phase (denoted by $\theta_{\sigma,\sigma;\Sigma^1}$):
$\theta_{\sigma,\sigma;\Sigma^1}= 2\times\frac{q\pi}{4}=\pi\,,
$ where $q=2$ is used  and the factor of $2$ is due to the fact that the full braiding process accumulates two times of half-braiding (exchange between $\sigma$ and $\sigma$ in the presence of the base loop $\Sigma^1$). If the base loop is provided by $\sigma$ instead, the topological invariant gives rise to the full braiding of another $\sigma$ around a $\Sigma^1$ as shown in Fig.~\ref{figure_3loop}(b), and the associated Berry phase is given by: 
$\theta_{\sigma,\Sigma^1;\sigma}=  \frac{q\pi}{4}=\frac{\pi}{2} \text{ mod }\pi\,,
$ 
where $\pi$ phase ambiguity arises from the possibility that $\Z_2$ gauge charge  may be attached to $\sigma$ such that there is $\pi$ phase contribution from the Aharonov-Bohm phase from the topological invariant $i\pi\int j^1\wedge d^{-1}\Sigma^1$. 
 
Likewise, the $\bar q$ term can also  be written in terms of the topological invariant:
$ -i\frac{\bar q\pi}{4}\int (*d^{-1}\sigma)\wedge (*d^{-1}\Sigma^1)\wedge (* \Sigma^1)
$. Pictorially, the topological invariant corresponds to the three-loop process shown in Fig.~\ref{figure_3loop}(c) where the symmetry flux $\sigma$ is a  base loop. The entire process leads to Berry phase (denoted by $\theta_{\Sigma^1,\Sigma^1;\sigma}$):
$\theta_{\Sigma^1,\Sigma^1;\sigma}= 2\times\frac{\bar q\pi}{4}=\pi\,,$ 
 where $\bar q=2$ is used for the $\mathsf{SEG}$ labeled by $ (0,2)$ in Table~\ref{table:z2_zk_2}. By choosing $\Sigma^1$ as the base loop, we may obtain the Berry phase accumulated by fully braiding $\Sigma^1$ around $\sigma$ with the base loop provided by another $\Sigma^1$ [see Fig.~\ref{figure_3loop}(d)]:
$\theta_{\sigma,\Sigma^1;\Sigma^1}=  \frac{\bar q\pi}{4}=\frac{\pi}{2} \text{ mod }\pi\,,
$ where $\pi$ phase ambiguity arises from the possibility that $\Z_2$ gauge charge  may be attached to $\sigma$ such that there is $\pi$ phase contribution from the Aharonov-Bohm phase from the topological invariant $i\pi\int j^1\wedge d^{-1}\Sigma^1$.

\textbf{In summary}, for the four $\mathsf{SEG}$s given by Table~\ref{table:z2_zk_2}, they support four different $\mathsf{SET}$ orders. All point-particles are either symmetry-neutral or carry integer $\Z_2$ symmetry charge. In other words, symmetry is not fractionalized and charge-loop braiding data is always trivial.  However, they can be experimentally distinguished by the mixed three-loop braiding process.  In total, we obtain five distinct $\mathsf{SET}$ orders with $\Z_2$ topological order and $\Z_2$ global symmetry. Likewise, for generic even $K$, there are  also   five $\mathsf{SET}$ orders.

  \subsection{$\mathsf{SET}$ orders in $\mathsf{SEG}(\Z_2,\Z_K)$ with $K\in\Z_{\rm odd}$}\label{sec:twist_z2z3_subsection}
We consider $K=3$ as an example. General odd $K$ is straightforward. In this case, there are two distinct $\mathsf{SEG}$s that are collected in Table~\ref{table:z2_zk_1} ($K = 3$) and the first subtable of Table \ref{table:z2_zk_2} ($K = 3$) respectively. For the first $\mathsf{SEG}$, the discussion is similar to that of $K=2$ in Table \ref{table:z2_zk_1}. We start with the action (\ref{eq:action_z2_z2_1_twist}) and the  background gauge field $A$ is   now constrained by Eq.~(\ref{eq:holo}) with $K_1=3$. Integrating out $b,a$ leads to $-i\frac{1}{2}\int A\wedge *j+i \pi \int j\wedge d^{-1} \Sigma\,$ where the first term indicates that the bosons (denoted by ``$e$'') that carry unit $\Z_2$ gauge charge also carry $1/2$ symmetry charge of $\Z_3$ group. However, there is no projective representation (with $\Z_2$ coefficient)  for $\Z_3$ symmetry group indicated by the trivial second group cohomology: $\mathcal{H}^2(\Z_3,\Z_2)=\Z_1$ (see Appendix~\ref{appendix_h2z3z2}), which means that this half-quantized symmetry charge cannot be  detected by   symmetry fluxes. The physical effect of this half-quantized symmetry charge is completely identical to that of $-1$ symmetry charge. 

More physically, let us perform an Aharonov-Bohm experiment by inserting symmetry fluxes (a loop) with flux $\Phi_A=0,\frac{2\pi}{3},\frac{4\pi}{3}$. The boson $e$ that moves around a symmetry flux with $\Phi_A$ will pick up a Berry phase $e^{i\frac{1}{2}\Phi_A}$ where $1/2$ is the symmetry charge carried by $e$. However, during this process, it is possible that a gauge flux $(\Phi_g=0,\pi)$ is dynamically excited and eventually attached   to the symmetry flux.  As a result, an additional Berry phase is accumulated: $e^{i\Phi_g}$, leading to the Berry phase  $e^{i\frac{1}{2}\Phi_A+\Phi_g}$. After repeating the experiments for each $\Phi_A$ sufficient times, the observer will eventually collect two data for each  symmetry flux. If $\Phi_A=0$, the Berry phase is either $0$ or $e^{i\pi}$; If $\Phi_A=\frac{2\pi}{3}$, the Berry phase is either $e^{i\frac{\pi}{3}}$ or $e^{i\frac{4\pi}{3}}$; If $\Phi_A=\frac{4\pi}{3}$, the Berry phase is either $e^{i\frac{2\pi}{3}}$ or $e^{i\frac{5\pi}{3}}$. It is clear that these observed data can be exactly obtained by considering the boson that carry unit gauge charge and $-1$ non-fractionalized symmetry charge whose Berry phase is given by $e^{-i\Phi_A+i\Phi_g}$. In other words, the half-quantized symmetry charge can not be distinguished from $-1$ symmetry charge. Therefore, for $\mathsf{SEG}$ in  Table~\ref{table:z2_zk_1} ($K = 3$), there is no symmetry fractionalization.

 For the second $\mathsf{SEG}$ (the first subtable of Table \ref{table:z2_zk_2} with $K=3$), since there doesn't exist nontrivial topological interactions between the two layers, this $\mathsf{SEG}$ is nothing but a simple stacking of a $\Z_2$ gauge theory and a direct product state with   $\Z_3$ symmetry. By definition, it is still a $\mathsf{SEG}$ but it doesn't have interesting symmetry-enriched properties. 
 
\textbf{In summary}, both $\mathsf{SEG}$s support the same $\mathsf{SET}$ order as shown schematically in Fig.~\ref{figure_example}(b). In this $\mathsf{SET}$ order,  the topological order is $\Z_2$-type. However, the $\Z_3$ symmetry  always trivially acts on the topological order due to the absence of both symmetry fractionalization and mixed three-loop braiding statistics. In other words, there is no interesting interplay beween $\Z_2$ topological order and $\Z_3$ symmetry. Likewise, for generic odd $K$, there is also only one $\mathsf{SET}$ order.

\section{Promoting $\mathsf{SEG}$ to $\mathsf{GT}^*$,   basis transformations, and the web of gauge theories}\label{sec:promotion}

  In the above discussions, we obtained many $\mathsf{SEG}$s, where the background gauge fields $\{A^I\}$ are treated as non-dynamical fields.  A caveat is that  basis transformations that mix $\{A^I\}$ and dynamical variables $\{a^I\}$ are strictly prohibited.  However, one may further give \emph{full} dynamics to the background gauge fields $\{A^I\}$, which leads to the mapping from $\mathsf{SEG}$s to $\mathsf{GT}^*$ as shown in Fig.~\ref{figure_tree}.   In other words, the symmetry twist now becomes   dynamical \cite{footnote_gauging}.  As a result, arbitrary basis transformations now can be applied.   It is legitimate to mix gauge fluxes and symmetry fluxes together to form a flux of a new gauge variable.  

\subsection{$\mathsf{SEG}(\Z_2,\Z_2)$}

  Let us consider  $\mathsf{SEG}(\Z_2,\Z_2)$ in Table~\ref{table:z2_zk_1} with $K=2$. The associated   dynamical gauge theory of $b,a,A,B$ (here, $b=b^1$\,,$a=a^1$ for this single layer case) can be written as:
  \begin{align}
  S=\frac{1}{2\pi} \int  \left(\begin{matrix}
B& b\\
\end{matrix}\right)  \left(\begin{matrix}
2&0\\
1&2\end{matrix}\right) \wedge d \left(\begin{matrix}
A\\
a\end{matrix}\right)\,,
  \end{align}
where the two-form gauge field $B$ is introduced to relax the holonomy of $A$ to $\mathrm{U(1)}$-valued in the path integral measure.  According to Eq.~(\ref{eq:GL_transformation}), one can apply the following two unimodular matrices to send the above theory to its canonical form:
\begin{align}
& W=\left(\begin{matrix}
1&-1\\
-1&2\end{matrix}\right)\,,\,\,\,
\Omega=\left(\begin{matrix}
1&0\\
2&1\end{matrix}\right)\,,  \\
&W \left(\begin{matrix}
2&0\\
1&2\end{matrix}\right)  \Omega^T= \left(\begin{matrix}
1&0\\
0&4\end{matrix}\right) 
 \end{align}
 which directly indicates that the resulting new gauge theory $\mathsf{GT}^*$ after giving full dynamics to the background gauge field is $\Z_4$ gauge theory (Fig.~\ref{figure_example}).

Likewise, for Table \ref{table:z2_zk_2},   the level matrix of the BF term is given by:
 \begin{align}
\left(\begin{matrix}
2&0&0\\
0&1&1\\
0&0&2\end{matrix}\right)
  \end{align}
 in the basis of $(b^1,b^2,B)$ and $(a^1,a^2,A)$. It can be diagonalized by using the following two unimodular matrices:
  \begin{align}
&W=\left(\begin{matrix}
1 &0&0\\
0&1&0\\
0&0&1\end{matrix}\right)\,,\,\,\,\Omega=  
\left(\begin{matrix}
1&0&0\\
0&1&0\\
0&-1&1\end{matrix}\right)\,,\\
&W \left(\begin{matrix}
2&0&0\\
0&1&1\\
0&0&2\end{matrix}\right) \Omega^T=\left(\begin{matrix}
2&0&0\\
0&1&0\\
0&0&2\end{matrix}\right)\,.\label{eq:canonical_z2z2}
  \end{align}
  As a result, the new 1-form gauge variables are given by the vector $(\tilde{a}^1,\tilde{a}^2,\tilde{A})^T$ where, 
  \begin{align}
  a^1=\tilde{a}^1\,,  a^2=\tilde{a}^2-\tilde{A}\,,A=\tilde{A}\,.
  \end{align}
From the canonical form (\ref{eq:canonical_z2z2}), it is clear that the resulting theory after  giving full dynamics to the background  gauge field  is $\Z_2\times \Z_2$ gauge theory. But we should also  examine how topological interaction terms transform. Since the second layer in the new basis is a trivial layer (level-1), we may neglect all topological interaction terms that include $\tilde{a}^2$.  Keeping this in mind, After the basis transformations, the topological interaction terms $\int \frac{iq}{4\pi^2}a^1\wedge a^2\wedge d a^2+\int \frac{i\bar q}{4\pi^2}a^2\wedge a^1\wedge d a^1$ are transformed to:
\begin{align}
\int \frac{iq}{4\pi^2}\tilde{a}^1\wedge \tilde{A}\wedge d \tilde{A}-\int \frac{i\bar q}{4\pi^2}\tilde{A}\wedge \tilde{a}^1\wedge d \tilde{a}^1\,.
\end{align}
Therefore, we reach the following conclusions. 
The resulting theory starting from $\mathsf{SEG}$ labeled by $(0,0)$ in Table \ref{table:z2_zk_2} is ``untwisted'' $\Z_2\times \Z_2$ gauge theory. The remaining $\mathsf{SEG}$s lead to twisted $\Z_2\times\Z_2$ gauge theory after giving dynamics to the background gauge field  (Fig.~\ref{figure_example}), which is also derived in \cite{3dset_chen} from a different point of view.

\subsection{$\mathsf{SEG}(\Z_2,\Z_3)$}

 For $\mathsf{SEG}$s in Table \ref{table:z2_zk_1},  $\mathsf{GT}^*$ is always $\Z_{2K}$ gauge theory which are  ``untwisted''. For $\mathsf{SEG}$s in Table \ref{table:z2_zk_2}, for even $K$, $\mathsf{GT}^*$s are   $\Z_2\times\Z_K$ gauge theories which have one untwisted version and three twisted versions, in a similar manner to $K=2$ discussed above.  But for odd $K$,   the resulting theory $\mathsf{GT}^*$ is still $\Z_{2K}$ gauge theory since the two groups are isomorphic: $\Z_2\times\Z_K\cong \Z_{2K}$ when $K\in\Z_{\rm odd}$. For example, for $K=3$:
\begin{align}
\left(\begin{matrix}  -1& 1\\
-3& 2 \end{matrix}\right)
\left(\begin{matrix}  2& 0\\
0& 3 \end{matrix}\right)
\left(\begin{matrix}  1& -3\\
1& -2 \end{matrix}\right)=\left(\begin{matrix}  1& 0\\
0& 6 \end{matrix}\right)\,.
\end{align}
Therefore, for odd $K$, the resulting gauge theory is the same as that in Table \ref{table:z2_zk_1}. In other words, after giving full dynamics to the background gauge field $A$, there is only one output: a $\Z_{2K}$ gauge theory  (Fig.~\ref{figure_example}). From this simple case, we   see there is an interesting pattern of many-to-one correspondence between $\mathsf{SEG}$s and $\mathsf{GT}^*$s.

 \subsection{$\mathsf{SEG}(\Z_2\times\Z_2,\Z_2)$}
 
\begin{figure}[t]
\centering
\includegraphics[width=7cm]{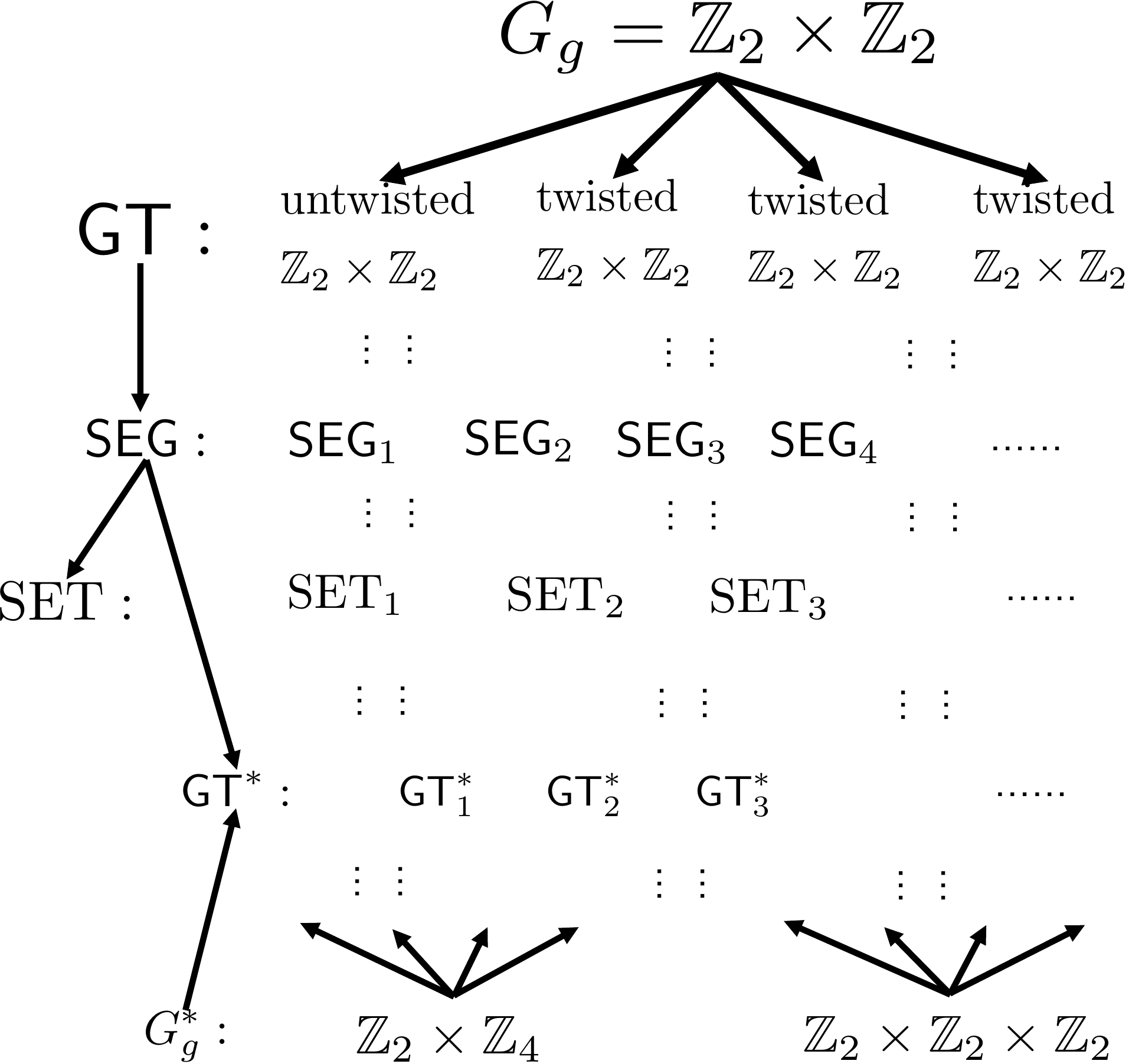}
\caption{ A skeleton of the web of gauge theories for $\mathsf{SEG}(\Z_2\times\Z_2,\Z_2)$.}
\label{figure_example_plus}
\end{figure}

 For $\mathsf{SEG}(\Z_2\times\Z_2,\Z_2)$, all $\mathsf{SEG}$s are collected in Table~\ref{table:z2z2_z2_1}. Before imposing symmetry, there are already four distinct gauge theories. Therefore, the resulting web of gauge theories is much more complex. A rough skeleton is shown in Fig.~\ref{figure_example_plus} where the resulting $\mathsf{GT}^*$ theories can be regrouped into two gauge groups $ {G}^*_{g1}=\Z_2\times\Z_4$ and $ {G}^*_{g2}=\Z_2\times\Z_2\times \Z_2$. 
 The first gauge group arises from the first and second subtables of Table~\ref{table:z2z2_z2_1} while the second gauge group arises from the third subtable of Table~\ref{table:z2z2_z2_1}. More concretely, let us consider the BF term of the first subtable after the background gauge field becomes  fully dynamical:
   \begin{align}
  \frac{1}{2\pi} \int  \left(\begin{matrix}
B& b^1 & b^2\\
\end{matrix}\right)  \left(\begin{matrix}
2&0&0\\
1&2&0\\
0&0&2\end{matrix}\right) \wedge d \left(\begin{matrix}
A\\
a^1\\a^2\end{matrix}\right)\,,
  \end{align}
where the two-form gauge field $B$ is introduced to relax the holonomy of $A$ to $\mathrm{U(1)}$-valued in the path integral measure.  According to Eq.~(\ref{eq:GL_transformation}), one can apply the following two unimodular matrices to send the above theory to its canonical form:
\begin{align}
& W=\left(\begin{matrix}
1&-1&0\\
-1&2&0\\
0&0&1\end{matrix}\right)\,,\,\,\,
\Omega=\left(\begin{matrix}
1&0&0\\
2&1&0\\
0&0&1\end{matrix}\right)\,,  \\
&W \left(\begin{matrix}
2&0&0\\
1&2&0\\
0&0&2\end{matrix}\right)  \Omega^T= \left(\begin{matrix}
1&0&0\\
0&4&0\\
0&0&2\end{matrix}\right) 
 \end{align}
which indicates that $ {G}^*_{g1}=\Z_2\times\Z_4$. Likewise, we have the following matrix calculation for the second subtable:
   \begin{align}
  \frac{1}{2\pi} \int  \left(\begin{matrix}
B& b^1 & b^2\\
\end{matrix}\right)  \left(\begin{matrix}
2&0&0\\
0&2&0\\
1&0&2\end{matrix}\right) \wedge d \left(\begin{matrix}
A\\
a^1\\a^2\end{matrix}\right)\,,
  \end{align}
and \begin{align}
& W=\left(\begin{matrix}
0&1&0\\
0&0&1\\
1&0&2\end{matrix}\right)\,,\,\,\,
\Omega=\left(\begin{matrix}
0&1&0\\
-1&0&1\\
2&0&-1\end{matrix}\right)\,,  \\
&W \left(\begin{matrix}
2&0&0\\
0&2&0\\
1&0&2\end{matrix}\right)  \Omega^T= \left(\begin{matrix}
2&0&0\\
0&1&0\\
0&0&4\end{matrix}\right) 
\end{align}
which still leads to ${G}^*_{g1}=\Z_2\times\Z_4$.

  For the third subtable,  the BF term is given by:
    \begin{align}
  \frac{1}{2\pi} \int  \left(\begin{matrix}
B& b^1 & b^2&b^3\\
\end{matrix}\right)  \left(\begin{matrix}
2&0&0&0\\
0&2&0&0\\
0&0&2&0\\
1&0&0&1\end{matrix}\right) \wedge d \left(\begin{matrix}
A\\
a^1\\a^2\\a^3\end{matrix}\right)\,,
  \end{align}
where the $4\times 4$ matrix can be diagonalized through:
\begin{align}
  & W=\left(\begin{matrix}
0&0&0&-1\\
0&1&0&0\\
0&0&1&0\\
1&0&0&0\end{matrix}\right)\,,\,\,\,
\Omega=\left(\begin{matrix}
0&0&0&-1\\
0&1&0&0\\
0&0&1&0\\
1&0&0&-1\end{matrix}\right)\,,  \\
&W  \left(\begin{matrix}
2&0&0&0\\
0&2&0&0\\
0&0&2&0\\
1&0&0&1\end{matrix}\right) \Omega^T=  \left(\begin{matrix}
1&0&0&0\\
0&2&0&0\\
0&0&2&0\\
0&0&0&2\end{matrix}\right)\,.
  \end{align}
  As a result, $ {G}^*_{g2}=\Z_2\times\Z_2\times\Z_2$.

\section{Summary and outlook}\label{sec:conclusions}

In this paper, we have studied the symmetry enrichment through topological quantum field theory description of three-dimensional topological phases.   All  phases constructed in this paper can be viewed as 3D  gapped quantum spin liquid candidates enriched by unbroken spin symmetry $G_s$.  Using the 5-step general procedure in Sec.~\ref{sec:general_procedures}, we have efficiently constructed  symmetry-enriched gauge theories   ($\mathsf{SEG}$) with gauge group $G_g=\Z_{N_1}\times\Z_{N_2}\times\cdots$ and symmetry group $G_s=\Z_{K_1}\times \Z_{K_2}\times\cdots$ as well as $G_s=\mathrm{U(1)}\times\Z_{K_1}\times\cdots$.  The relation between $\mathsf{SEG}$ and its parent gauge theory $\mathsf{GT}$ has been shown. We have also shown how to physically diagnose the ground state properties of $\mathsf{SEG}$s by investigating charge-loop braidings (patterns of symmetry fractionalization) and mixed multi-loop braiding statistics. By means of these physical detections, one can obtain a set of $\mathsf{SET}$ orders which represent the phase structures of ground states of   $\mathsf{SEG}$s. It is generally possible that two $\mathsf{SEG}$s may give rise to the same $\mathsf{SET}$ order. Finally, by providing full dynamics to the background gauge fields \cite{footnote_gauging}, the resulting new gauge theories $\mathsf{GT}^*$s can be obtained and  have been studied, all of which are summarized in a web of gauge theories (Fig.~\ref{figure_tree}). 
Throughout the paper,  many concrete examples have been studied in details. From those examples, we have seen that the general procedure provided in this paper is doable and efficient for the practical purpose of understanding 3D $\mathsf{SET}$ physics.

 We highlight some questions for future studies. 
 \textbf{(i)}  {Lattice models of $\mathsf{SEG}$s}.  Dijkgraaf-Witten models \cite{dw1990} and string-net models \cite{string_net} have been well studied. It is interesting to impose global symmetry (e.g., on-site finite unitary group) on these models in 3D.  Then, lattice models can be regarded as an ultra-violet definition of $\mathsf{SEG}$s.  Some progress on 2D $\mathsf{SET}$s has been made in Ref.~\cite{heinrich,set1}.  
 \textbf{(ii)} Material search  and the experimental fingerprint of the mixed three-loop braiding statistics. There are several   possible experimental realizations of $\Z_2$ spin liquids, such as the so-called Kitaev spin liquid state in the lattices in $\beta$- and $\gamma$-$\mathrm{Li_2IrO_3}$ \cite{kitaev_3d,exp_0,exp_1,exp_2,kitaev_kim,exp,kitaev_3d_1}. By further considering the unbroken $\Z_2$ Ising symmetry,  the resulting ground state should exhibit $\mathsf{SET}$ orders. As we studied in the paper,  the features of these $\mathsf{SET}$s are patterns of symmetry fractionalization and mixed three-loop braiding statistics. It is thus of  interest to theoretically propose an experimental fingerprint, especially, for the three-loop braiding statistics.
\textbf{(iii)}  {Anomalous $\mathsf{SEG}$s}. In our construction, by anomaly, we mean that global symmetry and gauge invariance cannot be compatible with each other. If both are preserved, the resulting $\mathsf{SEG}$ is anomaly-free as what we have calculated. As mentioned  in Sec.~\ref{sec:two}, the entries with ``N/A'' in   Table~\ref{table:z2z2_z2_1} means that there do not exist $\mathsf{SEG}$ descendants for the twisted gauge theory (with both nonzero $q$ and $\bar q$) in \emph{the} symmetry assignment (the first and second subtables) such that both global symmetry and gauge invariance are preserved simultaneously. In other words, either symmetry is broken or gauge invariance is violated. For the case in which symmetry is preserved but gauge invariance is violated, we conjecture it can be realized on the boundary of certain (4+1)D systems. More careful studies in the future along anomaly will be meaningful. 
 \textbf{(iv)} $\mathsf{GT}^*$s originated from $\mathsf{SEG}$s with $\mathrm{U(1)}$ symmetry.  In Sec.~\ref{sec:seg_with_u1} and Appendix, some examples of $\mathsf{SEG}$s with U(1) symmetry are studied. After $\mathrm{U(1)}$ symmetry group becomes a dynamical gauge group, the resulting theory $\mathsf{GT}^*$ should admit a mixed phenomenon generated by mixture of discrete gauge group and U(1) gauge group.  It will be interesting to study the properties of such a type of gauge theory and eventually build the web (i.e., Fig.~\ref{figure_tree}) of gauge theories for these cases. 
 \textbf{(v)}    {$\mathsf{SEG}$s with $\mathsf{Charles}$ symmetry \cite{3dset_ye}}.  $\mathsf{Charles}$ symmetry, which was introduced in \cite{3dset_ye}, is a 3D analog of 2D anyonic (topological) symmetry. A simple example is $\Z_3$ gauge theory where quasiparticle is permuted to its antiparticle while quasi-loop is permuted to its antiloop. And there is one species of defect-charge-loop composites. This is just one gauge theory by giving a gauge group and a $\mathsf{Charles}$ symmetry group. It will be interesting to investigate the possibility that there are more than one gauge theories enriched by $\mathsf{Charles}$. 
%\textbf{(iv)} The relation between $\mathsf{GT}^*$s and $\mathsf{SET}$s. We note that both of $\mathsf{GT}^*$s and $\mathsf{SEG}$s  are deduced from $\mathsf{SEG}$s, however through different points of view.  From   Fig.~\ref{figure_example}, it is clear that at least in the simplest examples we calculated, there is a one-to-one correspondence between $\mathsf{SET}$s and $\mathsf{GT}^*$s. Therefore, we conjecture that the one-to-one correspondence holds in general cases when $G_s$ does not include $\mathrm{U(1)}$ subgroups. However, the rigorous proof based on field theories is lacking and more concrete examples should be studied in the future.

  \section*{Acknowledgements}
We thank Ying Ran,  Xie Chen, Zheng-Cheng Gu, Meng Cheng, Chien-Hung Lin, and Michael Levin for enlightening discussions. P.Y. acknowledges Eduardo Fradkin's enrouragement and conversation during the preparation and also thanks Shing-Tung~Yau's  hospitality  at the Center of Mathematical Sciences and Applications at Harvard University where the work was done in part. Part of this work was done in Banff International Research Station, Banff, Calgary, Canada (P.Y.). S.Q.N. and Z.X.L. are supported by NSFC (Grant Nos.11574392), Tsinghua University Initiative Scientific Research Program, and the Fundamental Research Funds for the Central Universities, and the Research Funds of Renmin University of China (No.~15XNLF19).  This work was supported in part by the NSF through grant DMR 1408713 at the University of Illinois (P.Y.)

     \begin{widetext}
 \clearpage
 
 \appendix
\renewcommand{\thetable}{S\arabic{table}}

\renewcommand{\thefigure}{S\arabic{figure}}

 \section{General calculation of gauge theories with global symmetry}\label{sec:tech1}

 \subsection{$G_g=\Z_{N_1}\times \Z_{N_2}\times\Z_{N_3}$ with no symmetry}\label{section_useful_1}

 In this part, we present several details about $G_g=\Z_{N_1}\times\Z_{N_2}\times\Z_{N_3}$ and $G_s=\Z_{K_1}\times\Z_{K_2}\times\Z_{K_3}$. Each layer carries a unique symmetry charge. This case is relevant to those $\mathsf{SEG}$s even with only one gauge group but with two symmetry subgroups (via, e.g., setting $N_2=N_3=1$ and $K_1=1$). Most of derivations are similar to the previous cases except some subtle differences in the shift operations.  
 The gauge theory before imposing the global symmetry is given by:
\begin{align}S=\sum_{I=1}^3\frac{iN_I}{2\pi} \int b^I \wedge da^I +i \frac{\bar{\bar{q}}}{4\pi^2}\int a^1\wedge a^2 \wedge da^3\,. \label{GT:a1a2da3}
\end{align}
  The action  is invariant under the following gauge transformations parametrized by scalars $\{\chi^I\}$ and vectors $\{V^I\}$:
\begin{align}
a^I&\longrightarrow a^I +d\chi^I\,,\\
b^I&\longrightarrow b^I+dV^I-\frac{ \bar{\bar{q}}}{2\pi N^I} \epsilon^{IJ3} \chi^J \wedge da^3\,.\label{eq:modified_gauge_a1a2da3}
\end{align}
Let us investigate  the integral $\frac{1}{2\pi}\int_{\mathcal{M}^3}db^I$. Under the above modified gauge transformations (\ref{eq:modified_gauge_a1a2da3}), the integral will be changed by the amount below (for $I=1$, $\mathcal{M}^3=\mathcal{M}^1\times\mathcal{M}^2$ is considered):
\begin{align}
 \frac{1}{2\pi}\int_{\mathcal{M}^3}db^1\longrightarrow& \frac{1}{2\pi}\int_{\mathcal{M}^3}db^1 - \frac{ \bar{\bar{q}}}{4\pi^2 N_1} \int _{S^1}d\chi^2 \int_{M^2} da^3 = \frac{1}{2\pi}\int_{\mathcal{M}^3}db^1 - \frac{ \bar{\bar{q}} }{4\pi^2 N_1}  \times  2\pi \ell \times 2\pi \ell' \,,\label{eq:key_1_123}
\end{align}
where $\ell\,,\,\ell' \in  \Z$, and, the Dirac quantization condition (\ref{eq:dirac_a}) and homotopy mapping condition (\ref{eq:large}) are applied. In order to be consistent with the Dirac quantization condition (\ref{eq:dirac_b}), the change amount must be integral, namely, $\bar{\bar{q}}$ must be divisible by $N_1$. Similarly,   $ \bar{\bar{q}}$ is also divisible by $N_2 $ due to:
\begin{align}
 \frac{1}{2\pi}\int_{\mathcal{M}^3}db^2\longrightarrow& \frac{1}{2\pi}\int_{\mathcal{M}^3}db^2 + \frac{ \bar{\bar{q}}}{4\pi^2 N_2} \int _{S^1}d\chi^1 \int_{M^2} da^3=\frac{1}{2\pi}\int_{\mathcal{M}^3}db^2 + \frac{ \bar{\bar{q}}}{4\pi^2 N_2}  \times  2\pi \ell'' \times 2\pi \ell''' ,\label{eq:key_2_123}
\end{align}
where $  \ell''\,,\,\ell''' \in  \Z$. 
 Hence, $\bar{\bar{q}}=\frac{kN_1N_2}{ N_{12}}, k \in \Z$. 
Below, we want to show that $k$ has a periodicity $N_{123}$ (i.e., GCD of $N_1,N_2,N_3$)  and thereby $\bar{\bar{q}}$ is compactified: $\bar{\bar{q}}\sim \bar{\bar{q}}+\frac{N_{123} N_1N_2}{N_{12}}$. Let us consider the following redundancy due to shift operations:
 \begin{align}
\frac{1}{2\pi}\int db^1 \longrightarrow &\frac{1}{2\pi} \int db^1+\frac{ N_2 \tilde{K}_1}{4\pi^2 N_{12}} \int a^2\wedge da^3 \,, \label{eq:key_3_123}\\
\frac{1}{2\pi}\int d b^2\longrightarrow  &\frac{1}{2\pi}\int db^2-\frac{N_1  \tilde{K}_2}{4\pi^2 N_{12}}  \int  a^1\wedge da^3 \,,\label{eq:key_4_123}\\
\frac{1}{2\pi}\int d b^3\longrightarrow  &\frac{1}{2\pi}\int db^3+\frac{N_1  N_2\tilde{K}_3}{4\pi^2 N_3N_{12}} \int (d a^1\wedge a^2+a^1\wedge da^2) \,,\label{eq:key_4_123_plus}\\
 k\longrightarrow\,& \,k+\tilde{K}_1+\tilde{K}_2+\tilde{K}_3\,.
 \end{align}
Again, in order to be consistent with  Dirac quantization (\ref{eq:dirac_b}), the change amount of the integral $\frac{1}{2\pi}\int_{\mathcal{M}^3}db^I$ should be integral, namely:
 \begin{align}
  \frac{N_2 \tilde{K}_1}{4\pi^2 N_{12}} \int_{\mathcal{M}^3}   a^2\wedge da^3&\in \Z\,,\,\\
   \frac{N_1 \tilde{K}_2}{4\pi^2 N_{12}} \int_{\mathcal{M}^3}  a^1\wedge d a^3&\in \Z\,,\\
   \frac{N_1  N_2\tilde{K}_3}{4\pi^2 N_3N_{12}} \int_{\mathcal{M}^3}   (d a^1\wedge a^2+a^1\wedge da^2)&\in \Z\,.
\end{align} 
  We may apply the Dirac quantization condition (\ref{eq:dirac_a}) and the quantized Wilson loop $ \frac{N_I}{2\pi} \int_{\mathcal{M}^1} a^I \in\Z$ that is obtained via  equations of motion of $b^I$.  As a result, three constraints are achieved: $\tilde{K}_1/N_{12}\in\Z$, $\tilde{K}_2/N_{12}\in\Z$, $\tilde{K}_3/N_3\in\Z$. In deriving the result for $\tilde{K}_3$, Bezout's lemma is applied.  By using Bezout's lemma again, the minimal periodicity of $k$  is given by GCD of $N_{12}$ and $N_3$, which is   $N_{123}$.  As a result, we obtain the conditions on $\bar{\bar{q}}$ \emph{if} symmetry is not taken into consideration.
\begin{align}
\bar{\bar{q}}=k\frac{N_1N_2}{N_{12}} \text{  mod  } \frac{N_{123}N_1N_2}{N_{12}}\,,~~~ k\in\Z_{N_{123}}.\label{eq:key_quantized_no_symmetry_123}
\end{align}

 \subsection{$G_g=\Z_{N_1}\times \Z_{N_2}\times\Z_{N_3}$ with  $G_s=\Z_{K_1} \times \Z_{K_2} \times \Z_{K_3}$}\label{section_useful_2}

  To impose the symmetry, we add the following coupling term    in the action (\ref{GT:a1a2da3}):  
 \begin{align}
 \sum_{i}^3\frac{i}{2\pi} \int   A^{i}\wedge  db^{i}\,.\label{eq:symmetry_coupling_aada_123}
\end{align}  
 The change amounts of the integral $\frac{1}{2\pi}\int_{\mathcal{M}^3}db^I$ in Eqs.~(\ref{eq:key_1_123},~\ref{eq:key_2_123},~\ref{eq:key_3_123},~\ref{eq:key_4_123},~\ref{eq:key_4_123_plus}) should not only be integral [in order to be consistent with the Dirac quantization condition (\ref{eq:dirac_b})] but also be multiple of $K_i$ such that the coupling term (\ref{eq:symmetry_coupling_aada_123}) is gauge invariant \text{modular} $2\pi$.   More quantitatively,  with symmetry taken into account, from Eqs.~(\ref{eq:key_1_123},~\ref{eq:key_2_123}), we may obtain the quantization of $\bar{\bar{q}}$: $\bar{\bar{q}}=\frac{kN_1N_2K_1K_2}{\mathrm{GCD}(N_1K_1,N_2K_2)}$  with $k\in\Z$ such that the change amounts are multiple of $K_i$. Then, with these new quantized values, the shift operations (\ref{eq:key_3_123},~\ref{eq:key_4_123},~\ref{eq:key_4_123_plus}) are changed to:
 \begin{align}
\frac{1}{2\pi} \int db^1 \longrightarrow &\frac{1}{2\pi} \int db^1+ \tilde{K}_1\frac{  N_2K_1K_2}{4\pi^2\mathrm{GCD}(N_1K_1,N_2K_2)}
  \int a^2\wedge da^3 \,, \label{eq:key_5_123}\\
\frac{1}{2\pi}\int d b^2\longrightarrow  &\frac{1}{2\pi}\int db^2-\tilde{K}_2\frac{  N_1K_1K_2}{4\pi^2\mathrm{GCD}(N_1K_1,N_2K_2)} \int a^1\wedge da^3\,,\label{eq:key_6_123}\\
\frac{1}{2\pi}  \int d b^3\longrightarrow  &\frac{1}{2\pi} \int db^3+ \tilde{K}_3\frac{N_1  N_2K_1K_2 }{4\pi^2 N_3  \, \mathrm{GCD}(N_1K_1,N_2K_2)} \cdot\int (d a^1\wedge a^2+a^1\wedge da^2) \,.\label{eq:key_6_123_plus}
 \end{align}
After the integration over $\mathcal{M}^3$, the change amounts should be quantized at $K_1$ in Eq.~(\ref{eq:key_5_123}), $K_2$ in Eq.~(\ref{eq:key_6_123}), and $K_3$ in Eq.~(\ref{eq:key_6_123_plus}). 
 We may apply the Dirac quantization condition (\ref{eq:dirac_a}) and the quantized Wilson loop $ \frac{N_IK_I}{2\pi} \int_{\mathcal{M}^1} a^I \in\Z$ that is obtained via  equations of motion of $b^I$ in the presence of $A^I$ background.  As a result, three necessary and sufficient constraints are achieved: $\frac{\tilde{K}_1}{\mathrm{GCD}(N_1K_1,N_2K_2)}\in\Z$, $\frac{\tilde{K}_2}{\mathrm{GCD}(N_1K_1,N_2K_2)}\in\Z$, $\frac{\tilde{K}_3}{N_3K_3}\in\Z$. By using Bezout's lemma, the minimal periodicity of $k$  is given by   $\mathrm{GCD}$ of  $\mathrm{GCD}(N_1K_1,N_2K_2)$, $\mathrm{GCD}(N_1K_1,N_2K_2) $, and $N_3K_3$,  which is    $\mathrm{GCD}(N_1K_1,N_2K_2,N_3K_3)$. As a result, once symmetry is imposed, $\bar{\bar{q}}$ is changed from Eq.~(\ref{eq:key_quantized_no_symmetry_123}) to:
 \begin{align} 
\bar{\bar{q}}=&k\frac{N_1N_2K_1K_2}{\mathrm{GCD}(N_1K_1,N_2K_2)} \text{  mod  }  \frac{  N_1N_2K_1K_2 \,\mathrm{GCD}(N_1K_1,N_2K_2,N_3K_3)}{\mathrm{GCD}(N_1K_1,N_2K_2)} \,, \text{ with } k\in\Z_{\mathrm{GCD}(N_1K_1,N_2K_2,N_3K_3)}
\end{align}
which gives $\mathrm{GCD}(N_1K_1,N_2K_2,N_3K_3)$ $\mathsf{SEG}$s. Since $\mathrm{GCD}(N_1K_1,N_2K_2,N_3K_3)\geq \mathrm{GCD}(N_1 ,N_2,N_3)$, the allowed values of $\bar{\bar{q}}$ are  enriched by symmetry.

 \subsection{$G_g=\Z_{N_1}\times \Z_{N_2}\times \Z_{N_3}$ with  $G_s=\Z_{K_1}\times \Z_{K_2} \times \mathrm{U(1)}$}\label{appendix_sub_zn1zn2_k1k2u1_123}

 In this part, we consider $U(1)$ symmetry. We consider the following symmetry assignment  and add it in the action (\ref{GT:a1a2da3}):
 \begin{align}
\frac{i}{2\pi}  \sum_{i}^2\int   A^{i}\wedge  db^{i} +A^{U(1)} \wedge db^3\,.\label{eq:symmetry_coupling_aada_123_U(1)}
\end{align}  
where  the $\mathsf{U(1)}$ Wilson loop
 \begin{align}\int_{S^1} A_{U(1)} \in \mathbb{R} \label{U(1)wilsonloop}
 \end{align}
  meaning that the $\mathsf{U(1)}$ Wilson loop can be any real value.
 Under the gauge transformation (\ref{eq:modified_gauge_a1a2da3}), the change amounts of the integral $\frac{1}{2\pi}\int_{\mathcal{M}^3}db^I$ in Eqs.~(\ref{eq:key_1_123},~\ref{eq:key_2_123}) should be multiple of $K_1$ or $K_2$ such that the coupling terms (\ref{eq:symmetry_coupling_aada_123_U(1)}) is gauge invariant \text{modular} $2\pi$.   More quantitatively,  with symmetry taken into account, from Eqs.~(\ref{eq:key_1_123},~\ref{eq:key_2_123}), we may obtain the quantization of $\bar{\bar{q}}$: $\bar{\bar{q}}=\frac{kN_1N_2K_1K_2}{\mathrm{GCD}(N_1K_1,N_2K_2)}$  with $k\in\Z$ such that the change amounts are multiple of $K_i$. To remove the redundancy in the possible value of $\bar{\bar{q}}$, we do the shift operations as that from (\ref{eq:key_5_123}) to (\ref{eq:key_6_123_plus}). Similarly to the case above, after the integration over $\mathcal{M}^3$, the change amounts should be quantized at $K_1$ in Eq.~(\ref{eq:key_5_123}), $K_2$ in Eq.~(\ref{eq:key_6_123}), and zero in Eq.~(\ref{eq:key_6_123_plus}) due to the fact that the $\mathrm{U(1)}$ Wilson loop can be any real value.  As a result, three necessary and sufficient constraints are achieved: $\frac{\tilde{K}_1}{\mathrm{GCD}(N_1K_1,N_2K_2)}\in\Z$, $\frac{\tilde{K}_2}{\mathrm{GCD}(N_1K_1,N_2K_2)}\in\Z$, $\tilde{K}_3=0$. By using Bezout's lemma, the minimal period of $k$ is $GCD(N_1K_1,N_2K_2)$, i.e.
 
\begin{align} 
\bar{\bar{q}}=&k\frac{N_1N_2K_1K_2}{\mathrm{GCD}(N_1K_1,N_2K_2)} \text{  mod  }  N_1N_2K_1K_2 \,, \text{ with } k\in\Z_{\mathrm{GCD}(N_1K_1,N_2K_2)}
\end{align}
 which gives $\mathrm{GCD}(N_1K_1,N_2K_2)$ $\mathsf{SEG}$s.

 \subsection{$G_g=\Z_{N_1}\times \Z_{N_2}$ with  $G_s=\Z_{K} \times   \mathrm{U(1)} $-(I)}\label{appendix_subsection_zn1zn2_zku1_1}

%In the following, we consider the symmetry involving Abelian continuous group $\mathrm{U(1)}$ whose

 Here we consider the symmetry assignment and add it in the action (\ref{eq:aada_general_q}):
\begin{equation}
\frac{i}{2\pi} \int A^K \wedge db^1+ A^{U(1)} \wedge db^2
\label{coupling1:zKU(1)}
\end{equation}
which indicates that the first layer carries the discrete symmetry $\Z_K$ while the second layer  carries $\mathrm{U(1)}$.  To determine the possible values of $q$ in the presence of this global symmetry,  
 We observe that the change amounts of the integral $\frac{1}{2\pi}\int_{\mathcal{M}^3}db^1$ in Eq.~(\ref{eq:key_1}) should  be multiple of $K$ such that the first  
 coupling term in Eq.~(\ref{coupling1:zKU(1)}) is gauge invariant \text{modular} $2\pi$. But the key observation is that the $\mathrm{U(1)}$ Wilson loop (\ref{U(1)wilsonloop}) is any real value, therefore, to keep the second coupling term in Eq.~(\ref{coupling1:zKU(1)})  gauge invariant, the change amount $\frac{1}{2\pi}\int_{\mathcal{M}^3}db^2$ in Eq.~(\ref{eq:key_2}) should be strictly zero,  which would be only the case that $q=0$. Similarly,   $\bar{q}=0$. Therefore, $\mathsf{SEG}$ only happens when $q=\bar q=0$.

 \subsection{$G_g=\Z_{N_1}\times \Z_{N_2}$ with  $G_s=\Z_{K} \times   \mathrm{U(1)} $-(II)}\label{appendix_subsection_zn1zn2_zku1_2}
In this part, we consider the whole symmetry group $G_s$ at the same layer and add the following part  in the action (\ref{eq:aada_general_q}) where we first set $\bar{q}=0$:
\begin{equation}
\frac{i}{2\pi} \int A^K \wedge db^1+ A^{U(1)} \wedge db^1
\label{coupling2:zKU(1)}
\end{equation}

Similar to the case that the symmetry subgroup are assigned at different layers,  in order to keep to the second term in (\ref{coupling2:zKU(1)})   gauge invariant, the change amount of the integral $\frac{1}{2\pi} db^1$ should be strictly zero. Therefore, $q=0$.  For the similar reason, $\bar{q}=0$. This symmetry assignment also only happens when $q=\bar{ q}=0$.

\section{ Several examples}\label{appendix_examples}

 \subsection{$\mathsf{SEG}(\Z_2\times\Z_4,\Z_2)$}

In the main text, we illustrate the example of $\Z_2\times\Z_2$ gauge with $\Z_2$ symmetry. Here, we calculate another example: $G_g=\Z_2\times\Z_4$ with $G_s=\Z_2$.
 Before imposing symmetry, there are 4 gauge theories in total, denoted by $(q,\bar{q})$:(0,0),(0,4),(4,0) and (4,4).  In the first  subtable of Table~\ref{table:z2z4_z2_1},  the symmetry $\Z_2$ is assigned at the first layer where the $\Z_2$ gauge subgroup lives. From this table, it is clear that both $q$ and $\bar q$ have four choices, resulting in $4^2$ $\mathsf{SEG}$s. Among these four choices of, say, $q$, we may further regroup them into two groups: $\{0\text{ mod }16,\, 8\text{ mod }16 \}$ and $\{4\text{ mod }16,\, 12\text{ mod }16 \}$. The two choices in the former group are  $\mathsf{SEG}$ descendants of $\mathsf{GT}$ with $q=0\text{ mod }8$ before imposing symmetry.  The two choices in the latter  group are  $\mathsf{SEG}$ descendants of $\mathsf{GT}$ with $q=4\text{ mod }8$ before imposing symmetry. 
 In this sense, 
  this table is sharply different from the first subtable of Table~\ref{table:z2z2_z2_1} where some entries are marked by ``N/A''.

\begin{table}[h]\caption{$\mathsf{SEG}(\Z_2 \times \Z_4, Z_2)$.}\label{table:z2z4_z2_1}
\centering
\begin{tabular}{|c|c|c|c|c|c|}
\hline
    \multirow{4}{*}{ \begin{minipage}   {0.65in}Symmetry assignment\end{minipage}}  & \multicolumn{5}{c|}{ \multirow{4}{*}{ \includegraphics[width=5cm]{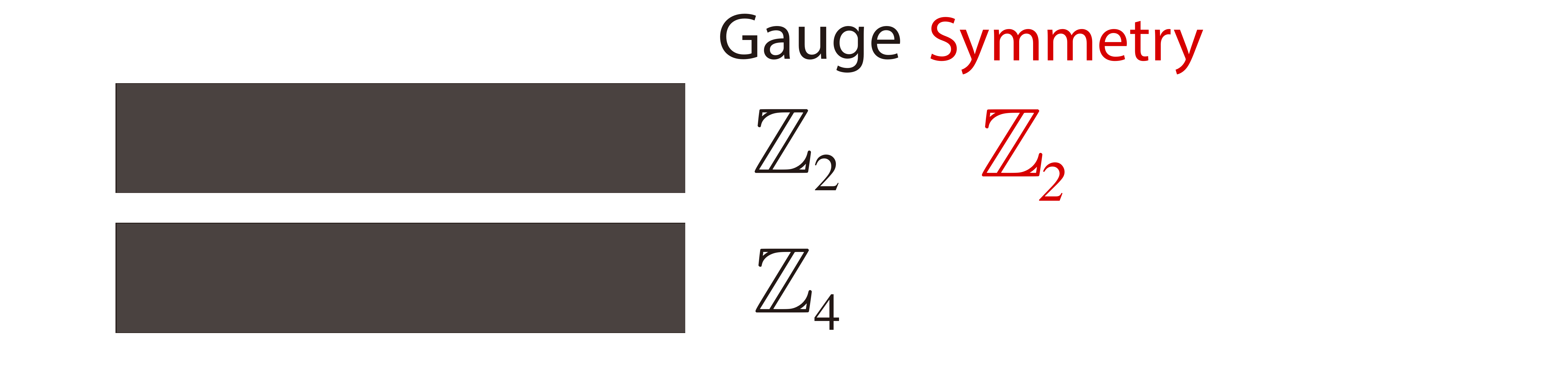}  }}   \\
               & \multicolumn{5}{l|}{  }   \\ 
               & \multicolumn{5}{l|}{  }   \\
               & \multicolumn{5}{l|}{  }   \\ 
\hline
 & \multicolumn{2}{c|}{$q/4\pi^2 a^1a^2da^2$}&\multicolumn{2}{c|}{$\bar{q}/4\pi^2 a^1a^2da^2$} &  \\
 \hline
 GT & 0 mod 8&4 mod 8& 0 mod 8 & 4 mod 8& $~~~~~~$\\
 \hline
 \multirow{2}{*}{$\mathsf{SEG}$} & 0 mod 16&4 mod 16&0 mod 16&4 mod 16&\multirow{2}{*}{$4^2$}\\
 &8 mod 16&12 mod 16&8 mod 16&12 mod 16& \\
 \hline
\end{tabular}
\begin{tabular}{|c|c|c|c|c|c|}
\hline
    \multirow{4}{*}{ \begin{minipage}   {0.65in}Symmetry assignment\end{minipage}}  & \multicolumn{5}{c|}{ \multirow{4}{*}{ \includegraphics[width=5cm]{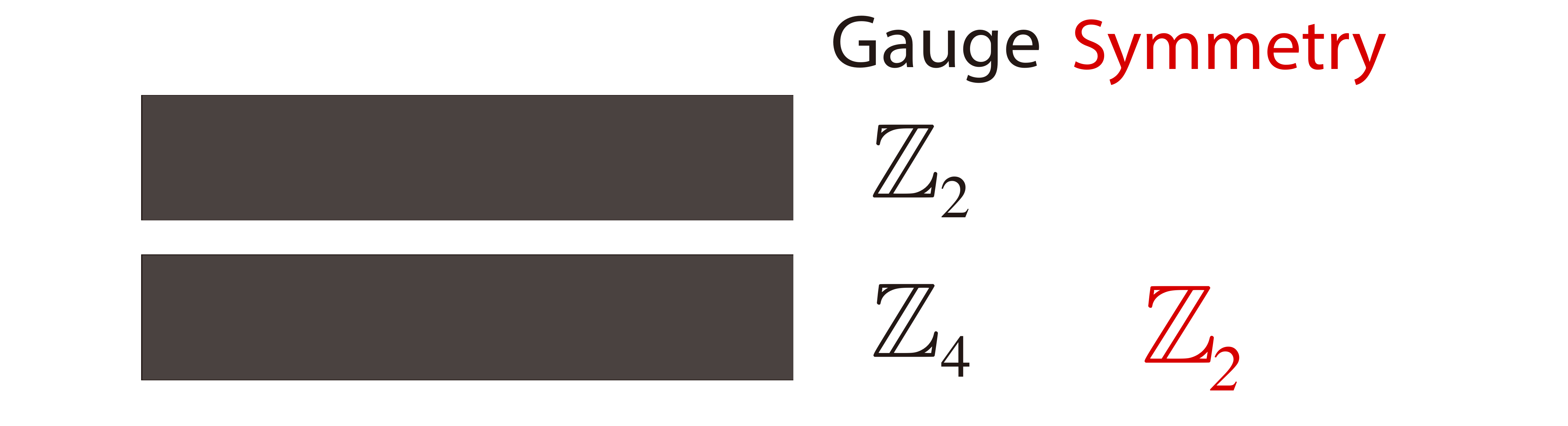}  }}   \\
               & \multicolumn{5}{l|}{  }   \\ 
               & \multicolumn{5}{l|}{  }   \\
               & \multicolumn{5}{l|}{  }   \\ 
\hline
 & \multicolumn{2}{c|}{$q/4\pi^2 a^1a^2da^2$}&\multicolumn{2}{c|}{$\bar{q}/4\pi^2 a^1a^2da^2$} & \\
 \hline
 GT & 0 mod 8&4 mod 8& 0 mod 8 & 4 mod 8& $~~~~~$\\
 \hline
 \multirow{2}{*}{$\mathsf{SEG}$} & 0 mod 16&N/A&0 mod 16&N/A&\multirow{2}{*}{$2^2$}\\
 &8 mod 16&&8 mod 16&& \\
 \hline
  
\end{tabular}
\begin{tabular}{|c|c|c|c|c|c|c|c|c|c|c|c|}
\hline
   \multirow{6}{*}{ \begin{minipage}   {0.65in}Symmetry assignment\end{minipage} } & \multicolumn{11}{c|}{\multirow{6}{*}{\includegraphics[width=4cm]{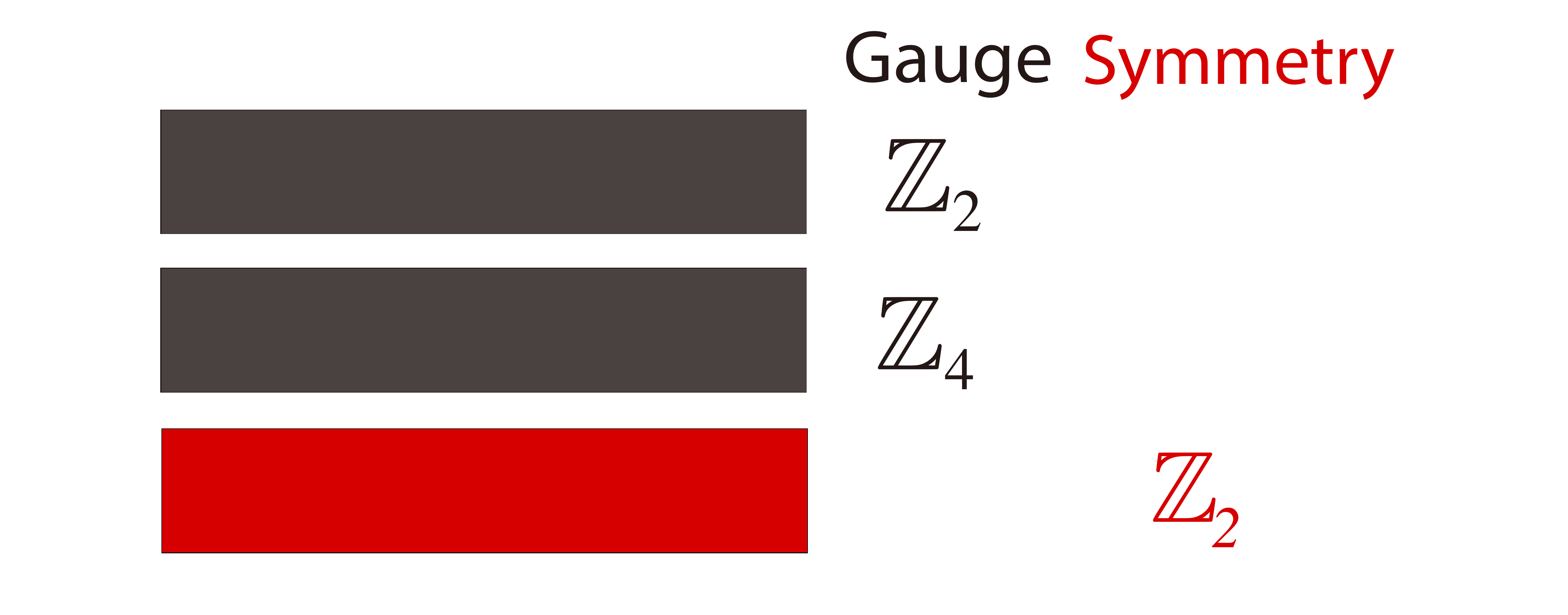}}}                                                                                                                                                                                                                                                                         \\
               & \multicolumn{11}{l|}{  }   \\ 
               & \multicolumn{11}{l|}{  }   \\
               & \multicolumn{11}{l|}{  }   \\ 
               & \multicolumn{11}{l|}{  }   \\  \hline
\multirow{2}{*}{$\mathsf{GT}$} &\multicolumn{2}{c|}{$ a^1a^2da^2 $}&\multicolumn{2}{c|}{$ a^2a^1da^1$}& $ a^1a^3da^3$ & $ a^3a^1da^1$ &$ a^2a^3da^3$&$ a^3a^2da^2$& $ a^1a^2da^3$ &$a^2a^3da^1$ &\\
\cline{2-12 }&0 mod 8&4 mod8&0 mod 8&4 mod 8&0 mod 2&0 mod 2&0 mod 4 &0 mod 4&0 mod 4& 0 mod 4&$~~$\\
\hline
\multirow{2}{*}{$\mathsf{SEG}$} &\multirow{2}{*}{0 mod 8}&\multirow{2}{*}{4 mod 8}&\multirow{2}{*}{0 mod 8}&\multirow{2}{*}{4 mod 8}&0 mod 4&0 mod 4&0 mod 8&0 mod 8& 0 mod 8&0 mod 8& \\
 &&&&&2 mod 4& 2 mod 4 &4 mod 8&4 mod 8 & 4 mod 8 &4 mod 8&~~~~~~~~$2^8$~~~~~~~~\\
\hline\hline
\end{tabular}
\end{table}

In the second subtable of Table~\ref{table:z2z4_z2_1}, the symmetry is assigned at the second layer where the $\Z_4$ gauge subgroup lives. The results are similar to the second table of Table \ref{table:z2z2_z2_1}, where some entries are marked by ``N/A''.   Totally, there are  $2^2$ $\mathsf{SEG}$s.

In the third subtable of Table~\ref{table:z2z4_z2_1}, the symmetry is assigned at the third layer where there is no gauge group.  This symmetry assignment  induces some new nonvanishing topological interactions involving the third layer. There are in total 8 kinds of topological interactions  \cite{footnote_8_topo}. Each topological interaction contains two choices of coefficients, rendering $2^8$ $\mathsf{SEG}$s.

\subsection{ $\mathsf{SEG}(\Z_{2},\Z_2 \times \Z_{2}$)}

In this part,  we consider $\mathsf{SEG}$s  whose symmetry group contains more than one cyclic subgroup. In this case, a lot of new ways of symmetry assignment exist.  Specifically, we consider  a relatively simple example: $\Z_2$ gauge theory with $\Z_2 \times \Z_2$ symmetry. In order to differentiate the two subgroups from each other, we introduce superscripts: $G_s=\Z_2^a\times \Z_2^b$.

                   In Table \ref{table:z2_z2z2_1}, the two symmetry subgroups are assigned to  the first and second layer, respectively. Before imposing symmetry, the coefficients $q,\bar{q}$  can only take value $0\text{ mod }2$, so all topological interaction terms identically vanish. This is exactly the fact that there is only one $\Z_2$ gauge theory.  After imposing symmetry, however,  the periods of both  $q, \bar{q}$ are enlarged from $2$ to $8$. Within one period,  they  can take either $0$ or $4$,  resulting in $2^2$ different $\mathsf{SEG}$s. Another  $2^2$ $\mathsf{SEG}$s can be obtained by simply exchanging the subscripts $a,b$.

\begin{table*}[h]\caption{$\mathsf{SEG}$ $(\Z_2,  \Z_2 \times \Z_2)$.  \small The superscripts $a$ and $b$ are added to distinguish the two $\Z_2$ subgroups. The two symmetry subgroups are carried by the two layers respectively. There are two \emph{independent} ways of symmetry assignment obtained by exchanging the auxiliary superscripts $a\longleftrightarrow b$. }\label{table:z2_z2z2_1}
\begin{tabular}{|c|c|c|c|}
\hline\hline
                   \multirow{4}{*}{ \begin{minipage}   {0.65in}Symmetry assignment\end{minipage}}  & \multicolumn{3}{c|}{ \multirow{4}{*}{ \includegraphics[width=5cm]{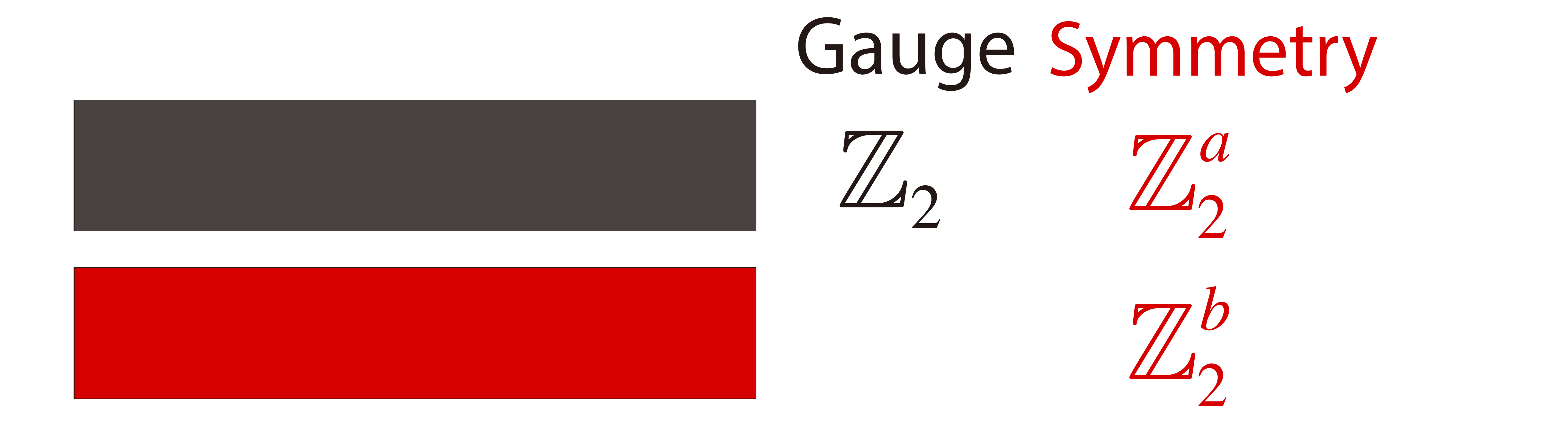}  }}     \\
                    & \multicolumn{3}{l|}{  }   \\ 
                    & \multicolumn{3}{l|}{  }   \\
                    & \multicolumn{3}{l|}{  }   \\ 
                   \hline
\multirow{2}{*}{$\mathsf{GT}$}   & $ q/4 \pi^2 a^1a^2da^2$ & $ \bar{q} /4 \pi^2 a^2a^1da^1$  &  \\ \cline{2-4} 
                      & 0 mod 2                                  & 0 mod 2                              &    \\ \hline
\multirow{4}{*}{$\mathsf{SEG}$} &               \multirow{4}{*}{}                  &         \multirow{4}{*}{}               &\multirow{4}{*}{}     \\ 
                      & 0 mod 8                                  & 0 mod 8                  &        $2^2$      \\ 
                       & 4 mod 8                                  & 4 mod 8                  &               \\ 
                        &                               &              &             \\ \cline{2-4} 
                       \hline\hline
\end{tabular}
\end{table*}

In Table~\ref{table:z2_z2z2_2}, we assign the two symmetry subgroups at the second and third layer, both of which are trivial layers. In this case, as there are three layers, we need to consider 8 different topological interactions as collected in the table. As explained also in the main text, there are only two linearly independent three-layer topological interaction terms since $a^3a^1da^2$ is $a^1a^2da^3+a^2a^3da^1$ up to a total derivative.
 Again, before imposing symmetry, coefficients of any kinds of topological terms identically vanish.  After symmetry is considered, it turns out that  these 8 topological interactions generate   $2^8$  different $\mathsf{SEG}$s.  
 In addition, in Table~\ref{table:z2_z2z2_3}, two ways to assign the two symmetry subgroups in the same layer are considered. In the first subtable, there is only one $\mathsf{SEG}$.  But in the second subtable,   the calculation shows that there are $2^2$ $\mathsf{SEG}$s.

 \begin{table}\caption{$\mathsf{SEG}$ $(\Z_2,  \Z_2 \times \Z_2)$ \small The superscripts $a$ and $b$ are added to distinguish the two $\Z_2$ subgroups. The gauge group is carried by the first layer, while the two symmetry subgroups by the second and third layers respectively.} \label{table:z2_z2z2_2}
\begin{tabular}{|c|c|c|c|c|c|c|c|c|c|}
\hline\hline
   \multirow{6}{*}{ \begin{minipage}   {0.65in}Symmetry assignment\end{minipage} } & \multicolumn{9}{c|}{\multirow{6}{*}{\includegraphics[width=5cm]{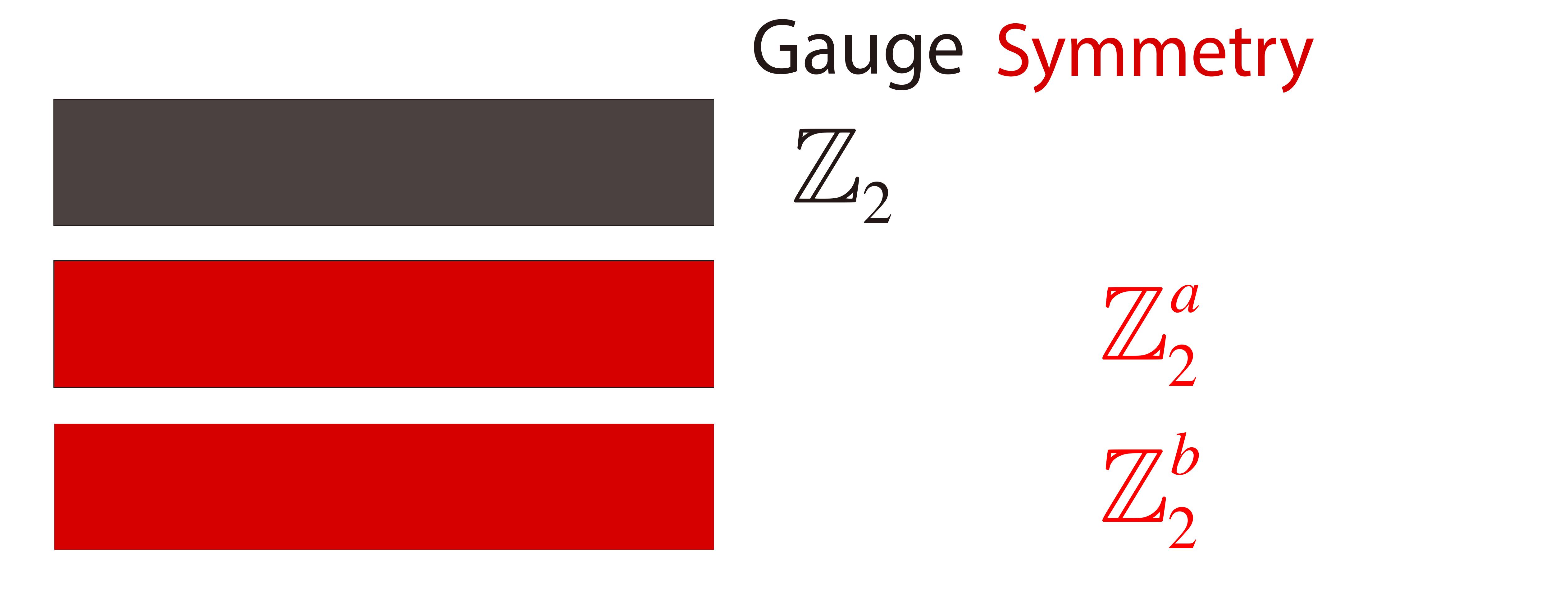}}}                                                                                                                                                                                                                                                                         \\
               & \multicolumn{9}{l|}{  }   \\ 
               & \multicolumn{9}{l|}{  }   \\
               & \multicolumn{9}{l|}{  }   \\ 
               & \multicolumn{9}{l|}{  }   \\
               & \multicolumn{9}{l|}{  }   \\  \hline
\multirow{2}{*}{$\mathsf{GT}$} &{$ a^1a^2da^2 $}&{$ a^2a^1da^1$}& $ a^1a^3da^3$ & $ a^3a^1da^1$ &$ a^2a^3da^3$&$ a^3a^2da^2$& $ a^1a^2da^3$ &$a^2a^3da^1$ &\\
\cline{2-10 }&0 mod 2&0 mod 2&0 mod 2&0 mod 2&0 mod 2&0 mod 2&0 mod 2&0 mod 2&\\
\hline
\multirow{2}{*}{$\mathsf{SEG}$} &\multirow{2}{*}{}0 mod 4&\multirow{2}{*}{}0 mod 4&0 mod 4&0 mod 4&\multirow{2}{*}{}0 mod 4&{0 mod 4}&{0 mod 4}&{0 mod 4}& \\
\cline{2-9} & 2 mod 4&2 mod 4&2 mod 4&2 mod 4&2 mod 4 &2 mod 4& 2 mod 4&2 mod 4 & $2^8$\\
\hline\hline
\end{tabular}

\end{table}

\begin{table}\caption{$\mathsf{SEG}$ $(\Z_2,  \Z_2 \times \Z_2)$ \small  $G_s=\Z_2\times\Z_2$ is carried entirely by either the first layer (the first subtable) or the second layer (the second subtable).}\label{table:z2_z2z2_3}

\begin{tabular}{|c|c|c|c|}
\hline\hline
                   \multirow{4}{*}{ \begin{minipage}   {0.65in}Symmetry assignment\end{minipage}}  & \multicolumn{3}{c|}{ \multirow{4}{*}{ \includegraphics[width=5cm]{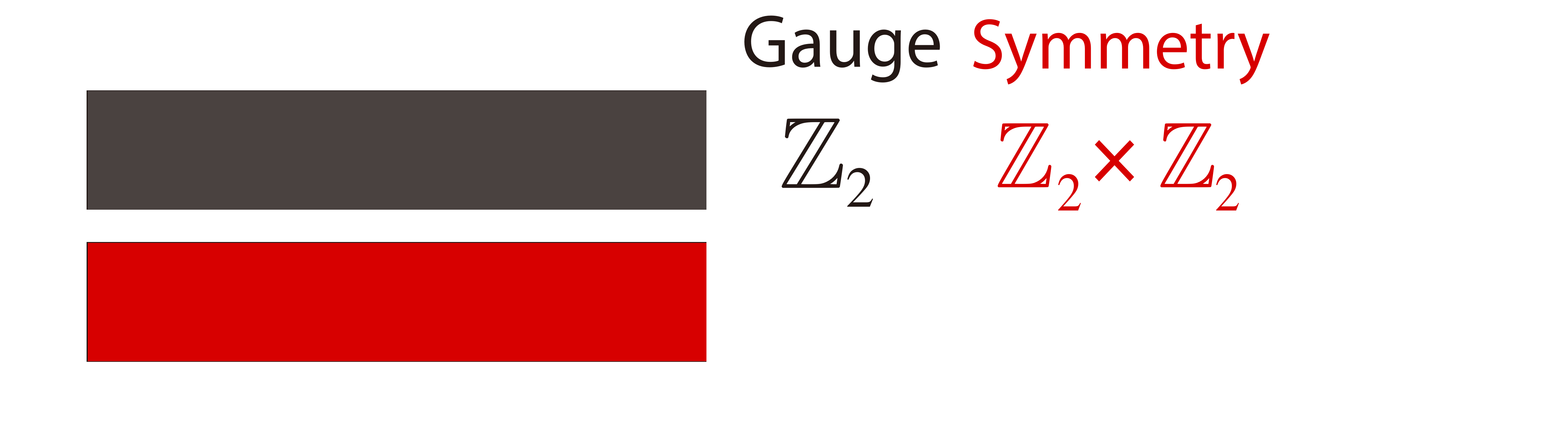}  }}     \\
                    & \multicolumn{3}{l|}{  }   \\ 
                    & \multicolumn{3}{l|}{  }   \\
                    & \multicolumn{3}{l|}{  }   \\ 
                   \hline
\multirow{2}{*}{$\mathsf{GT}$}   & $ q/4 \pi^2 a^1a^2da^2$ & $ \bar{q} /4 \pi^2 a^2a^1da^1$  &  \\ \cline{2-4} 
                      & 0 mod 2                                  & 0 mod 2                              &   \\ \hline
$\mathsf{SEG}$ &              0 mod 4               &       0 mod 4              &  1     \\ 
              &                     &                &       \\ 
                    \hline\hline
\end{tabular} 
\begin{tabular}{|c|c|c|c|}
\hline\hline
                   \multirow{4}{*}{ \begin{minipage}   {0.65in}Symmetry assignment\end{minipage}}  & \multicolumn{3}{c|}{ \multirow{4}{*}{ \includegraphics[width=5cm]{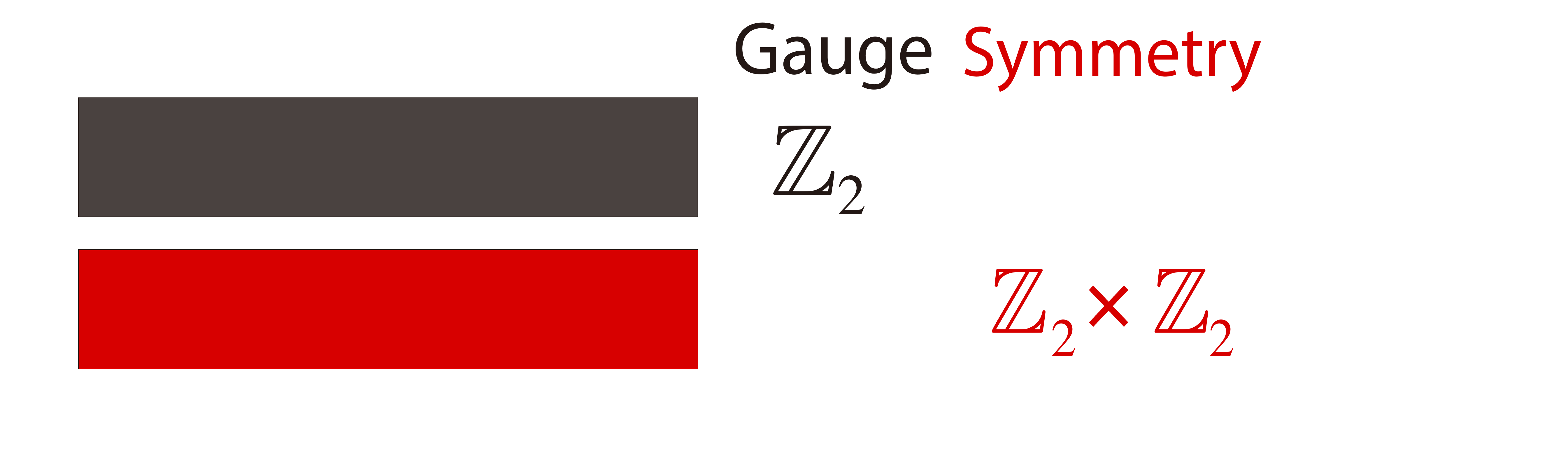}  }}     \\
                    & \multicolumn{3}{l|}{  }   \\ 
                    & \multicolumn{3}{l|}{  }   \\
                    & \multicolumn{3}{l|}{  }   \\ 
                   \hline
\multirow{2}{*}{$\mathsf{GT}$}   & $ q/4 \pi^2 a^1a^2da^2$ & $ \bar{q} /4 \pi^2 a^2a^1da^1$  & \\ \cline{2-4} 
                      & 0 mod 2                                  & 0 mod 2                              &   \\ \hline
\multirow{2}{*}{$\mathsf{SEG}$} &              0 mod 4               &       0 mod 4              &   $2^2$     \\ 
&  2 mod 4&  2 mod 4 &   \\
                    \hline\hline
\end{tabular}
\end{table}

   %%%%% 

 \subsection{$\mathsf{SEG}(\Z_{N}, \mathrm{U(1)})$}\label{appendix_sub_zn_u1_12}

  To impose the $\mathrm{U(1)}$ symmetry to the $\Z_N$ gauge theory, there are two ways, i.e. two symmetry assignments. The first one is to assign the symmetry at the same layer as that where $\Z_N$ gauge lives. For this symmetry assignment, it is equivalent to $\mathsf{SET}$ $N_1=N$,$N_2=1$,$K=1$ in the Appendix \ref{appendix_subsection_zn1zn2_zku1_1},  so there is only one $\mathsf{SEG}(\Z_N,\mathrm{U(1)})$. The other way is to assign it at another layer whose BF term is level-one, which is equivalent to $\mathsf{SET}$  $N_1=N$,$N_2=1$,$K=1$ in the Appendix \ref{appendix_subsection_zn1zn2_zku1_2}, so there is also only one $\mathsf{SEG}(\Z_N,\mathrm{U(1)})$.

    \subsection{$\mathsf{SEG}(\Z_{N},\Z_{K}\times \mathrm{U(1)})$}

For the $\Z_{N}$ gauge enriched by $\Z_{K}\times \mathrm{U(1)}$ symmetry, there are five symmetry assignments in Table~\ref{table:assignment_zN_U(1)}. Four of them only involve two layers which all have only one $\mathsf{SEG}$.  The fifth symmetry assignment gives rise to $[\text{GCD}(N,K)]^3$. As we would see below, two roots of $[\text{GCD}(N,K)]^3$ come from the stacking of $\mathsf{SEG}(\Z_{N}, \Z_{K})$   and a direct product state with $\mathrm{U(1)}$ symmetry (n.b., $\mathrm{U(1)}$ $\mathsf{SPT}$in 3D is always trivial). The third root comes from the nontrivial interaction $a^1a^2da^3$ which correlates all layers together. Note that since the layers where the symmetry are assigned are level-one, exchanging the $\Z_K$ and $\mathrm{U(1)}$ symmetry does not lead to anything new.

\begin{table}[H]\caption{The five symmetry assignments of $\Z_{N} $ Gauge with $\Z_{K}\times \mathrm{U(1)}$ symmetry and the number of corresponding $\mathsf{SEG}$.}\label{table:assignment_zN_U(1)}
\centering
\begin{tabular}{|c|c|c|c|c|c|}
\hline
\hline
&I&II&III&IV&V\\
\multirow{4}{*}{}& \multirow{4}{*}{\includegraphics[width=1.2 in]{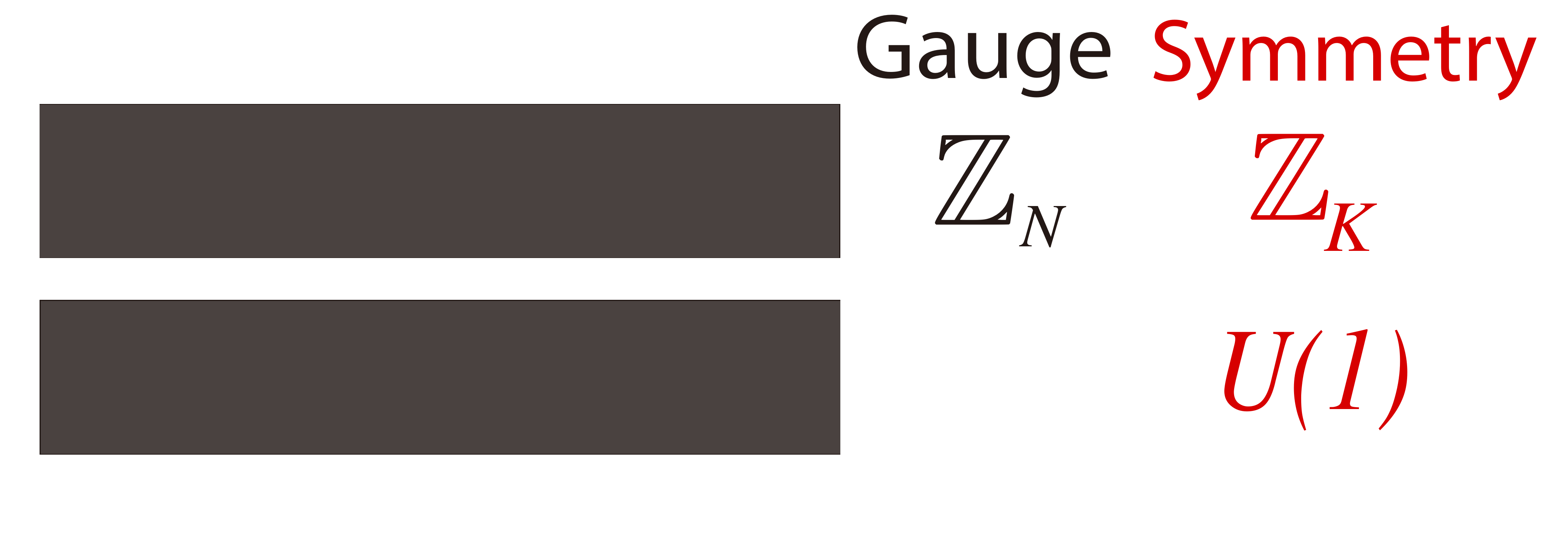}} &\multirow{4}{*}{\includegraphics[width=1.3 in]{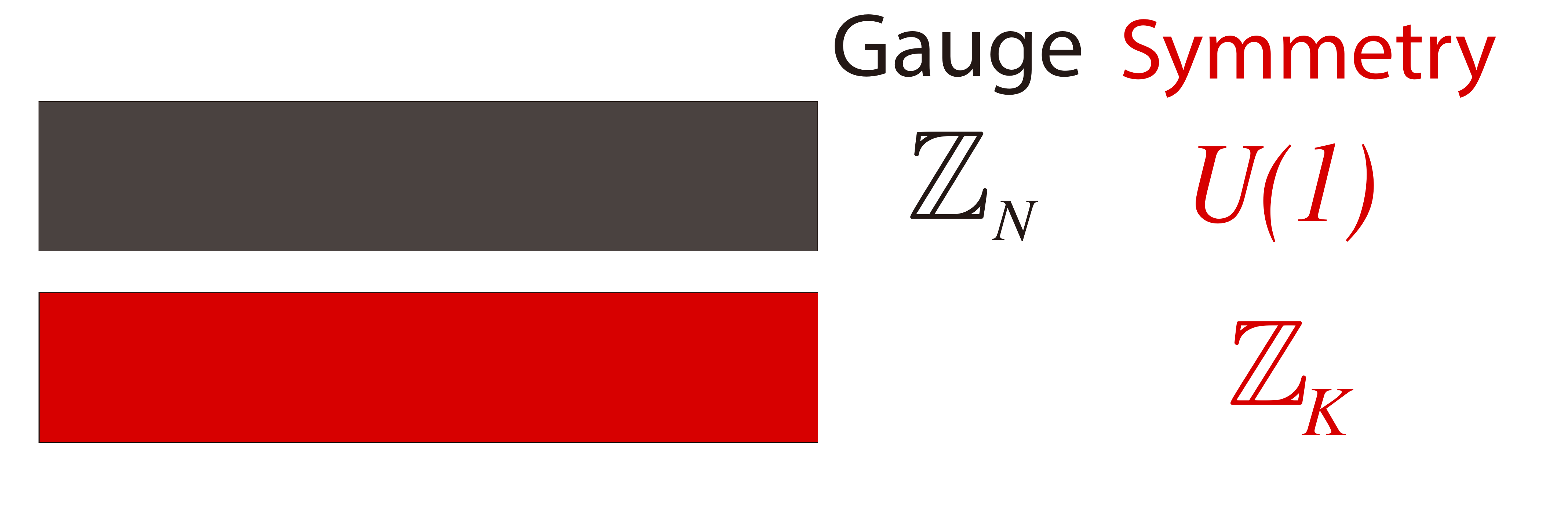}}&\multirow{4}{*}{\includegraphics[width=1.3 in]{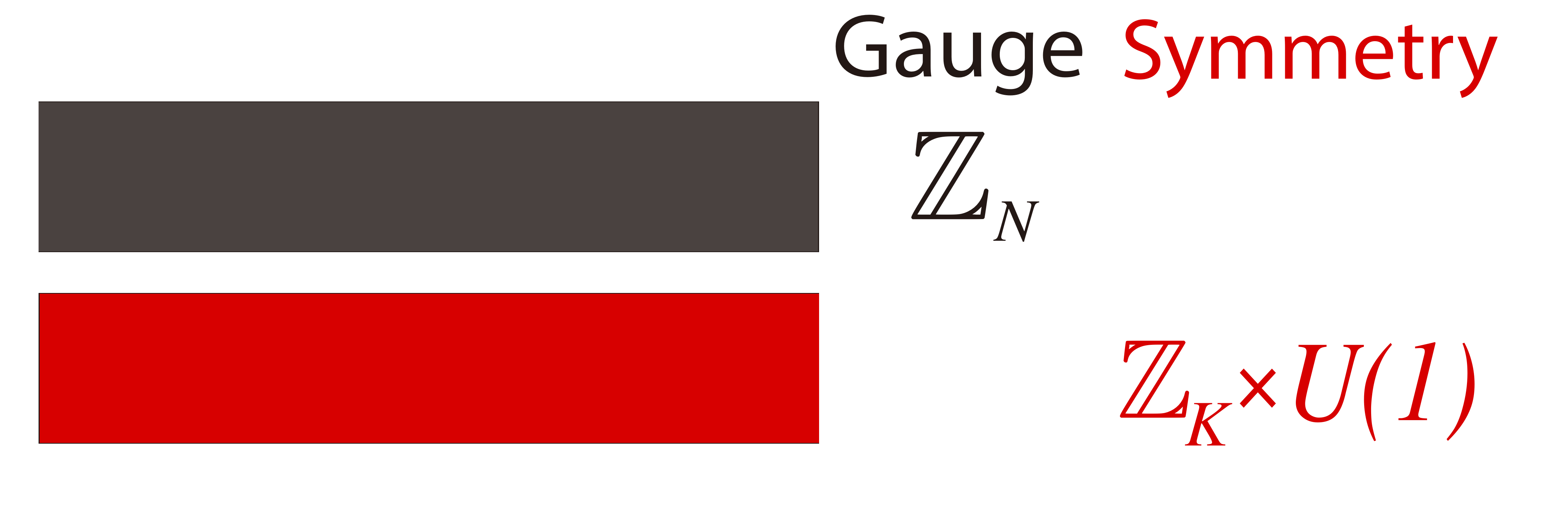}}&\multirow{4}{*}{\includegraphics[width=1.2 in]{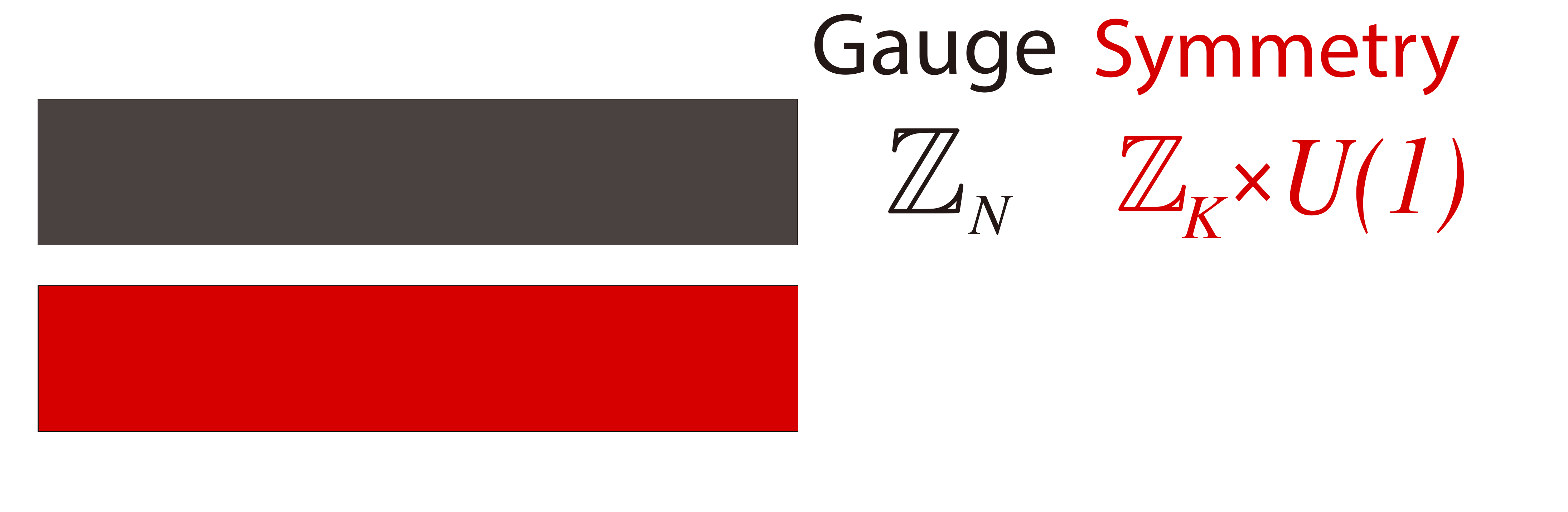}}&\multirow{5}{*}{\includegraphics[width=1.2 in]{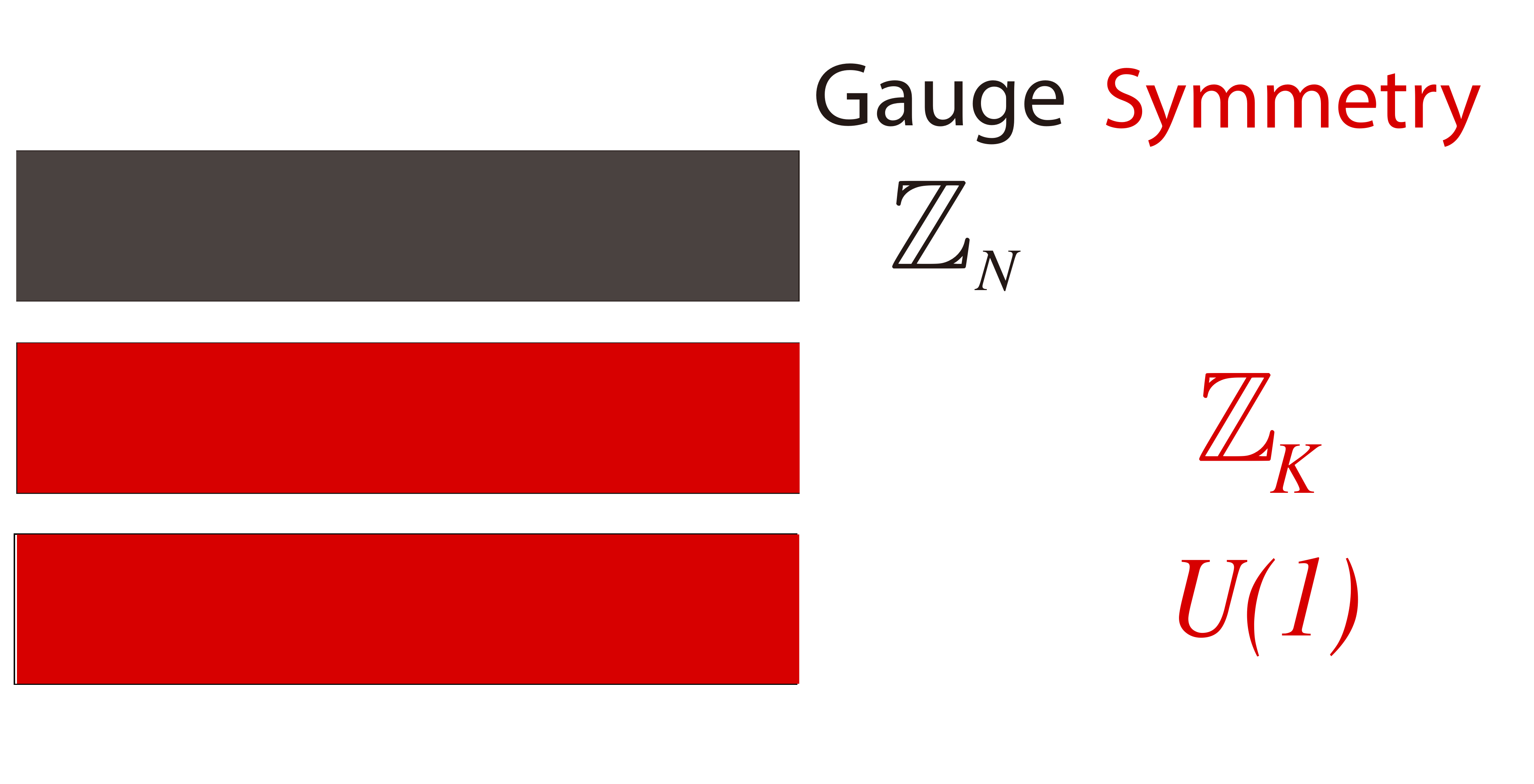}} \\
Symmetry  &&&&&\\
Assignment&&&&&\\
&&&&&\\
\hline
\multirow{2}{*}{$\mathsf{SEG}$}&\multirow{2}{*}{1}&\multirow{2}{*}{1}&\multirow{2}{*}{1}&\multirow{2}{*}{1}&\multirow{2}{*}{$[\text{GCD}(N,K)]^3$}\\
&&&&&\\
\hline
\hline
\end{tabular}
\end{table}
 We main focus on the symmetry assignment V in Table~\ref{table:assignment_zN_U(1)}.  As there are three layers that we have to take into account, there are in total 8 $aada$ type topological interaction terms.  To count the total number of    $\mathsf{SEG}$s in this symmetry assignment, we have to determine the period of of the coefficients of these eight topological interaction terms. Below we consider each of them separately because each alone can determine a set of root $\mathsf{SEG}$. 
   \begin{enumerate}
\item For topological interaction $ a^1a^2da^2$ or $ a^2a^1da^1$,  the theory reduces to that of stacking $\mathsf{SEG}(\Z_N,\Z_K)$ and $\mathrm{U(1)}$ $\mathsf{SPT}$in three dimensions. From the calculation in Appendix \ref{appendix_sub_n1n2_k1k2_1} by setting $N_1=N,N_2=1,K_1=1,K_2=K$, there are $\text{GCD}(N,K)$ different root $\mathsf{SEG}(\Z_N,\Z_K)$s from $a^1a^2da^2$ and another  $\text{GCD}(N,K)$ root $\mathsf{SEG}(\Z_N,\Z_K)$s from $a^2a^1da^1$. From Ref.~\cite{spt6} there is only one $\mathrm{U(1)}$ $\mathsf{SPT}$in three dimensions. Therefore, there are $\text{GCD}(N,K)$ different $\mathsf{SEG}(\Z_N,\Z_K\times \mathrm{U(1)})$s from the topological interaction $a^1a^2da^2$ and another $\text{GCD}(N,K)$ root $\mathsf{SEG}(\Z_N,\Z_K\times \mathrm{U(1)} )$s from $a^2a^1da^1$.    
\item      For topological interaction $a^1a^3da^3$ or $a^3a^1da^1$, the $\mathsf{SEG}(\Z_N,\Z_K\times \mathrm{U(1)})$ reduces to the stacking of $\mathsf{SEG}(\Z_N,\mathrm{U(1)})$ and $\Z_K$ $\mathsf{SPT}$in three dimensions. From the result in Appendix \ref{appendix_sub_zn_u1_12} (when $\Z_N$ and $\mathrm{U(1)}$ are not in the same layer), we know that there is only one $\mathsf{SEG}(\Z_N,\mathrm{U(1)})$ and from Ref.~\cite{spt6}, there is only one $\Z_K$ SPT. Therefore, there is only one root $\mathsf{SEG}(\Z_N,\Z_K\times \mathrm{U(1)})$ from $a^1a^3da^3$ and also only one from $a^3a^1da^1$. 
\item For topological interaction $a^2a^3da^3$ or $a^3a^2da^2$, the $\mathsf{SEG}(\Z_N,\Z_K\times \mathrm{U(1)})$ reduces to the stacking of $\Z_N$ gauge theory and $\Z_K\times\mathrm{U(1)}$ $\mathsf{SPT}$in three dimension. It is known that there is only one $\Z_N$ gauge theory and from Ref.~\cite{spt6}, there is only one $\Z_K\times \mathrm{U(1)}$ SPT. Therefore, there is only one root $\mathsf{SEG}(\Z_N,\Z_K\times \mathrm{U(1)})$ from $a^2a^3da^3$ and also only one from $a^3a^2da^2$. 
\item For topological interaction $a^1a^2da^3$,  the symmetry assignment V is equivalent to $\mathsf{SET}$ $N_1=N$,$N_2=N_3=1$ and $K_1=1,K_2=K$ in Appendix \ref{appendix_sub_zn1zn2_k1k2u1_123}. Therefore,  there are $\text{GCD}(N,K)$ $\mathsf{SEG}(\Z_N, \Z_K \times \mathrm{U(1)})$s in total. For another three-layer topological interaction $a^2a^3da^1$,  it is equivalent to exchange the layer index as $1 \longleftrightarrow 3$, $2 \longleftrightarrow 1$, $3\longleftrightarrow 2$ in Appendix. \ref{appendix_sub_zn1zn2_k1k2u1_123}.  Employing the similar procedure as those for $a^1a^2da^3$, we find that the $q=0$, and so there is only one $\mathsf{SEG}(\Z_N,\Z_K\times \mathrm{U(1)})$. 
\end{enumerate}

In summary, for symmetry assignment V, each of $a^1a^3d^3$,$a^3a^1da^1$,$a^2a^3da^3$,$a^3a^2da^2$ and $a^2a^3da^1$ contributes  only one root $\mathsf{SEG}(\Z_N,\Z_K\times \mathrm{U(1)} )$ and each of $a^1a^2da^2$, $a^2a^1da^1$ and $a^1a^2da^3$ contributes $\text{GCD}(N,K)$ root $\mathsf{SEG}(\Z_N,\Z_K\times \mathrm{U(1)})$, so in total there are $[\text{GCD}(N,K)]^3$ $\mathsf{SEG}(\Z_N,\Z_K\times \mathrm{U(1)})$s for the symmetry assignment in Table~\ref{table:assignment_zN_U(1)}.

\subsection{$\mathsf{SEG}(\Z_{N_1} \times \Z_{N_2},\mathrm{U(1)})$}

Without $\mathrm{U(1)}$ symmetry, there are in total $(N_{12})^2$  $\Z_{N_1}\times \Z_{N_2}$ gauge theories, where $N_{12}$ is the greatest common divisor of $N_1$ and $N_2$. With the $\mathrm{U(1)}$ symmetry, there are three symmetry assignments, as shown in Table~\ref{table:assignment_zNzN_U(1)}. For the assignment I and II, it  is equivalent to $\mathsf{SET}$ $K=1$ in Appendix \ref{appendix_subsection_zn1zn2_zku1_1}. so there is only one $\mathsf{SEG}$ whose parent gauge theory is untwisted $\Z_{N_1}\times\Z_{N_2}$ gauge theory.

\begin{table}[H]\caption{The symmetry assignments of $\Z_{N_1} \times \Z_{N_2}$ gauge with $\mathrm{U(1)}$ symmetry and the numbers of corresponding gauge theory and symmetry enriched gauge theory. }
\label{table:assignment_zNzN_U(1)}
\centering
\begin{tabular}{|c|c|c|c|}
\hline
\hline
&I&II&III \\
\multirow{4}{*}{} &\multirow{4}{*}{\includegraphics[width=1.2 in]{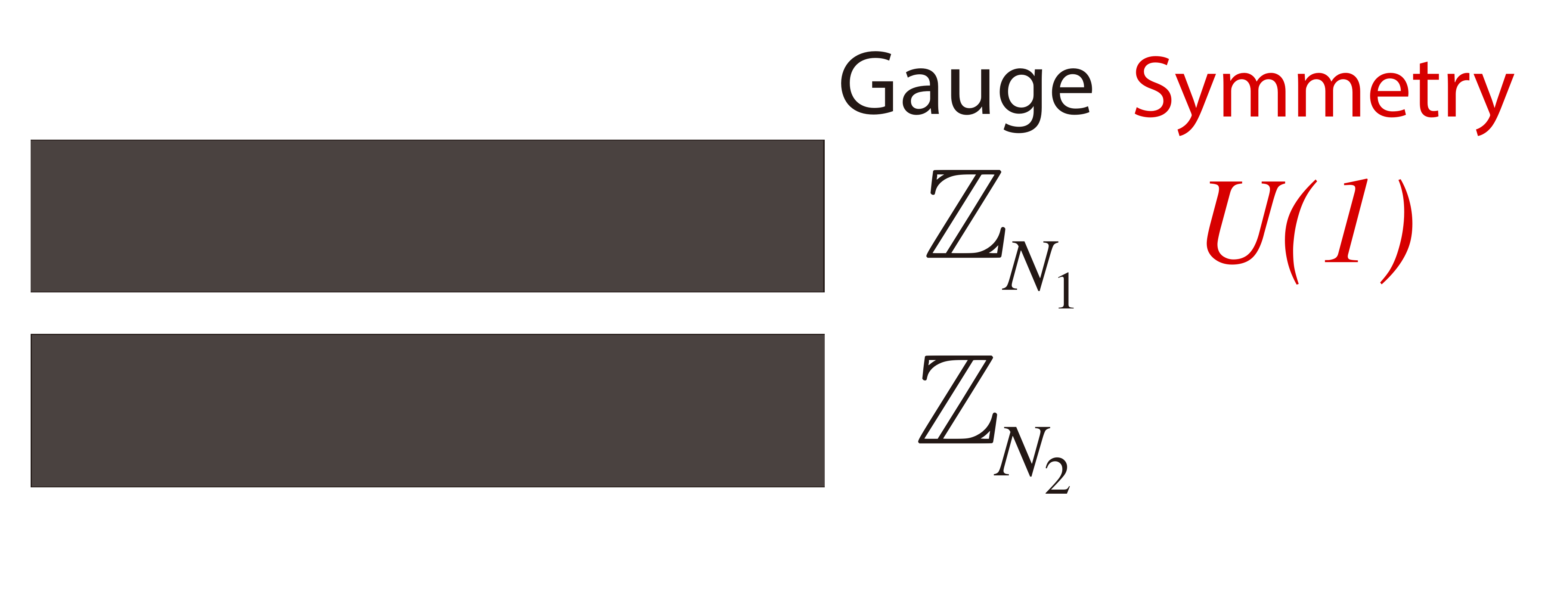}} &\multirow{4}{*}{\includegraphics[width=1.2 in]{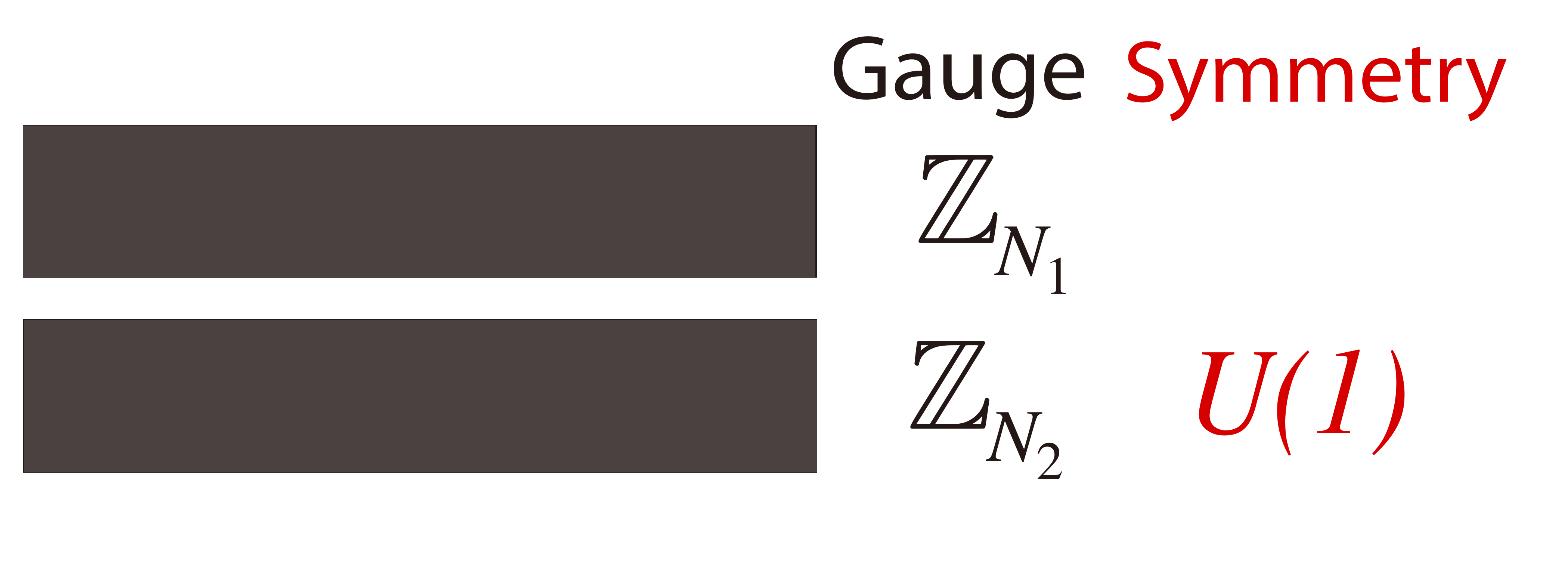}}  &\multirow{4}{*}{\includegraphics[width=1.2 in]{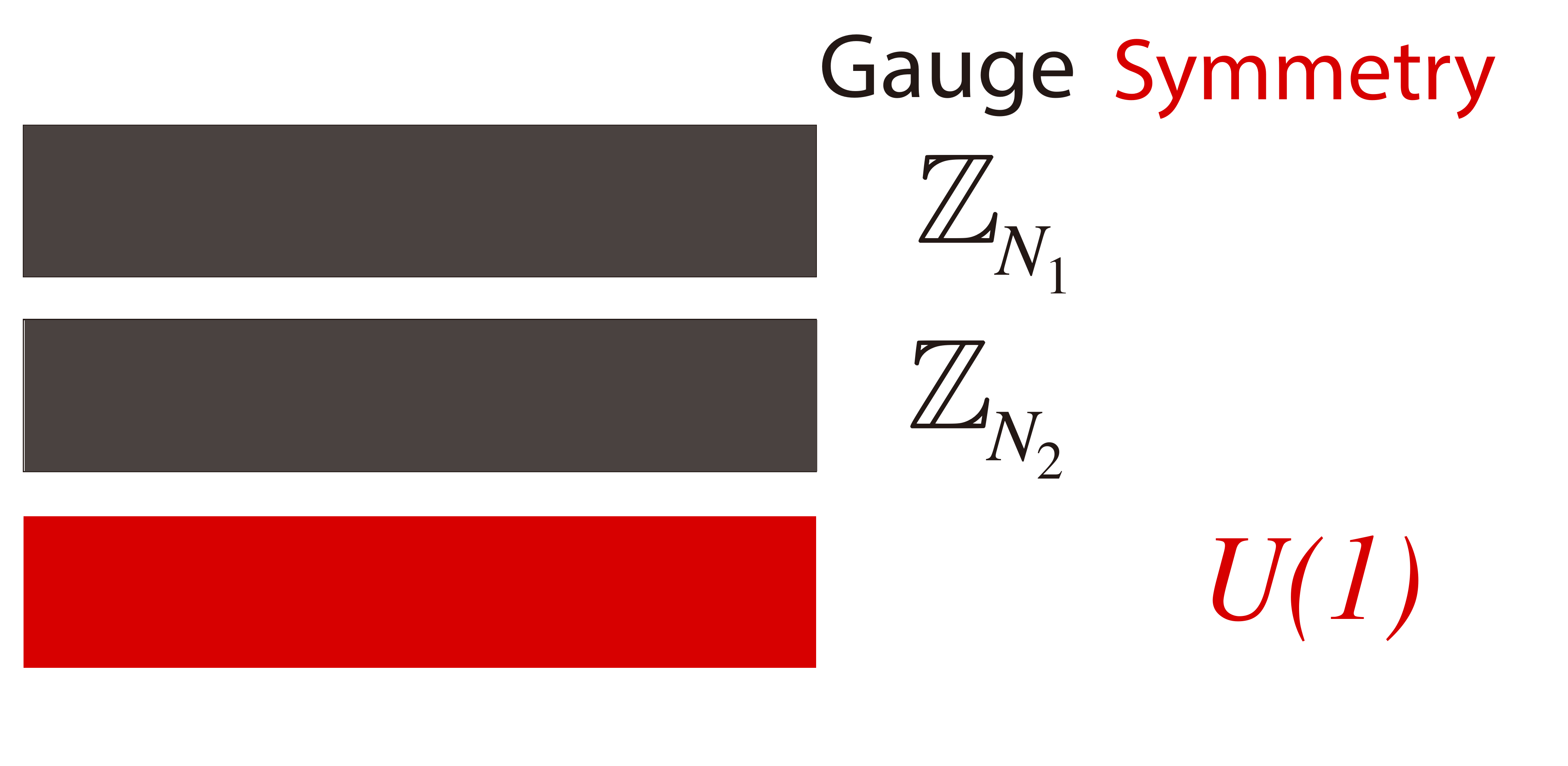}}  \\
Symmetry & && \\
assignment & & & \\
&&&\\
\hline
\multirow{2}{*}{$\mathsf{SEG}$} &\multirow{2}{*}{1} & \multirow{2}{*}{1} & \multirow{2}{*}{$(N_{12})^3$}\\
&&&\\
\hline \hline
\end{tabular}
\end{table}

  For the assignment III, the number of $\mathsf{SEG}(\Z_{N_1} \times \Z_{N_2}, \mathrm{U(1)})$ is $(N_{12})^3$ compared to the $(N_{12})^2$ gauge theories.   
For topological interaction $a^1a^2da^2$ or $a^2a^1da^1$,  the root $\mathsf{SEG}(\Z_{N_1} \times \Z_{N_2},\mathrm{U(1)})$ are just stacking the $\Z_{N_1} \times \Z_{N_2}$ root gauge theories and $\mathrm{U(1)}$ $\mathsf{SPT}$in three dimension. We know that there are $(N_{12})^2$ $\Z_{N_1} \times \Z_{N_2}$ gauge theories and only one $\mathrm{U(1)}$ $\mathsf{SPT}$in three dimensions. Therefore there are $N_{12}$ root $\mathsf{SEG}(\Z_{N_1}\times \Z_{N_2},\mathrm{U(1)})$ from $a^1a^2da^2$ and another  $N_{12}$ root $\mathsf{SEG}(\Z_{N_1}\times \Z_{N_2},\mathrm{U(1)})$ from $a^2a^1da^1$.  

  For the choice of interaction $a^1a^3da^3$, $a^3a^1da^1$, $a^2a^3da^3$ or $a^3a^2da^2$,  there is only one $\mathsf{SEG}(\Z_{N_1}\times \Z_{N_2},\mathrm{U(1)})$ for all cases.
  
  For topological interaction $a^1a^2da^3$, it is equivalent to $\mathsf{SET}$ $N_3=K_1=K_2=1$ in Appendix~\ref{appendix_sub_zn1zn2_k1k2u1_123}, so there are $N_{12}$ $\mathsf{SEG}(\Z_{N_1}\times \Z_{N_2},\mathrm{U(1)})$s. But for $a^2a^3da^1$, it is equivalent to   exchange the layer index as $1 \longleftrightarrow 3$, $2 \longleftrightarrow 1$, $3\longleftrightarrow 2$ in Appendix~\ref{appendix_sub_zn1zn2_k1k2u1_123}. Employing the similar procedure as those for $a^1a^2da^3$, we find that $q=0$ and so there is only one $\mathsf{SEG}(\Z_N,\Z_K\times \mathrm{U(1)})$.

In summary,   for symmetry assignment III, each of $a^1a^3d^3$,$a^3a^1da^1$,$a^2a^3da^3$,$a^3a^2da^2$ and $a^2a^3da^1$ contributes  only one root $\mathsf{SEG}(\Z_{N_1}\times \Z_{N_2}, \mathrm{U(1)})$ and each of $a^1a^2da^2$, $a^2a^1da^1$ and $a^1a^2da^3$ contributes $N_{12}$ root $\mathsf{SEG}(\Z_{N_1}\times \Z_{N_2}, \mathrm{U(1)})$, so in total there are $(N_{12})^3$ $\mathsf{SEG}(\Z_{N_1}\times \Z_{N_2}, \mathrm{U(1)})$ for the symmetry assignment III in Table~\ref{table:assignment_zNzN_U(1)}.
     
%     
%\section{Calculation of $\mathcal H^2(\Z_2,\Z_2)$}\label{appendix_h2z2z2}
%XXXXXX
%XXXXX
%
%XXX
%
%
%XX
%
%
%xx

\section{Calculation of $\mathcal H^2(\Z_2,\Z_2)=\Z_2$ and  $\mathcal H^2(\Z_3,\Z_2)=\Z_1$ }\label{appendix_h2z3z2}

In this Appendix, we calculate the second group cohomology $\mathcal H^2(G_s,G_g)$ which describes topologically distinct patterns of  $G_s$ symmetry fractionalization in the charge of $G_g$ (which is abelian) gauge field. Mathematically, $\mathcal H^2(G_s,G_g)$ is a set of equivalent classes of 2-cocycles $\omega_2(g_1,g_2)$, where $g_1,g_2\in G_s$ and $\omega_2(g_1,g_2)$ are $G_g$ valued. The 2-cocycles are solutions of the 2-cocycle equations:
\begin{align}
d\omega_2(g_1,g_2,g_3) &= \omega_2(g_2,g_3)\omega_2^{-1}(g_1g_2,g_3)\omega_2(g_1,g_2g_3)\omega_2^{-1}(g_1,g_2) \nonumber\\
&=1.
\end{align}
 If $G_g=\mathbb Z_2$ , then $\omega_2(g_1,g_2)$ takes value $\pm1$. Two 2-cocycles $\omega_2'(g_1,g_2)$ and $\omega_2(g_1,g_2)$ are equivalent if they differ by a 2-coboundary $\omega_2'(g_1,g_2)=\omega_2(g_1,g_2)\Omega_2(g_1,g_2)$, with
\begin{eqnarray}\label{cb}
\Omega_2(g_1,g_2)={\Omega_1(g_1)\Omega_1(g_2)\over\Omega_1(g_1g_2)},
\end{eqnarray}
where $\Omega_1(g)$ are $G_g$ variables. A 2-cocycle is said to be trivial if it is equivalent to $\omega_2(g_1,g_2)=1$ for all $g_1,g_2\in G_s$.

In the following we adopting the canonical gauge choice\cite{Chenlong} such that $\omega_2(E,g)=\omega_2(g,E)\equiv1$. To ensure that this is still the case after a gauge transformation, namely, to ensure $\omega_2'(g,E)=\omega_2(g,E)\Omega_2(g,E)=1$ still holds, $\Omega_1(E)\equiv1$ is required. 

Now we calculate two simple examples  $\mathcal H^2(\mathbb Z_2,\mathbb Z_2)$ and  $\mathcal H^2(\mathbb Z_3,\mathbb Z_2)$ using above definition. 

{\bf Cohomology $\mathcal H^2(\mathbb Z_2,\mathbb Z_2)$}. If $G_s=\mathbb Z_2=\{E,Q\}$, then there is only one 2-cocycle equation,
\[
d\omega_2(Q,Q,Q)=\omega_2(Q,Q)\omega_2^{-1}(E,Q)\omega_2(Q,E)\omega_2^{-1}(Q,Q)=1.
\]
Since $\omega_2(E,Q)=\omega_2(Q,E)=1$, above equation gives no constraint for the variable $\omega_2(Q,Q)$. Since $G_g=\mathbb Z_2$, $\omega_2(Q,Q)$ is a free $\mathbb Z_2$ variable and can freely take values $\pm1$.
On the other hand, the 2-coboundary 
\[
\Omega_2(Q,Q)={\Omega_1(Q)\Omega_1(Q)\over \Omega_1(E)}=1
\]
is trivial, so there is no gauge degrees of freedom under the canonical gauge condition. This means that $\omega_2(Q,Q)=1$ and $\omega_2(Q,Q)=-1$ stand for two different classes of 2-cocycles, which yields the result
\[
\mathcal H^2(\mathbb Z_2, \mathbb Z_2)=\mathbb Z_2.
\]

{\bf Cohomology $\mathcal H^2(\mathbb Z_3,\mathbb Z_2)$}. If $G_s=\mathbb Z_3=\{E,P,P^2\}$, substituting $g_1,g_2,g_3$ by $P,P^2$, we obtain eight equations, two of which are independent. The first two equations are 
\begin{eqnarray}
\omega_2(P,P)\omega_2^{-1}(P^2,P)\omega_2(P,P^2)\omega_2^{-1}(P,P)=1,\nonumber\\
\omega_2(P,P^2)\omega_2^{-1}(P^2,P^2)\omega_2(P,E)\omega_2^{-1}(P,P)=1.\nonumber
\end{eqnarray}

We obtain,
\begin{eqnarray*}
&&\omega_2(P,P^2)=\omega_2(P^2,P),\\
&&\omega_2(P,P)\omega_2(P^2,P^2)=\omega_2(P,P^2).
\end{eqnarray*}

If we let $\omega_2(P,P)=\sigma,\  \omega_2(P^2,P^2)=\eta$, where $\sigma,\eta$ are $G_g=\mathbb Z_2$ variables, then $\omega_2(P,P^2)=\sigma\eta$.

On the other hand, from equation (\ref{cb}), we obtain, 
\begin{eqnarray}
&&\Omega_2(P,P)={\Omega_1(P)\Omega_1(P)\over \Omega_1(P^2)}=\Omega_1(P^2),\nonumber\\
&&\Omega_2(P,P^2)=\Omega_2(P^2,P)=\Omega_1(P)\Omega_1(P^2),\nonumber\\
&&\Omega_2(P^2,P^2)={\Omega_1(P^2)\Omega_1(P^2)\over \Omega_1(P)}=\Omega_1(P).\nonumber
\end{eqnarray}

If we chose $\Omega_1(P^2)=\sigma$, $\Omega_1(P)=\eta$, then we obtain a new 2-cocyle
\[
\omega_2'(g_1,g_2)=\omega_2(g_1,g_2)\Omega_2(g_1,g_2)=1
\] 
for all $g_1,g_2\in \mathbb Z_3$. Thus we have shown that these 2-cocyles are trivial, namely, 
\[
\mathcal H^2(\mathbb Z_3, \mathbb Z_2)=\mathbb Z_1.
\]

\clearpage
 
 \end{widetext}


\begin{thebibliography}{139}
\bibitem{footnote_deconfined} It should be noted that confined phases may also support topological orders. An example is   shown in a parton construction in \cite{3dset_ye} where topological order is given by untwisted $\Z_2\times\Z_6$ gauge group.   Throughout the paper, only deconfined regime of gauge theories will be taken into consideration. In deconfined phases, the gauge fluxes and gauge particles naturally behave as emergent excitations of topological orders.
\bibitem{footnote_anomaly} For an anomaly-free effective field theory in condensed matter sense, these two conditions are necessary but their sufficiency is problematic as the UV definition of a continuum field theory should be justified by lattice models with local interactions. Although $\mathsf{GT}$s have Dijkgraaf-Witten lattice model correspondence, symmetry enriched ones, i.e., $\mathsf{SEG}$s should be more carefully investigated. The future studies of lattice models of $\mathsf{SEG}$s will be interesting.  Throughout the paper, we assume sufficiency holds but with the above caveat.
\bibitem{footnote_gauging} Unless otherwise specified, throughout the paper, we try to avoid to use the terminology like ``gauging'', ``fully gauging'', ``partially gauging'', ``weakly gauging'', ``orbifolding''. In literatures, these terms are sometimes   mixed with ``symmetry twist'' (i.e., response theory). By ``giving full dynamics to...'' in this paper, we really mean that, mathematically, a new path-integral measure $D A^1D A^2\cdots$ is introduced.
\bibitem{footnote_4loop} In general, mixed version of four-loop braiding process should be considered if the total subgroups of symmetry and gauge are no less than 4. In this paper, mixed three-loop process is already enough for all examples we consider.
\bibitem{footnote_zeromode} If $\text{Det} \Lambda=0$, then the zero eigenvalues give rise to   gapless photon modes described by the sub-leading Maxwell action term.
\bibitem{footnote_symmetry}In this paper, one main goal is to present the general methods of constructing  path-integral formalism of $\mathsf{SEG}$s  and diagnose $\mathsf{SET}$ orders, rather than completeness of classification. We only consider the cases where  topological currents carry unit  symmetry charge while generically they can carry charge-2 or more. Fortunately, for $\Z_2$ symmetry considered in the main text, $1$ is sufficient since $2\sim 0$ in $\Z_2$ symmetry. Furthermore, we  also do not discuss the cases in which there exist more than one layers that couple to the same background gauge field. In principle, if all these cases are involved, we can obtain the complete classification based on the current working assumptions, which will be leaved to future work.
\bibitem{footnote_8_topo} As a side note, there are only two linearly independent three-layer topological interaction terms since $a^3a^1da^2$ is $a^1a^2da^3+a^2a^3da^1$ up to a total derivative.
\end{thebibliography}
\end{document}